\documentclass[preprint,showpacs,preprintnumbers]{revtex4}
\usepackage{amsmath}
\usepackage{graphicx}
\usepackage{dcolumn}
\usepackage{bm}
\usepackage{epsfig}
\usepackage[T1]{fontenc}
\usepackage{ae,aecompl}
\usepackage{epstopdf, epsfig}
\usepackage{color}
\usepackage{appendix}
\usepackage{upgreek}
\usepackage{amsfonts}
\usepackage{mathrsfs}
\usepackage{wasysym}
\usepackage{amssymb}
\usepackage{color}
\usepackage{graphicx}
\usepackage{xspace}
\usepackage{enumerate}
\usepackage{tikz,fp}
\usepackage{tikz-cd}

\usepackage{pstricks}
\usepackage{hyperref}
\usepackage{dsfont}
\usepackage{pifont}
\usepackage{multirow}
\usepackage{balance}
\usepackage{makecell}
\usepackage[version=4]{mhchem}
\usepackage{picins}
\usepackage[figuresright]{rotating}
\usepackage{booktabs}
\usepackage{array}
\usepackage{mathtools}
\usepackage{verbatim}
\usepackage{url}
\usepackage{siunitx}
\usepackage{wrapfig}

\usepackage{lipsum}

\usepackage[absolute,overlay]{textpos}

\begin{document}

\title{Advances in the kinetics of heat and mass transfer in near-continuous complex flows*}
\author{Aiguo Xu$^{1,2,3}$\footnote{
Corresponding author. E-mail: Xu\_Aiguo@iapcm.ac.cn}, 
Dejia Zhang$^{1,4,5}$, 
Yanbiao Gan$^6$}
\affiliation{
1, National Key Laboratory of Computational Physics, 
Institute of Applied Physics and Computational Mathematics, 
P. O. Box 8009-26, Beijing 100088, P.R.China  \\
2, State Key Laboratory of Explosion Science and Technology, 
Beijing Institute of Technology, Beijing 100081, China \\
3, HEDPS, Center for Applied Physics and Technology, 
and College of Engineering, Peking University, Beijing 100871, China \\
4, State Key Laboratory for GeoMechanics and Deep Underground Engineering, 
China University of Mining and Technology, Beijing 100083, P.R.China\\
5, National Key Laboratory of Shock Wave and Detonation Physics, 
Mianyang 621999, P.R.China\\
6, Hebei Key Laboratory of Trans-Media Aerial Underwater Vehicle, School of Liberal Arts and Sciences, 
North China Institute of Aerospace Engineering, Langfang 065000, P.R.China
}
\date{\today }

\begin{abstract}

The study of macro continuous flow has a long history. Simultaneously, the exploration of heat and mass transfer in small systems with a particle number of several hundred or less has gained significant interest in the fields of statistical physics and nonlinear science.
However, due to absence of suitable methods, the understanding of mesoscale behavior 
situated between the aforementioned two scenarios, which challenges the physical function of traditional continuous fluid theory and exceeds the simulation capability of microscopic molecular dynamics method, remains considerably deficient.
This greatly restricts the evaluation of effects of mesoscale behavior and impedes the development of corresponding regulation techniques. To access the mesoscale behaviors, there are two ways: from large to small and from small to large.
Given the necessity to interface with the prevailing macroscopic continuous modeling currently used in the mechanical engineering community, 
our study of mesoscale behavior begins from the side closer to the macroscopic continuum, that is from large to small.
Focusing on some fundamental challenges encountered in modeling and analysis of near-continuous flows, we review the research progress of discrete Boltzmann method (DBM). 
The ideas and schemes of DBM in coarse-grained modeling and complex physical field analysis are introduced.
The relationships, particularly the differences, between DBM and traditional fluid modeling as well as other kinetic methods are discussed.
After verification and validation of the method, some applied researches including the development of various physical functions associated with discrete and non-equilibrium effects are illustrated. Future directions of DBM related studies are indicated.

\end{abstract}

\preprint{}
\maketitle

\section{Introduction}

Fluid is a general term encompassing both gases and liquids. The two fluids we are most familiar with are the atmosphere and water: the atmosphere surrounds the entire Earth, and 70\% of the Earth's surface is covered by water.
Since ancient times, water and air, on the one hand, have nurtured life.
On the other hand, it also brings natural disasters such as typhoons, floods and droughts to human beings from time to time.
Human daily life, production and other activities are inseparable from fluid.
In the struggle with nature and through production practice, human beings gradually accumulate experience and enhance understanding of fluid behavior.
The continuous summarization and enhancement of these understandings gradually formed the discipline of fluid mechanics.
The air and water we perceive in our daily lives appear to be continuous, and this perception is based on a macroscopic understanding of continuous flow.
However, with the emergence of molecular kinetic theory, people gradually realized that the sense of continuity is because the distances between molecules are exceedingly small compared to the scales we typically perceive in our daily lives.

In addition to the seemingly continuous flow of water and air, there exist various flows in our daily lives and industrial production that exhibit distinct discrete characteristics.
Examples include dust explosions, particle flow, traffic flow, and the movement of crowds.
For a uniform description, we treat people and cars as particles.
These flows are perceived as discrete because the distance between particles is no longer negligible relative to the scale of our everyday perception.
However, when observed on a significantly larger scale (for example, exceeding 1,000 times the distance between particles), these flows can also be considered continuous.
It can be seen that the fluid system itself is objective, but the description of continuous and discrete is dependent on the scale of the behavior we are concerned with. Hsue-shen Tsien, a famous Chinese scientist, first suggested that according to the Knudsen (Kn) number, the fluid behavior could be divided into continuous flow, slip flow, transition flow and free molecular flow~\cite{Tsie1946JAS}.
In addition, it should be noted that unlike microscopic molecular collisions, collisions between particles composed of a large number of molecules are dissipative.

The development of scientific theories typically follows a progression from simplicity to complexity.
The theory of fluid mechanics begins with a macroscopic continuous flow such as air and water in which collisions between microscopic particles do not dissipate.
Considering the stage of study and space limitation of current review, the continuous and discrete flows discussed in the following parts are ``simple'' cases in which there is no dissipation in the collisions of microscopic particles which constitute the fundamental units.
Interestingly, and somewhat, the study of heat and mass transfer in these ``simple'' cases has been, and is still going through a long process.

\begin{figure*}[htbp]
\center\includegraphics*
[width=0.8\textwidth]{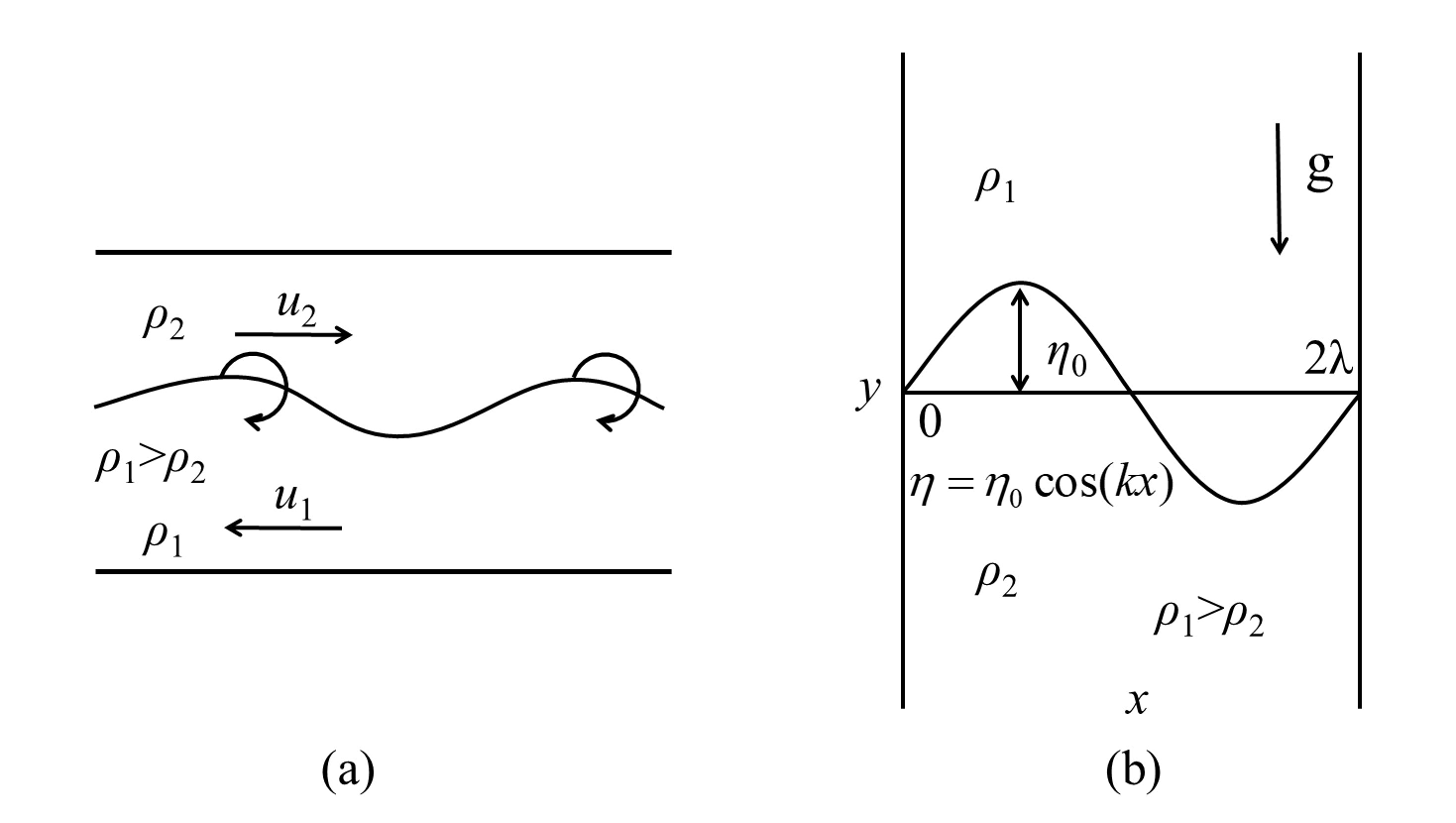}
\caption{ (a) Sketch of the formation mechanism of KHI.
(b) Sketch of the formation mechanism of RTI.
} \label{fig001}
\end{figure*}

To show that the study of these simple scenarios also contains a wealth of  yet unexplored complex behaviors, we begin with an everyday phenomenon: the fluid (interface) instability.
Fluid (interface) instability refers to the phenomenon that after the fluid interface is disturbed, the interface cannot automatically return to its original state under the action of the system's own force, and the disturbance amplitude increases with time.
Two  common types of fluid instabilities are the Kelvin-Helmholtz instability (KHI)~\cite{Kelvin1871PM, Helmholtz1868} and Rayleigh-Taylor instability(RTI)~\cite{Rayleigh1882PLMS,Taylor1950PRSLS}.
KHI occurs when there's a difference in tangential velocity on both sides of the interface, leading to phenomena like rolling sea waves and cirrus clouds.
RT instability refers to the phenomenon of interface instability that can be caused by any disturbance when acceleration is directed from a medium with high density to one with low density.
Due to the relativity of motion, another equivalent image of RT instability is the phenomenon of interface instability caused by the acceleration of a denser medium by a less dense medium.
Common examples of RT instability, such as interface instability when milk is poured into a glass of water, volcanic eruptions, mushroom clouds, etc.
Figure \ref{fig001} shows the mechanical causes of the formation of KHI and RTI.
In addition, there is a class of fluid (interface) instability known as Richtmyer-Meshkov instability (RMI)~\cite{Richtmyer1960CPAM, Meshkov1969SFD}.
It refers to the phenomenon that the interface disturbance  cannot naturally recover by the system itself under the action of shock wave, and the disturbance amplitude increases gradually with time.
In fact, RMI can be regarded as a special case of RTI when the acceleration takes the form of a pulse.
At the same time, during the evolution of RTI and RMI, a tangential velocity difference on both sides of the interface will inevitably induce the KHI.

\begin{figure*}[htbp]
\center\includegraphics*
[width=0.8\textwidth]{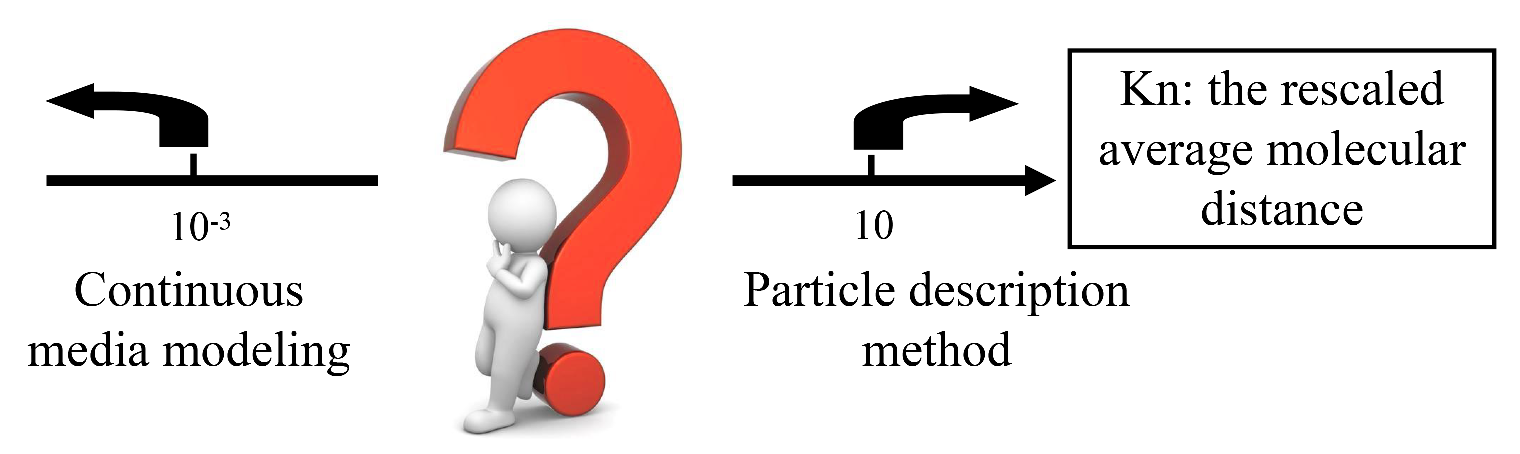}
\caption{ The mesoscale dilemma between macroscopic continuous and microscopic particle descriptions.
} \label{fig002}
\end{figure*}

The three kinds of fluid instabilities, RTI, RMI, and KHI, have been the focus of high energy density physics laboratories and inertial confinement fusion (ICF) laboratories around the world~\cite{Liu2022POP,Liu2023POP,Lan2022MRE,Chen2022MRE,
Lan2021PRL,Qiao2021PRL}.
In fact, fluid instability phenomenon widely exists in nature and engineering technology fields, such as astrophysics, energy and power, aerospace, chemical engineering and materials science~\cite{Gan2011PRE,Lin2017PRE,Lin2019CTP,Lin2021PRE,Chen2018POF,
Chen2022FOP,ChenLu2021FOP,Zhang2021POF,Li2022FOP,Lai2023CAF,Gan2019FOP}.
As a physical phenomenon that must occur when certain conditions are met, fluid instability is also a double-edged sword.
On one hand, it can accelerate the mixing of substances, which is conducive to the mixing and combustion of liquid fuels in internal combustion engines, aeroengines and supersonic ramjets.
It also plays a critical role in initiating explosives.
On the other hand, it is also a key factor that seriously affects the success of ICF ignition and some equipment performance.
It poses a potential threat to the safety of explosives.
The guiding principle here is straightforward: if it is beneficial, strengthen it; if it is harmful, try to suppress it.
The basis of effective utilization and suppression is to have a clear understanding of the characteristics, mechanisms and laws in the process of its occurrence and development.
These systems often have the following characteristics: (i) Despite their macroscopic scale, they encompass a large number of intermediate-scale spatial structures and kinetic patterns. The existence and evolution of these structures and models greatly affect the physical properties and functions of the system. (ii) Such systems often have numerous internal interfaces, including material interface and mechanical interface (shock wave, rarefaction wave, detonation wave, etc.). (iii) The internal force and response processes within the system are exceedingly complex.
Furthermore, the study of these systems is faced with the following problems: The large-scale slow behavior of these systems can be reasonably described by the Navier-Stokes (NS) equations.
However, in the description of flow behavior in some low-pressure rarefaction regions, the description of internal structure of shock wave or detonation wave, and the description of non-equilibrium behavior caused by fast changing flow or reaction, NS equations are insufficient in physical function.
At the same time, for the flow behavior of interest, microscopic Molecular Dynamics (MD) simulations are often powerless due to the limitations of applicable (spatiotemporal) scales.
As shown in Fig. \ref{fig002}, for the mesoscale dilemma where the macroscopic continuous NS models suffer from inadequate physical capabilities, while micro MD simulations are unable to reached the scale, the corresponding research is extremely weak due to the lack of suitable models and methods.
Compared with the macroscopic continuous case, the typical characteristics of mesoscale behavior are as follows: The discrete effect is more significant, and the non-equilibrium effect is more significant.
Currently, mesoscale kinetic models related to discrete and non-equilibrium phenomena are far from being fully understood, which significantly hinders the assessment of various effects in mesoscale behavior and the development of corresponding control techniques.

Similar problems are faced in the following situations: In the field of aerospace, spacecraft traverses the rarefied atmosphere during its launch and reentry.
With the increase of the height in the rarefied region, the fluid behavior predicted by NS deviates increasingly from experimental observations.
Even at the same altitude, the reasonability of NS describing practical flow behaviors in different parts of the same spacecraft may also be different.~\cite{Tsie1946JAS, Chen2017book,Li2011SCPMA}.
In the field of micro-porous flow and microfluidics, behaviors such as heat transfer, mass transfer, and phase changes within microchannels often show significant deviations from macroscopic NS descriptions~\cite{Karniadaskis2005book,Keerthi2018Nature, Quesada2019Micromachines,Kavokine2021ARFM}.
Descriptions of complex flow processes, such as jet atomization, bubble collapse under shock, and droplet fragmentation under laser ablation, challenge the rationality of the NS model~\cite{Jiang2010PECS,Ding2017JFM, Liang2018SCPMA, Ding2018POF, Luo2019JFM, Zhang2020FOP}.
In combustion environments, the phase transition and mixing of liquid and gas~\cite{Jiang2022book,Wang2018book}, as well as complex flow processes like droplet vaporization under laser ablation and laser ablation propulsion, pose challenges to the rationality and physical capabilities of NS modeling~\cite{Xiao2017Science,Liu2023POP,Rinderknecht2018PPCF}.
The internal non-equilibrium phenomena of plasma system are very rich and complex, and hydrodynamic description is an important means to study the behavior of plasma.
However, the in-depth study of a series of strong non-equilibrium behaviors poses a challenge to the rationality and capabilities of NS modeling~\cite{Vidal1993PoFB1993, Bond2017JFM,Rinderknecht2018PPCF, Song2023POF,Yao2020MRE,Cai2021MRE}.
It has been pointed out that in ICF, the non-continuous and highly non-equilibrium kinetic behaviors may significantly affect the success of ignition~\cite{Rinderknecht2018PPCF,Yao2020MRE,Cai2021MRE}.

The physical reason for the above dilemma is that the traditional NS model is based on the assumption of continuous media and near equilibrium approximation.
However, near sharp interfaces such as shock waves and boundary layers, the average molecular spacing is no longer a negligible small quantity relative to the scale of the structure we are concerned about.
In terms of describing fast changing modes, the thermodynamic relaxation time is no longer a negligible small quantity compared to the time scale of the flow (or reaction) we are concerned with.
During the process of flow (or reaction), the system no longer has enough time to return to near thermodynamic equilibrium.
In addition, traditional fluid modeling only focuses on/describes the spatiotemporal evolution of the conserved moments of the distribution function (density, momentum, and energy).
Subsequently, we will further discuss that this approach is inadequate for describing micro-mesoscale structures and fast-changing modes, and its shortcomings become more pronounced as the degree of discreteness and non-equilibrium increases.

There are only two ways to advance to the mesoscale science: from macro to meso and from micro to meso.
The former involves scaling down from larger to smaller, while the latter involves scaling up from smaller to larger. Given that the fluid models predominantly used in engineering applications such as aerospace, ICF and some other equipments are mainly macroscopic continuous NS models, the issue we are facing is as follows: How to seamlessly connect mesoscale modeling with macroscopic continuous modeling?
Therefore, in the discussions within this article, we adopt a scheme that gradually increases the degree of discreteness and non- equilibrium from macro to meso, with a primary focus on behaviors closer to the macroscopic side within the mesoscale range.

\section{\label{sec:level2} State of art for complex flow}

\subsection{\label{sec:Problems-challenges}Problems and challenges}

The discontinuity (or discreteness) of a fluid system is closely related to the degree of Thermodynamic Non-Equilibrium (TNE)
\footnote{
The Thermodynamic Non-Equilibrium (TNE) in DBM literature is called Thermal Non-Equilibrium in some other literatures. Which is right, or more reasonable?\\
Here, the relationship between thermology, thermodynamics and statistical physics is involved.
Thermology is certainly before thermodynamics. However, without statistical physics and subsequent thermodynamics, many problems in thermology cannot be explained clearly. In a sense, thermodynamics is an upgraded version of thermology.
``The biggest difference between thermodynamics and thermology in ordinary physics series is that in thermology, more emphasis is placed on the introduction and measurement of thermodynamic quantities, heat engines and their efficiency; while in thermodynamics, it is raised to the level of logical reasoning and deriving calculation formulas."
See Huichuan SHEN, Preface of ``Fine Solution of Exercises in Thermal Physics" (in Chinese).
\url{https://www.zhihu.com/question/400424184} \\
The case under DBM study, to be more specific, belongs to the kinetic theory section of nonequilibrium statistical physics part of Statistical Physics. You can find the following three characteristics in DBM study: 
(i) Both the heat and force are involved.
(ii) It is concerned with the process of time evolution.  Steady state is a special case: time is still evolving, but the outcome is no longer changing.
(iii) The strength and means of deviating from equilibrium state and the resulting effects are described by the non-conserved kinetic moments of $(f-f^{eq})$. The kinetic moments correspond to macroscopic quantities, which relates DBM to thermodynamics.
This is the consideration that we used Thermodynamic Non-Equilibrium, instead of Thermal Non-Equilibrium, at that time.\\
Which is right, or more reasonable?
-- Both are right, both are reasonable. It's just that the perspective of consideration/interpretation is slightly different.
}.
Knudsen (Kn) number, on the one hand, can be regarded as a rescaled average molecular distance, describing the discontinuity (or discreteness) of the system.
On the other hand, it can also be regarded as a rescaled thermodynamic relaxation time to describe the degree of TNE of the system.
The concepts of non-equilibrium flow and discontinuous flow (discrete flow) overlap from some sense in physical connotation.
The study of non-equilibrium complex flows has made great progress, but still faces some fundamental scientific challenges, such as:

(i) Cross-scale modeling and simulation has been a research hotspot for about 15-20 years.
During this period, it has experienced fluctuations in popularity.
The main methodology is to modify the macroscopic fluid/solid equations based on (quasi) continuity and (near) equilibrium assumption~\cite{Agarwal2001POF}.
That is, the physical quantities used to describe the system's state and behavior are still based on those used in traditional fluid/solid mechanics equations.
A very natural fundamental scientific question arises: As the degree of dispersion/non-equilibrium increases, is it really no problem to focus only on the few physical variables used in a continuous model (e.g. NS)?

(ii) It is a typical feature that complex structures and behaviors generally take on different characteristics when viewed from different angles.
``How to extract more valuable information'' combined with ``how to analyze it'' determines our research capabilities and depth.
Among them, the technical key is as below: when dealing with systems that are becoming increasingly complex, how to achieve an intuitive geometric correspondence for describing complex system states and behaviors?

(iii) Turbulence research remains an enduring and important research focus due to its significance.
Turbulent mixing is an important aspect of hydrodynamic instability research.
It can be thought that as the vortex and other structures concerned become smaller and smaller, the average molecular distance is no longer a negligible small quantity relative to the scale of the structure or behavior concerned, that is, the discreteness becomes more pronounced, and discrete effects become more significant.
However, the early concepts and theoretical framework of turbulence are established on the continuum theory~\cite{Fluid-Mechanics2008}.
So, the third fundamental scientific question is as follows: Is the turbulence concept derived from pure mathematical deductions based on a macroscopic continuous image entirely consistent with that in pursuit of physical origin?

(iv) At present, the studies on non-linear systems~\cite{Zhang2015PRL,Zhou1999PRL} and heat and mass transfer in small systems have made rich progress~\cite{Tang2015Energy,Zong2023Nanoscale,Yang2023PRB,Zhao2006PRL, Li2015PRL,Wang2020PRL,Li2012ROMP,Wang2010PRL,Zhao2014PRL}.
Heat and mass transfer in small systems exhibit some behaviors that seem ``anomalous'' (different from the macroscopic continuous cases) and have attracted significant attention in the fields of statistical physics and nonlinear science~\cite{Maasilta2014Phys}.
However, it's worth noting that the small systems extensively studied in the literature of statistical physics and nonlinear science are often much smaller than the mesoscale non-equilibrium situations of interest in the mechanical engineering field.
The situation we face is that engineering applications largely rely on macroscopic continuous modeling, but also encounter unsatisfactory or even unsolvable problems.
The question we face is how to seamlessly connect mesoscale modeling with macroscopic continuous modeling?

(v) In the past years, the main idea of cross-scale modeling and simulation is based on reductionism.
As a scientific method, reductionism is significant in uncovering the internal mechanisms of complex systems and phenomena, determining causality between the whole and its parts, and achieving precise modeling of complex systems.
However, a typical characteristic of complex systems is that as complexity increases, the system may exhibit emergent and novel behavioral features that cannot be derived from any constituent or coexistence rules from simpler or simplest cases. (Emergence, the idea that the whole is greater than the sum of its parts, is a central theme in complexity science.)
So, the fifth fundamental scientific question that DBM needs to face is: Is the reductionist approach alone sufficient for cross-scale modeling and simulation?

(vi) The sixth basic science problem relates to the most basic parameter of non-equilibrium flow description/cognition, that is the non-equilibrium strength.
Given the complexity of non-equilibrium behavior, any definition of non-equilibrium strength depends on the research perspective.
In other words, there may be the following situations: The non-equilibrium strength is increasing from a certain perspective, while from another perspective,  it appears to be decreasing. How to describe the non-equilibrium strength of complex flows accurately?

\subsection{Ideas and schemes}

The problems discussed above are all problems where the complexity of the problem challenges the physical function of the model.
The problem of physical function cannot be solved by improving algorithm accuracy.
Multi-scale/cross-scale modeling and simulation are efforts to address these and similar problems.

\begin{figure*}[htbp]
\center\includegraphics*
[width=0.5\textwidth]{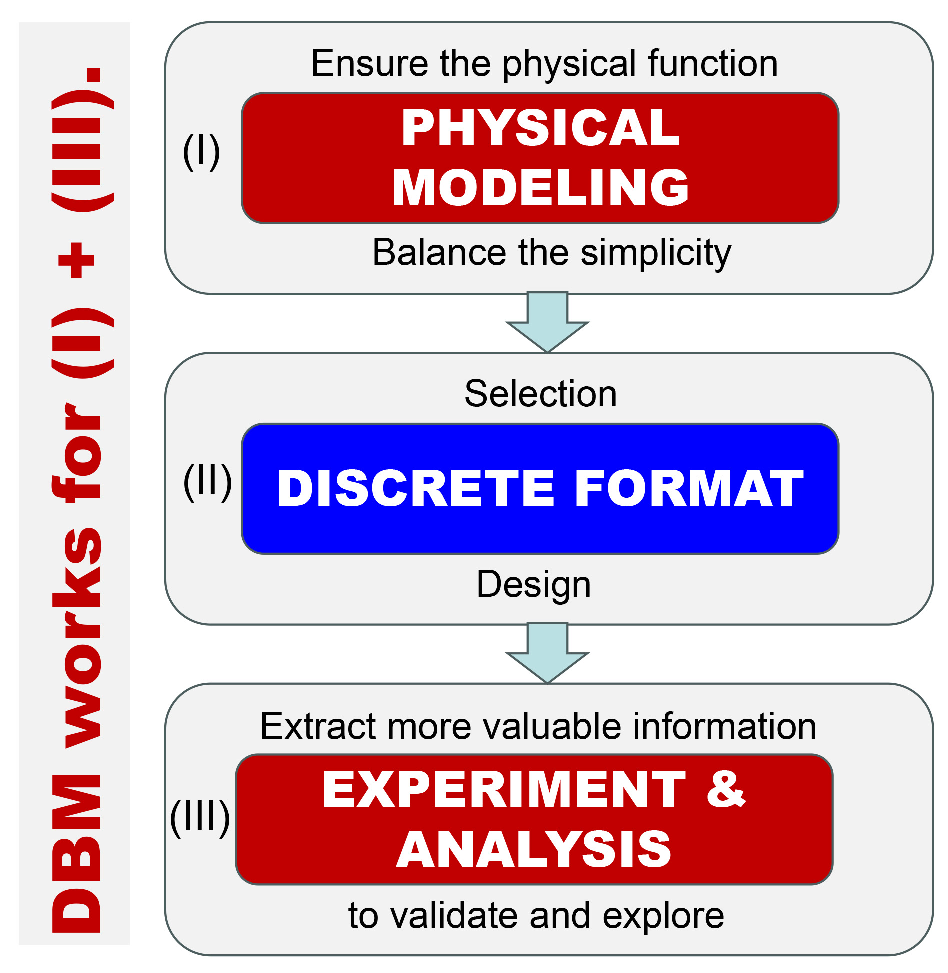}
\caption{ The three main steps of complex flow simulation research.
} \label{fig003}
\end{figure*}

As shown in Fig. \ref{fig003}, simulation study of complex flow includes three major steps,  (i) physical modeling, (ii) discrete format selection/design, (iii) numerical experiment and complex physical field analysis.
As users of discrete formats, the focus of the physical research groups generally lies in the two ends of (i) and (iii), leaving the design of the optimal discrete format in step (ii) to more specialized computational mathematicians.
The core tasks of the physical research group are as follows: (i) Ensure the validity of the physical model (theoretical model) for the problem under investigation while maintaining simplicity, and (ii) try to extract more valuable physical information for the complex physical field from the massive data.

People's understanding of fluids begins with continuous and (near) equilibrium flow.
As the degree of discreteness and nonequilibrium increases, the complexity of system behavior sharply increases, which poses challenges for the description and control of corresponding flow behavior.
Our understanding of fluid behavior has two progression modes: one is the macroscopic cognitive progression mode centered on conserved quantities, and the other is the microscopic cognitive progression mode centered on molecular dynamics.
People's understanding of fluid behavior begins with macroscopic continuous images.
The order of scientific research is generally: from conserved quantities, to spatiotemporally slow variables, followed by gradually increasing fast variables.
Early research on conserved quantities and slow variables prepares cognitive and technical foundations for gradually increasing fast variable research in the later stages.
The space slowly changes, corresponding to the large structure, and the relative average molecular spacing is small enough to be negligible. At this point, the assumption of continuity is reasonable.
A slow change in time indicates that the system has enough time to recover to the thermodynamic equilibrium state, and the thermodynamic (near) equilibrium assumption is reasonable at this time. Traditional fluid modeling is based on and describes the three conservation laws of mass, momentum, and energy.
However, as the rate of spatial change increases, that is, with a decrease in the focus structure, the relative average molecular spacing gradually or becomes less negligible, and the discrete effect becomes more significant.
Alternatively, as the rate of change over time increases, during the flow or reaction process, the system may not have enough time to return to thermodynamic equilibrium, resulting in a higher degree of thermodynamic non-equilibrium.
At this point, the rationality of the two basic assumptions of traditional fluid modeling is increasingly challenged. How to model and analyze has become a problem to be solved.
The path from large to small scale has encountered a theoretical rationality bottleneck.

Molecular dynamics is theoretically complete, but it requires extremely high computational resources, so the applicable spatiotemporal scale is very small.
The behaviors we focus on often occur at spatiotemporal scales that are beyond the reach of molecular dynamics.
So, on the path from small to large scale, we encountered a bottleneck in scale.

Statistical physics is a bridge that connects microscopic and macroscopic aspects.
It utilizes a probabilistic approach based on the understanding of the microstructure of matter and interactions between microscopic particles to provide a microscopic explanation for the physical properties and macroscopic laws of objects composed of a large number of particles.
It is also known as statistical mechanics.
Statistical physics is both a theory of physics and a methodology. The microscopic particles described in its framework can be molecules in gas, atoms in crystals, photons in laser beams and other conventional microscopic particles, as well as stars in the Milky Way galaxy, cars on highways, sheep in flocks, people in social groups, and so on.
The kinetic theory in non-equilibrium statistical physics is the foundation of macroscopic fluid mechanics theory.
Among them, the kinetic theory based on the Boltzmann equation is a relatively complete part of non-equilibrium statistical physics~\cite{Chen2010book}.
In response to the ``mesoscale'' dilemma of insufficient physical functionality of macro models and limited applicability of micro models, as shown in Fig. \ref{fig002}, developing kinetic models based on the Boltzmann equation has become the primary approach.

\subsection{ Statistical physics: coarse-grained modeling and phase space description }

The fundamental equation of non-equilibrium statistical physics is the Liouville equation:
\begin{equation}
\frac{{\partial F}}
{{\partial t}} + \sum\limits_{i = 1}^N {\left( {{\mathbf{\dot{q}}}_i \cdot \frac{\partial }
{{\partial {\mathbf{q}}_i }} + {\mathbf{\dot{p}}}_i \cdot \frac{\partial }
{{\partial {\mathbf{p}}_i }}} \right)F}  = 0,
\label{Eq.Liouville}
\end{equation}
It uses the $N$-body distribution function $F = F\left( {{\mathbf{q}}_1 ,{\mathbf{q}}_2 , \cdots ,{\mathbf{q}}_N ,{\mathbf{p}}_1 ,{\mathbf{p}}_2 , \cdots ,{\mathbf{p}}_N ,t} \right)$ and its evolution equation to describe the state and behavior of the system, where  $({\mathbf{q}}_i ,{\mathbf{p}}_i )$ are the generalized coordinate and generalized momentum of the $i$th particle.
The Bogoliubov-Born-Green-Kirkwood-Yvon (BBGKY) hierarchy chain is equivalent to Liouville equation.
Equivalent to the description of molecular dynamics without any simplification, it can be regarded as the holographic description based on the molecular level, which is the starting point of theoretical thinking~\cite{Chen2010book} .
For a macroscopic system, $N$ is the order of magnitude of the Avogadro constant, i.e. $10^{23}$.
In general, the Liouville equation (\ref{Eq.Liouville}) is fundamentally unsolvable and cannot be directly applied.
So we have to simplify it.
It is impossible and, in most cases, unnecessary to capture all the details of a system's state and behavior simultaneously.
We need to break down complex problems, according to the research needs, to grasp the main contradiction.
The direction of simplification involves making the governing equation depend on a distribution function with fewer bodies.
What about discarded information?
According to the mean-field approach, a correction term is added to the simplified governing equation to compensate for the lost information, ensuring that the kinetic properties to be studied remain unchanged.

Specifically, since the $N$-body distribution function evolution equation (\ref{Eq.Liouville}) cannot be processed, we first thought of simplifying it to the ($N$-1)-body distribution function evolution equation:
\begin{equation}
\frac{{\partial f_{N - 1} }}
{{\partial t}} + \sum\limits_{i = 1}^{N - 1} {\left( {{\mathbf{\dot{q}}}_i  \cdot \frac{\partial }
{{\partial {\mathbf{q}}_i }} + {\mathbf{\dot{p}}}_i  \cdot \frac{\partial }
{{\partial {\mathbf{p}}_i }}} \right)f_{N - 1} }  = C_{N - 1} ,
\label{2}
\end{equation}
where $C_{N-1}$ is a correction term describing the contribution of the $N$-body distribution function, whose function is to recover some of the information lost due to the simplification of the model, so as to ensure that the kinetic properties we want to study will not change due to the simplification of the model. Since the evolution equation of the ($N$-1)-body distribution function is still unmanageable, we simplify it to the ($N$-2)-body distribution function evolution equation, and so on.
Finally, the following 2-body distribution function evolution equation is obtained:
\begin{equation}
\frac{{\partial f_2 }}
{{\partial t}} + \sum\limits_{i = 1}^2 {\left( {{\mathbf{\dot{q}}}_i  \cdot \frac{\partial }
{{\partial {\mathbf{q}}_i }} + {\mathbf{\dot{p}}}_i  \cdot \frac{\partial }
{{\partial {\mathbf{p}}_i }}} \right)f_2 }  = C_2 ,
\label{3}
\end{equation}
Among them, the modified term $C_2$ describes the contribution of the 3-body distribution function $F_3$.
 It's important to note the fact that the probability of 3 particles colliding together is much less than the probability of 2 particles colliding together at the same time.
Thus, the effect of the 3-body interaction is much smaller than the effect of the 2-body interaction, we can approximate $F_3$ as zero. Equation (3) simplifies to:
\begin{equation}
\frac{{\partial f_2 }}
{{\partial t}} + \sum\limits_{i = 1}^2 {\left( {{\mathbf{\dot{q}}}_i  \cdot \frac{\partial }
{{\partial {\mathbf{q}}_i }} + {\mathbf{\dot{p}}}_i  \cdot \frac{\partial }
{{\partial {\mathbf{p}}_i }}} \right)f_2 }  = 0.
\label{4}
\end{equation}
Let us make two further assumptions:
(i) The interaction between particles depends only on the distance between them, not on the direction:
\begin{equation}
U_{12} \left( {{\mathbf{q}}_1 ,{\mathbf{q}}_2 } \right) = U_{12} \left( {\left| {{\mathbf{q}}_1  - {\mathbf{q}}_2 } \right|} \right).
\label{5}
\end{equation}
(ii)The 2-body distribution function can be written in the form of the product of two monomer distribution functions, i.e,
\begin{equation}
F_2 ({\mathbf{q}}_1 ,{\mathbf{p}}_1 ;{\mathbf{q}}_2 ,{\mathbf{p}}_2 ;t) = F_1 ({\mathbf{q}}_1 ,{\mathbf{p}}_1 ;t)F_1 ({\mathbf{q}}_2 ,{\mathbf{p}}_2 ;t).
\label{6}
\end{equation}
Then, we finally obtain the evolution equation of the monomer distribution function $F_1 \equiv f$. That is the Boltzmann equation,
\begin{equation}
\frac{{\partial f}}
{{\partial t}} + {\mathbf{v}} \cdot \frac{{\partial f}}
{{\partial {\mathbf{r}}}} + {\mathbf{a}} \cdot \frac{{\partial f}}
{{\partial {\mathbf{v}}}} = Q\left( {f,f} \right),
\label{7}
\end{equation}
where
\begin{equation}
Q\left( {f,f} \right) = \int\limits_{ - \infty }^{ + \infty } {} \int\limits_0^{4\pi }  ( f^{'}f^{'}_{1} - ff_{1}) g \sigma d\Omega d \mathbf{u}_{1},
\label{8}
\end{equation}
describes the rate of change of the distribution function caused by molecular collisions.

From Liouville equation to Boltzmann equation, the coarse-grained physical modeling has been experienced ($N$-1) times.
It can be seen that compared with Liouville equation, Boltzmann equation is already a highly coarse-grained physical model.
As the foundation of non-equilibrium statistical physics, the Liouville equation is in principle equivalent to molecular dynamics without any simplification, and is a ``holographi'' description of $N$-body systems, applicable to both solid and fluid systems. The Boltzmann equation, however, is limited to fluid systems.

In statistical physics, there are two commonly used phase Spaces: $\mu$ space and $\Gamma$ space.
For a system composed of $N$ identical particles without internal degrees of freedom, if a 6-dimensional phase space is formed based on generalized coordinates and generalized momenta (or velocities) describing the microscopic states of particles, then a state of the system is a distribution of $N$ points in the space.
Figure 4 shows the schematic of three common phase spaces, where position or velocity space shown in Fig.4(a) can be regarded as a sub-space of $\mu$  space shown in Fig.4(b), and $\mu$ space shown in Fig.4(b) can be regarded as a sub-space of $\Gamma$ space shown in Fig.4(c).
Figure 4(b) shows two states of the $N$ particle system. If the generalized coordinates and generalized momenta of these $N$ particles are used as the basis to open phase space, i.e., the $\Gamma$ space, then a state of the system appears as a single point in the space.
Figure 4(c) shows two states of the $N$ particle system.
Obviously, the usual coordinate space and velocity space are subspaces of two kinds of phase spaces.
In coordinate (velocity) space, it is impossible to know the velocity (coordinate) information of a particle by a distribution alone, As shown in Figure 4(a).
In the coordinate space, the particle velocity information needs to be roughly obtained by comparing the position information of the two moments before and after.
In velocity space, the position information of a particle needs the position information of adjacent two moments and the time step to be roughly known.

\begin{figure*}[htbp]
\center\includegraphics*
[width=1.0\textwidth]{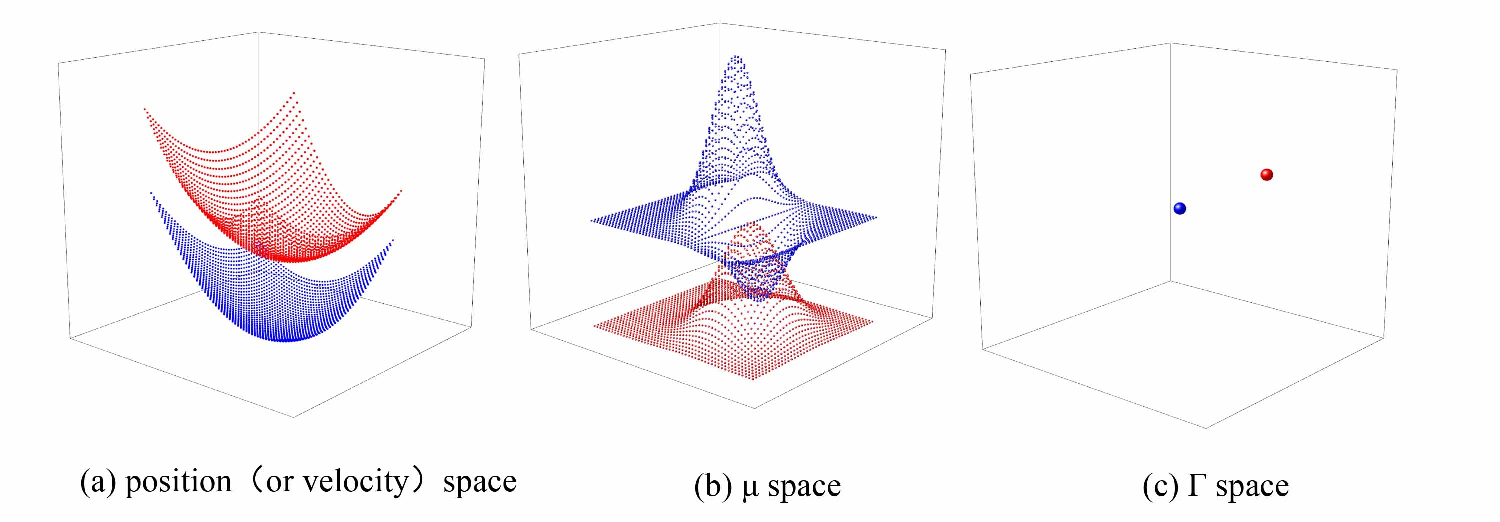}
\caption{ Schematic of phase space description method.
} \label{fig004}
\end{figure*}

The process from Boltzmann equation to hydrodynamic equations is also a physical coarse-graining modeling process that grasps the main contradiction according to the research demand.
In macroscopic fluid theory, fluid states are described using density $\rho$, temperature $T$, velocity $\mathbf{u}$, and pressure $p$.
The fundamental governing equations are the Navier-Stokes equations:
\begin{equation}
\left\{ {\begin{array}{*{20}c}
   {\frac{{\partial \rho }}
{{\partial t}} + \nabla  \cdot (\rho {\mathbf{u}}) = 0}  \\
   {\frac{{\partial (\rho {\mathbf{u})}}}
{{\partial t}} + \nabla  \cdot (\rho {\mathbf{uu}}{\text{ + }}p{\mathbf{I}}) =  - \nabla  \cdot {\bm{\Pi }}}  \\
   {\frac{{\partial E}}
{{\partial t}} + \nabla  \cdot [( E + p){\mathbf{u}}] =  - \nabla  \cdot ({\mathbf{q}} + {\bm{\Pi }} \cdot {\mathbf{u}})}  \\
 \end{array} } \right.
\label{9}
\end{equation}
Where $\mathbf{I}$ is the unit tensor.
In kinetic theory, the state of a system is described using the distribution function $f(\mathbf{x},\mathbf{v},t)$ and its kinetic moments.
Among them, the three conserved moments of the distribution function $f$, (density $\rho$, momentum $\rho \mathbf{u}$, and energy $E$) and the two non-conserved moments (viscous stress $\bm{\Pi}$ and heat flux $\mathbf{q}$) are described into traditional fluid theory.
The relationship between these moments and the distribution function $f$ is as follows:
\begin{equation}
{\mathbf{W}} = \int {f{\text{ }}{\bm{\Psi }}{\text{ }}d{\mathbf{v}}} ,
\label{10}
\end{equation}
where $\mathbf{v}$ is the molecular velocity, and
\begin{equation}
\begin{aligned}
{\mathbf{W}} = \left[ {\rho ,\rho {\mathbf{u}},E = (D+n)\rho T/2 + \rho {\mathbf{u}} \cdot {\mathbf{u}}/2} \right]^T ,
\label{11}
\end{aligned}
\end{equation}
\begin{equation}
{\bm{\Psi }} = \left[ {1,{\mathbf{v}},{\mathbf{v}} \cdot {\mathbf{v}}/2 } \right]^T ,
\label{12}
\end{equation}  
\begin{equation}
{\bm{\Pi }} = \int {\left( {f - f^{eq} } \right)\left( {{\mathbf{v}} - {\mathbf{u}}} \right)} \left( {{\mathbf{v}} - {\mathbf{u}}} \right)d{\mathbf{v}},
\label{eq.13}
\end{equation}
\begin{equation}
{\mathbf{q}} = \int {\left( {f - f^{eq} } \right)\frac{1}
{2}\left( {{\mathbf{v}} - {\mathbf{u}}} \right)} \left( {{\mathbf{v}} - {\mathbf{u}}} \right) \cdot \left( {{\mathbf{v}} - {\mathbf{u}}} \right)d{\mathbf{v}}
\label{eq.14}
\end{equation}
where $f^{eq}$ is the corresponding equilibrium distribution function.
Here, $D$ is the spatial dimension and $n$ represents the extra degree of freedom.
The higher order non-conserved moments of the distribution function $f$ are not included in the description of traditional fluid theory, which is naturally a double-edged sword: on the one hand, it brings the simplicity of traditional fluid theory, but at the same time, it also brings constraints to describe the more discrete and non-equilibrium cases.
In a more reasonable kinetic description, the emergence of more high-order moments of $f$  is an inevitable requirement driven by the requirement/original intention of ``the complexity of system behavior increases dramatically + the physical description/control ability does not decrease''.
\emph{With the increase of the degree of discreteness and the degree of thermodynamic non-equilibrium, the complexity of the system behavior increases sharply, and more physical quantities are needed to describe the state and behavior.}
A lot of research can be done without increasing the number of physical quantities, but the obvious consequence is a sharp decline in the actual control of system behavior~\cite{Xu2022CMK}.
This point seems to have not received sufficient attention!

It should be noted here that, like NS, Burnett and other hydrodynamic equations, DBM describes a physical image, providing a series of physical constraints on the model for studying physical problems, naturally expressed by equations and relations, but without specific discrete formats.

To establish the connection between the macroscopic and microscopic scales, we must try to cross the threshold of mesoscale.
The complexity of ``mesoscale'' behavior is far greater than expected by reductionist approach. When dealing with a complex system, the general cognitive process is to start with a general overview of macroscopic (system) scale behavior, and then gradually explore the mechanism of its microscopic (unit) scale, and gradually establish the relationship between system scale behavior and unit scale behavior.
However, directly establishing such a connection is often extremely challenging.
The research shows that there may be a general dominant principle in the mesoscale between the unit scale and the system scale, that is, the coordination of different control mechanisms in competition.
This is highly consistent with the idea that in addition to reductionism paradigm, emergentism paradigm is also needed in physics.
Under the initiative of scientists like Academician Jinghai Li and many others, the field of \emph{meso science}, which focuses on multi-scale and cross-scale research, is attracting more interest with time~\cite{Li2013book,Huang2018SR}.

\section{\label{sec:level3} Discrete Boltzmann method }

The nature of an entity comprises both specific individual characteristic and commonality.
As the individual characteristic gradually diminishes, what remains is more common.
The further one moves away from the original specific system, the stronger the universality becomes.
For a model to fully recover all the properties of the system, it must be as complex as the actual system.
Perfect in theory, but too complicated to handle, it becomes impractical.
Therefore, it is necessary to ``choose something to do and ignore others'' and simplify the model to the point where it can be handled.
When it comes to mechanistic research, further simplification is required to the extent that the mechanism can be analyzed clearly, which is exactly what some mechanism studies do at the beginning.
Research based on simple models provides cognitive basis and technical basis for subsequent research on more complex or realistic situations.
Knowledge from practice ultimately returns to practice.
The research of complex systems requires multi-level and multi-perspective methods and cognition.

Statistical physics is a bridge between micro and macro scales, and the kinetic theory is a bridge between micro and macro descriptions of fluid systems.
Coarse-grained modeling is a basic means to grasp the main contradiction according to the research needs of non-equilibrium statistical physics.
The related study on Boltzmann equation is a relatively mature part in the theory of kinetics. In the context of fluid dynamics, ``cross-scale'' typically refers to scaling across Knudsen numbers.

Complex flows are typically associated with unsteady behavior, but the study of steady-state behavior is also an important aspect.
From a descriptive perspective, steady-state conditions can be achieved by setting the time derivative terms to zero in the governing equations.
In contrast, the usual DBM is designed for unsteady flow~\cite{Xu2022CMK}, but steady-state DBM and its application are also important contents of DBM research.

For the sake of clarity, we refer to DBM with a nonzero time span as time-dependent DBM and DBM that only focuses on the behavior at a single moment as time-independent DBM.
The simplest case of time-independent DBM is steady-state DBM~\cite{Zhang2023POF}  .
In this review, we will first introduce time-dependent DBM designed for typical unsteady situations and then provide a brief overview of steady-state DBM.
The physical behavior description functions of the two are complementary. But the design of the latter is significantly different from that of the former.

\subsection{Brief review of non-steady DBM \label{Brief-review}}

\begin{figure*}[htbp]
\center\includegraphics*
[width=0.8\textwidth]{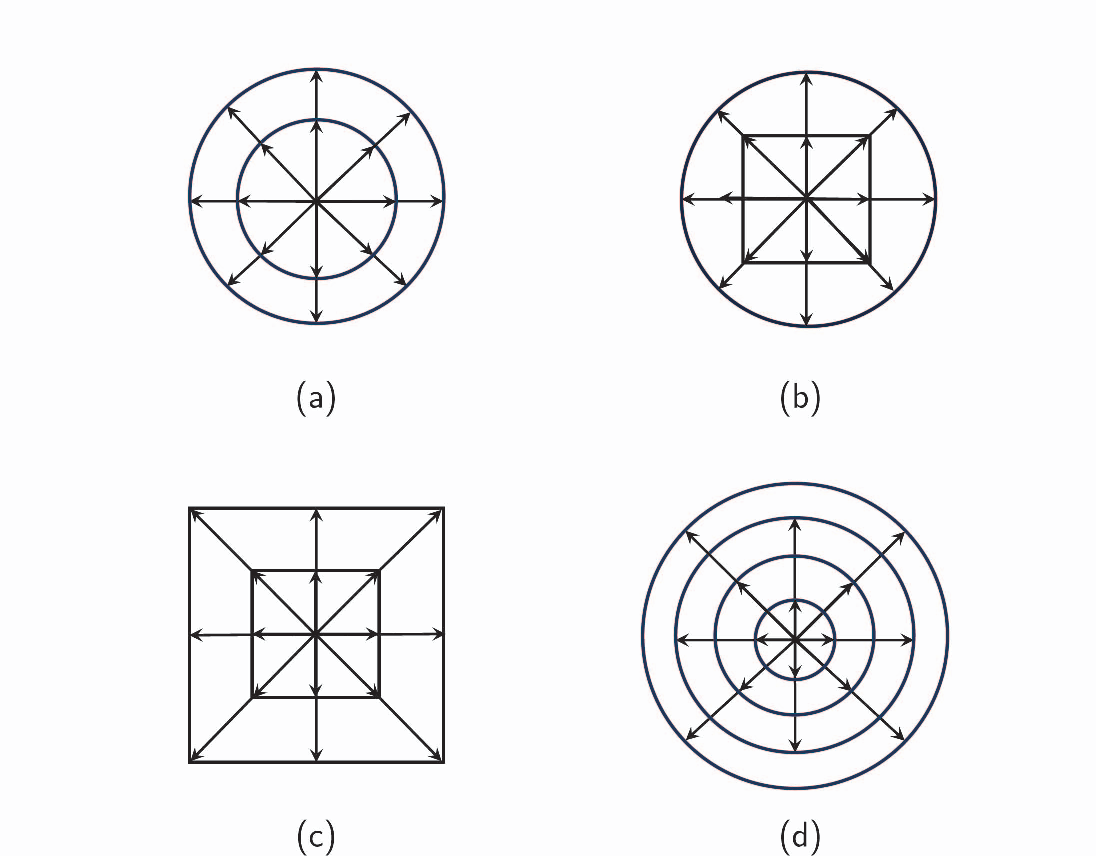}
\caption{ Schematic of several commonly used two-dimensional discrete velocity sets.
} \label{fig005}
\end{figure*}

The Discrete Boltzmann Method (DBM) is proposed to (partially) solve a series of fundamental problems, especially those presented in section \ref{sec:level2}.
It is a kind of (coarse-grained) physical model construction method and complex physical field analysis method developed based on the discrete Boltzmann equation,
\begin{equation}
\frac{\partial f_{i}}{\partial t} + \mathbf{v}_{i} \cdot \frac{\partial f_{i}}{\partial \mathbf{x}} + (\text{force} \  \text{term})_{i} = (\text{collision}\   \text{term})_{i}
\label{eq.15}
\end{equation}
where the subscript ``\emph{i}'' of the distribution function $f$ is the index of the discrete velocity, corresponding to the discrete velocity $\mathbf{v}_{i}$.
Obviously, the discrete velocity selection rule is the key technology in DBM modeling. 
Figure \ref{fig005} shows the schematic of several commonly used discrete velocity sets in current literature. 
\footnote{Please note that this is by no means a standard or optimal choice that can be applied to every situation. The selection of the optimal discrete velocities depends not only on the discrete formats of time integral and spatial derivative, but also on the specific fluid behavior. This is still an open topic in computational mathematics and is beyond the scope of this article.}

In terms of physical model construction, DBM includes two steps of coarse-grained physical modeling: modification and simplification of Boltzmann equation and discretization of particle velocity space.
The basic principle of coarse-grained physical modeling is that the behaviors of the system to be studied cannot be changed due to the simplification of the model.
In kinetic theory, in addition to the distribution function itself, system properties are described by the kinetic moments of the distribution function.
So, both before and after simplifying the collision term and discretizing the particle velocity space, the control equation should provide the same moments of the distribution function that correspond to the desired kinetic properties.
This is represented mathematically as:
\begin{equation}
\int {} f{\text{ }}{\bm{\Phi }}\left( {\mathbf{v}} \right)d{\mathbf{v}} = \sum\limits_i {} f_i {\bm{\Phi }}\left( {{\mathbf{v}}_i } \right).
\label{eq.16}
\end{equation}
Due to the lack of analytical solutions for the kinetic moments of the distribution function $f$, except for the three conserved moments, it is not convenient to obtain the physical constraints required for discrete velocity selection using equation (\ref{eq.16}).
So, we turned to examine all the kinetic moments of the equilibrium distribution function $f=f^{eq}$ involved in the calculation of the left side of equation (\ref{eq.16}).
Furthermore, it is required that the results of these kinetic moments remain unchanged when converted to summation for calculation. Mathematically, this can be expressed as:
\begin{equation}
\int {} f^{(0)} {\bm{\Psi }}\left( {\mathbf{v}} \right)d{\mathbf{v}} = \sum\limits_i {} f_i ^{(0)} {\bm{\Psi }}\left( {{\mathbf{v}}_i } \right)
\label{eq.17}
\end{equation}
Please note that moving from equation (\ref{eq.16}) to equation (\ref{eq.17}) provides sufficient conditions for the required physical constraints.
Both $\bm{\Phi}$ and $\bm{\Psi}$ are column vectors, whose elements correspond to the kinetic moments that should keep values.
The elements of the former $\bm{\Phi}$ are partial elements of the latter $\bm{\Psi}$.
The latter  $\bm{\Psi}$ contain elements with higher powers of particle velocity.
As the degree of discreteness/non-equilibrium of the system increases, elements with higher particle velocity powers are gradually added to $\bm{\Psi}$.

\begin{figure*}[htbp]
\center\includegraphics*
[width=0.6\textwidth]{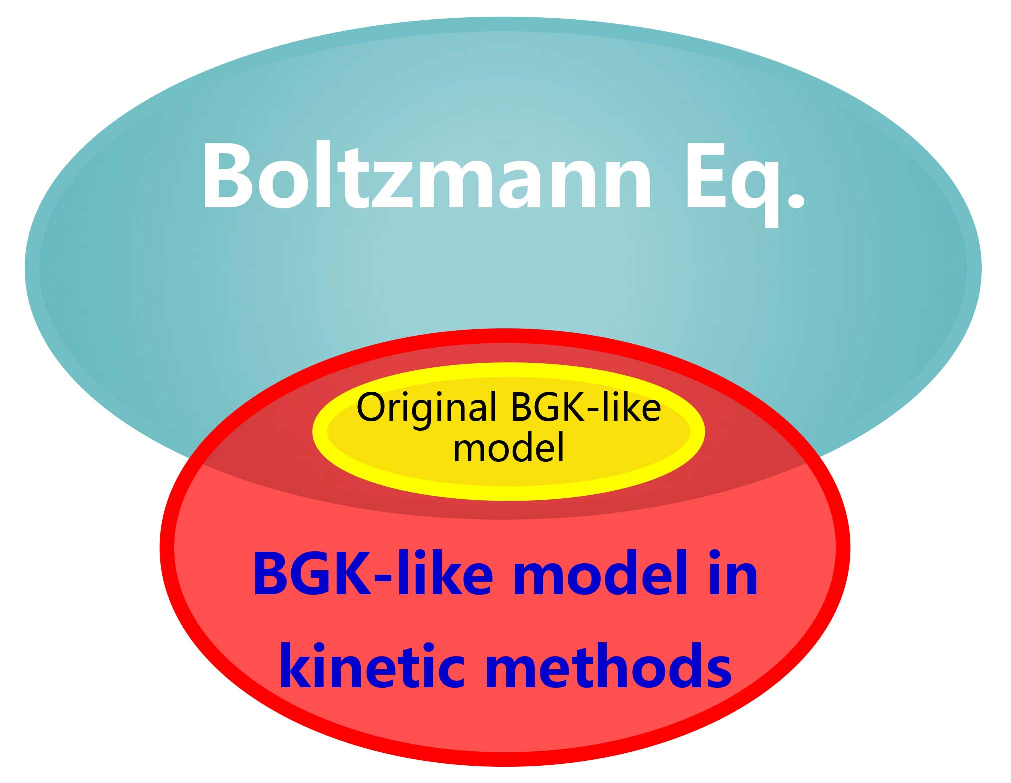}
\caption{ Schematic diagram of the application scopes of the Boltzmann equation, original BGK-like model, and BGK-like model in kinetic methods.
} \label{fig006}
\end{figure*}

There are a few points to highlight here.
(i) Many behaviors of complex flows exceed the descriptive power of the original Boltzmann equation.
The Boltzmann equation in DBM is actually a modified Boltzmann equation based on specific situations combined with mean field theory, and its applicability may, in certain aspects, exceed that of the original Boltzmann equation, as shown in Fig. \ref{fig006}~\cite{Xu2022CMK, Gan2022JFM}.
So, DBM is actually a coarse-grained modeling and analysis method that combines kinetic theory with mean field theory
\footnote{
Physics study has two research paradigms, reductionism and emergentism. The latter is used to describe the part that cannot be included in the former. Statistical physics is a one of the disciplines that do not rely on reductionism. 
However, in the study of fluid physics, it is often necessary to use kinetic theory to investigate the mesoscopic kinetic behavior ignored by the traditional macroscopic modeling, so the role of kinetic theory here is similar to 
that of reductionism.
DBM modeling is originated from a reasonable combination of kinetic theory and mean field theory, where the latter is used to describe the part that the former cannot describe and consequently its role here is similar to that of  emergentism.
}.
(ii) When dealing with the external force term of the Boltzmann equation, in order to avoid the difficulty of making differentiation to the discrete velocity, the DBM modeling approach is to first approximate the external force term, use the property that the equilibrium distribution function is derivable to the particle velocity.
Once the differentiation of particle velocities is accomplished, the expression is then rewritten in terms of discrete velocities.
(iii) The model equations of DBM include the system behavior evolution equation(s) and discrete velocity constraint equations, where the evolution equations, in addition to discrete Boltzmann equation, depending on the specific situation, may also include phase field evolution equation, chemical reaction evolution equation, electromagnetic field evolution equation, etc.  
(iv) DBM only gives the physical constraints that need to be followed when choosing discrete velocities, but does not contain the specific discretization format. In order to reduce the influence of the dramatic change of flow field on the discrete velocity distribution function, the discrete velocity constraint equations can be written in the form of central moments and used to select the discrete velocities on the fluid element, i.e. the discrete peculiar velcities
${{\bf{c}}_i} = \left( {{{\bf{v}}_i} - {\bf{u}}} \right)$.

\begin{figure*}[htbp]
\center\includegraphics*
[width=0.8\textwidth]{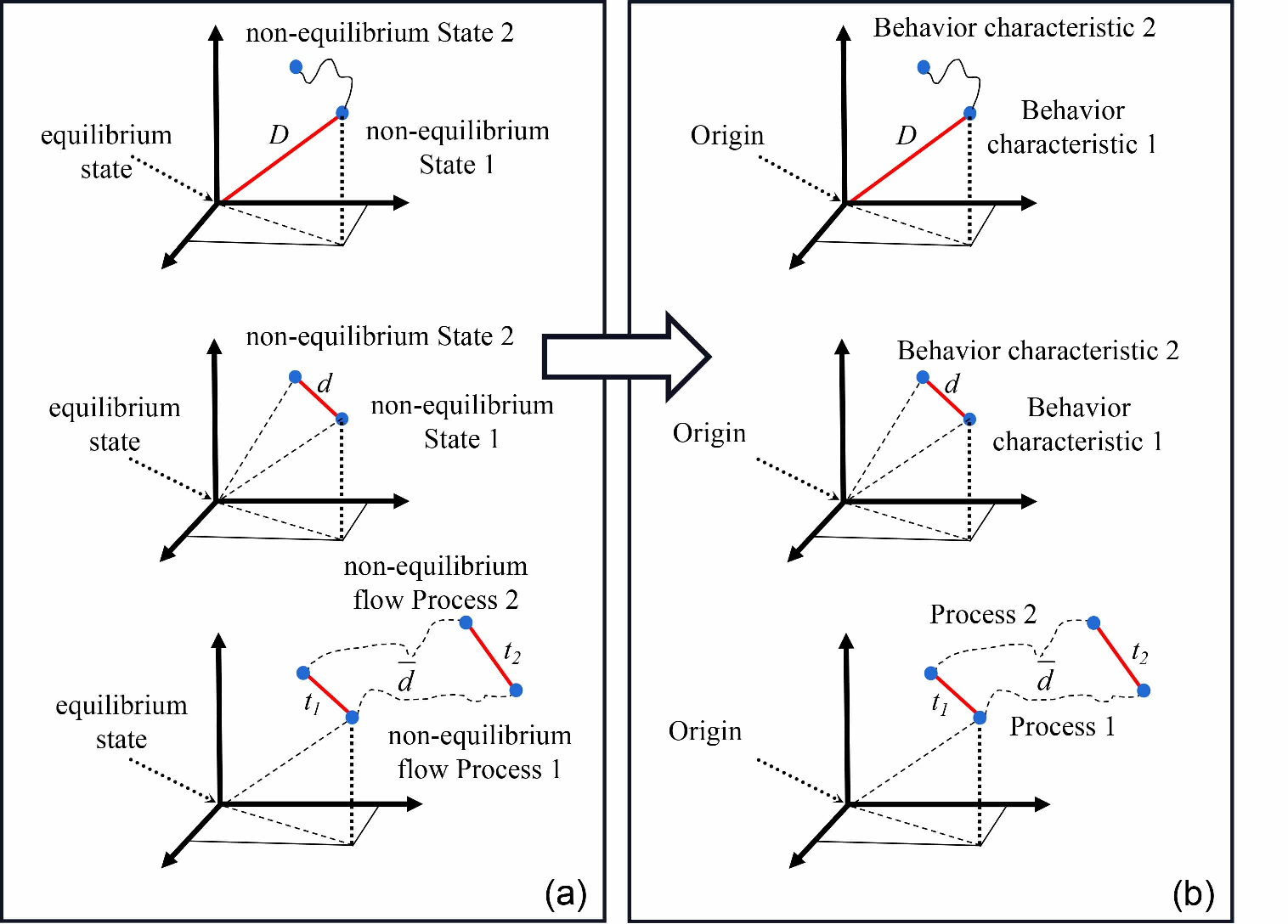}
\caption{ Phase space description methods: from non-conserved moments of $(f-f^{(0)})$  to any set of system characteristics.
} \label{fig007}
\end{figure*}

In terms of complex physical field analysis method, DBM is a specific application and further development of statistical physics coarse-grained description method, non-equilibrium behavior description method, and phase space description method within the framework of discrete Boltzmann equation.
From a historical perspective, DBM has evolved from the physical modeling branch of the Lattice Boltzmann Method (LBM), with some abandoning and new addition.
It preserves the use of discrete velocity, but is no longer limited to specific discrete format.
It provides only the most necessary physical constraints for the discrete format. It is no longer based on the continuity assumption and near equilibrium approximation of traditional fluid modeling, and no longer uses the ``lattice gas'' image of standard LBM.
It adds new detection, presentation, description, and analysis schemes for non-equilibrium states and resulting effects based on phase space, and introduces more information extraction techniques and complex physical field analysis techniques over time.
Figure \ref{fig007} depicts the expansion from the phase-space description of the non-conserved kinetic moments,
\begin{equation}
\begin{gathered}
  {\bm{\Delta }}_n^*  = {\mathbf{M}}_n^* \left( {f - f^{(0)} } \right) \hfill \\
  {\text{      = }}\int {d{\mathbf{v}}\left( {f - f^{(0)} } \right)\underbrace {({\mathbf{v}} - {\mathbf{u}})({\mathbf{v}} - {\mathbf{u}}) \cdots ({\mathbf{v}} - {\mathbf{u}})}_{n{\text{-th order tensor}}}}  \hfill \\
  {\text{    }} = \sum\limits_i {\left( {f_i  - f_i ^{(0)} } \right)} \underbrace {({\mathbf{v}}_i  - {\mathbf{u}})({\mathbf{v}}_i  - {\mathbf{u}}) \cdots ({\mathbf{v}}_i  - {\mathbf{u}})}_{n{\text{-th order tensor}}} \hfill \\
  \label{eq.18}
\end{gathered}
\end{equation}
to phase space description method based on any set of behavioral features,
\begin{equation}
{\bm{{\rm X}}} = \left\{ {X_1 ,X_2 ,X_3 , \cdots } \right\},
\end{equation}
In Eq. (\ref{eq.18}), $\mathbf{u}$ is the local flow velocity.
In the phase space or its subspace, we can define the non-equilibrium strength (or the strength of corresponding behavioral features) of the corresponding perspective by the distance from the state point to the coordinate origin, such as $D$ in Fig. \ref{fig007}.
Using the concept of distance between two points to describe the differences between two non-equilibrium states (or corresponding behavioral features), such as $d$ in Fig. \ref{fig007}, and so on.
Therefore, a more complete name for DBM is the Discrete Boltzmann \emph{modeling and analysis} method~\cite{Xu2022CMK}.
The use of non-conserved kinetic moments of $(f-f^{(0)})$ to detect and describe the specific way in which a system deviates from its thermodynamic equilibrium state and the various effects caused by this are key techniques for DBM in complex physical field analysis.

The behavior of actual systems is often complex.
Coarse-grained modeling is a process of losing information, but this losing information has a bottom line.
The bottom line is that the properties under investigation cannot be changed due to the simplification of the model.
DBM, based on the research needs of the problem, selects a perspective to study a set of kinetic properties of the system, and requires the kinetic moments describing these properties to be preserved during the model simplification process.
As the degree of discreteness and non-equilibrium increases, the complexity of system behavior increases sharply.
Then  incorporating some higher-order kinetic moments into the descriptive perspective becomes an inevitable outcome to ensure that the control ability does not decrease.
Therefore, \emph{using more physical quantities to describe the system state and behavior is a typical feature that distinguishes DBM from traditional fluid modeling and other current kinetic methods}.
From the perspective of Kinetic Macro Modeling (KMM),\footnote{Corresponding to the physical function of DBM is the Extended Hydrodynamic Equations (EHE), which include not only the conservation moment evolution equations corresponding to the three conservation laws of mass, momentum, and energy, but also some of the most closely related non-conservation moment evolution equations. For convenience of description, we refer to the modeling method of deriving EHE based on kinetic equations as Kinetic Macro Modeling (KMM).} this is a requirement for obtaining more accurate constitutive relationships.
From the perspective of the kinetic theory, this is a requirement for obtaining more accurate distribution function $f$ under given conditions.
Different perspectives lead to the same goal.

The starting point of DBM lies within the mesoscale range, approaching the macroscopic side.
Due to the stage nature of development, the primary situations currently considered by DBM still fall within the realm where the Chapman-Enskog (CE) multiscale analysis theory is valid.
Therefore, the CE multi-scale analysis theory is a mathematical guarantee for the reasonable and effective DBM approach~\cite{Xu2022CMK}.

\subsection{Rules for selecting discrete velocities}

By simultaneously taking the density moment, momentum moment, and energy moment on both sides of the Boltzmann equation, a set of fluid dynamic equations that are consistent with the NS form can be obtained,
\begin{equation}
\left\{ {\begin{array}{*{20}c}
   {\frac{{\partial \rho }}
{{\partial t}} + \nabla  \cdot (\rho {\mathbf{u}}) = 0}  \\
   {\frac{{\partial (\rho {\mathbf{u})}}}
{{\partial t}} + \nabla  \cdot (\rho {\mathbf{uu}}{\text{ + }}p{\mathbf{I}}) =  - \nabla  \cdot {\bm{\Delta }}_2^* }  \\
   {\frac{{\partial E}}
{{\partial t}} + \nabla  \cdot [( E + p){\mathbf{u}}] =  - \nabla  \cdot ({\bm{\Delta }}_{3,1}^*  + {\bm{\Delta }}_2^*  \cdot {\mathbf{u}})}  \\
 \end{array} } \right.
\end{equation}
For convenience of description, here we refer to this as the generalized NS.
The advantage of this generalized NS is that there is no approximation.
Its viscous stress,
\begin{equation}
\begin{aligned}
{\bm{\Delta }}_2^*  &= {\mathbf{M}}_2^* \left( {f - f^{(0)} } \right) = {\mathbf{M}}_2^* \left( {f^{(1)}  + f^{(2)}  + f^{(3)}  +  \cdot  \cdot  \cdot } \right) \\
&= {\bm{\Delta }}_2^{*(1)}  + {\bm{\Delta }}_2^{*(2)}  + {\bm{\Delta }}_2^{*(3)}  +  \cdot  \cdot  \cdot
\end{aligned}
\end{equation}
and heat flux,
\begin{equation}
\begin{aligned}
{\bm{\Delta }}_{3,1}^*  &= {\mathbf{M}}_{3,1}^* \left( {f - f^{(0)} } \right) = {\mathbf{M}}_{3,1}^* \left( {f^{(1)}  + f^{(2)}  + f^{(3)}  +  \cdot  \cdot  \cdot } \right) \\
&= {\bm{\Delta }}_{3,1}^{*(1)}  + {\bm{\Delta }}_{3,1}^{*(2)}  + {\bm{\Delta }}_{3,1}^{*(3)}  +  \cdot  \cdot  \cdot
\end{aligned}
\end{equation}
include contributions from various orders of non-equilibrium distribution functions $f^{(j)}$, where $j=1,2,3,\cdot \cdot \cdot$, corresponding to the discrete/non-equilibrium order described by the power of Kn number.
the subscript ``3,1'' signifies that a 3rd-order tensor has undergone 1 contraction operation to become a 1st-order tensor, which is a vector.
The meanings of other subscripts below are similar, and so on.
However, the limitation of this approach is that it cannot provide specific expressions for stress and heat flux directly.
To derive analytical expressions for constitutive relationships, Chapman and Enskog developed the CE multi-scale analysis method later named after them. CE multiscale analysis is actually a generalized Taylor expansion and analysis method. The independent variable here is the Kn number.
Not only the distribution function $f$, but also the temporal and spatial derivatives are Taylor expanded at the point,  $\text{Kn}=0$ (i.e., continuity, thermodynamic equilibrium).
More physical image explanations for CE multiscale analysis can be found in reference~\cite{Xu2022CMK}.
The CE multi-scale analysis informs us the following:

(i) The Boltzmann equation, under the quasi-continuous and near-equilibrium conditions, corresponds to the system of fluid dynamics equations which are the NS equations.
The viscous stress and heat flux considering only the first order non-equilibrium $f^{(1)}$ are the NS stress $\bm{\Pi }$ and NS heat flux $\mathbf{q}$ in Eqs. (\ref{eq.13}) and (\ref{eq.14}), respectively. 
As the degree of discretization/non-equilibrium increases, the contribution of second-order $f^{(2)}$ or even higher-order non-equilibrium $f^{(j)}(j>2)$ should be considered in viscous stress and heat flow.

(ii) Among all the kinetic moments of the distribution function $f$, only the three conserved moments (density, momentum, and energy) and two non-conserved moments (viscous stress and heat flux) enter the NS description.
The remaining non-conserved moments have not entered the NS theory, which is a double-edged sword: on the one hand, they bring the simplicity of traditional fluid mechanics theory; however, on the other hand, it creates obstacles for NS to describe situations with higher degrees of discreteness/non- equilibrium, which is the physical reason why NS cannot describe well situations with higher degrees of discreteness/non-equilibrium.

(iii) The ability to recover corresponding levels of macroscopic hydrodynamic equations (such as NS, Burnett equations, etc.) is only part of the physical function of DBM; Corresponding to the physical functions of DBM is the Extended Hydrodynamic Equations (EHE), which include not only the three common hydrodynamic equations but also some of the most closely related non-conservative moment evolution equations.
The extended part (i.e., the evolution equations for relevant non-conserved moments) becomes increasingly necessary as the degree of discreteness/non-equilibrium increases.

(iv) As the degree of discreteness/non-equilibrium characterized by Kn number increases, the complexity of DBM simulation increases more slowly compared to KMM simulation (i.e. deriving and solving EHE), thus allowing it to go further.
If the order of the Kn number to be considered increases by 1, then the number of kinetic moments in DBM that need to be preserved increases by 2, while the complexity of deriving EHE from KMM sharply increases.
Furthermore, in the case of multiple media, due to the existence of different media flow velocities and average flow velocity, and the definition of temperature depends on the referenced flow velocity, the form of EHE is not unique.
These different forms of EHE correspond to descriptions from different perspectives of complex flow dynamics in multiple media.
It can be seen that in the case of multiple media, the correspondence between DBM and physically equivalent EHE is one-to-several.

DBM simulation does not require knowing the specific form of extremely complex EHE with physical equivalence, but we can borrow KMM's ideas and physical images to quickly screen out the kinetic moments involved in the more accurate stress and heat flux calculations (without strictly deriving equations, only checking the order of the kinetic moments). This part of the kinetic moments are converted from integral to summation for calculation, and the results need to remain unchanged. At this point, we have obtained the most necessary physical constraints to follow for the selection of discrete velocity: the linear equations of $f_{i}^{(0)}$ represented by equation (\ref{eq.17}).

\subsection{Description of non-equilibrium behavior and effect \label{Description}}

Non-equilibrium strength/degree/extent/intensity is one of the most fundamental parameters for describing and recognizing non-equilibrium flows.
In addition to commonly used Kn numbers, spatial gradients of macroscopic physical quantities such as density, flow velocity, temperature, pressure, etc., DBM uses physical quantities based on non-conserved kinetic moments of $(f-f^{(0)})$ to describe the way and magnitude of system deviation from equilibrium.
Furthermore, DBM uses the (independent components of) non-conserved moments of $(f-f^{(0)})$ [e.g., $\{ \bm{\Delta}_{n}^{*}, n=2,(3,1),3,(4,2),(5,3),\cdot \cdot \cdot \}$] as the bases to open phase space, providing an intuitive geometric correspondence for complex system states and behavioral characteristics, as shown in Fig. \ref{fig007}.
It is easy to understand that each non-conserved moment itself and any independent component of $(f-f^{(0)})$ describe the way and magnitude of the system deviating from equilibrium from its own perspective, thus offering a unique view of non-equilibrium features.
These non-equilibrium strengths from different perspectives are interrelated, complementary, and cannot be replaced by each other.
Together, they form a more complete description.
It should be emphasized that as a later and more fundamental description method, DBM naturally inherits all traditional methods of describing non-equilibrium behavior.

The definition of any non-equilibrium strength depends on the research perspective.
The non-equilibrium strength may be increasing from a certain perspective; while from another perspective, it may be decreasing.
This is one of the concrete manifestations of the complexity of non-equilibrium behavior. Studying non-equilibrium strength and effects from only one perspective is often one-sided or even incorrect.
Being incorrect refers to the misconception that the conclusions obtained are independent of the research perspective and are universal.
In view of this, DBM introduces a multi-perspective non-equilibrium description scheme based on non-conserved moments of $(f-f^{(0)})$, gradients of macroscopic quantities, thermodynamic relaxation time $\tau$, Kn number, morphological description, etc., to provide a cross-positioning of non-equilibrium intensity of complex flows from multiple perspectives.
For the convenience of description, the concept of non-equilibrium strength vector is further introduced, where each component of the vector represents a non-equilibrium intensity from a different perspective.
For example, the composition of a non-equilibrium strength vector may be
\begin{equation}
{\mathbf{S}} = \left\{ {D_2 ,D_{3} ,|\bm{\nabla} \rho| ,|\bm{\nabla} T|,\tau ,{\text{Kn}}, \cdots } \right\},
\end{equation}
where
\begin{equation}
D_2  = \sqrt {\Delta _{2,xx}^{*2}  + \Delta _{2,xy}^{*2}  + \Delta _{2,yy}^{*2} } ,
\end{equation}
is the distance in the $\bm{\Delta}_{2}^{*}$ sub-space of non-conservative moment phase space, and is the non-equilibrium strength from the perspective of $\bm{\Delta}_{2}^{*}$.
$T$ is the local temperature.
In some cases, $D_{2}$ is defined as
\begin{equation}
D_2  = \sqrt {\Delta _{2,xx}^{*2}  + \Delta _{2,xy}^{*2}  + \Delta _{2,yx}^{*2}  + \Delta _{2,yy}^{*2} }
.
\end{equation}
Similar but different options exist for defining non-equilibrium strengths such as $D_{3}$.
These different choices correspond to different research perspectives.

Just as the description of phase space does not require its coordinates with the same unit, the various components of the non-equilibrium intensity vector also do not need to have the same unit.
The non-equilibrium strength from different perspectives, as a set of behavioral characteristics, can also be described using the phase space method, providing an intuitive geometric correspondence for the non-equilibrium strength of complex flows from different perspectives.
Of course, as a coarse-grained modeling and description method, the physical accuracy of DBM, including the precision of describing behavioral characteristics such as non-equilibrium intensities, should be adjusted based on specific research requirements~\cite{Xu2022CMK}.

\subsection{Additional remarks on DBM}

At a Report on the 2nd U.S.-Japan Joint Seminar, held at Santa Barbara, California, from 7-10 August, 1996, the late eminent scholar, Mr. Chang-Lin Tien, et al. commented the LBM as follows: ``Many physical phenomena and engineering problems may have their origins at molecular scales, although they need to interface with the macroscopic or `human scales'.
The difficulty arises in bridging the results of these models across the span of length and time scales. The lattice Boltzmann method attempts to bridge this gap''.
More details are referred to Ref.~\cite{Tien1997MTE}.
Many people's understanding on LBM is directly or indirectly affected by this or similar comments.

In fact, starting from its predecessor, the lattice gas (cellular automata) method, there are two types of LBMs in the literature: coarse-grained physical model construction method~\cite{Barry2010book, Lee1952PR, Succi2001book} and numerical solving method for certain control equations~\cite{He2009book, Guo2013book, Huang2015book}.
The latter accounts for the vast majority of existing literature, to the extent that LBM has almost become the abbreviation or synonym for the latter.
However, the LBM method mentioned by Mr. Chang-Lin Tien in the above review was clearly a cross scale physical model construction method, which falls into the former category.
Coarse-grained physical modeling is a fundamental means for non-equilibrium statistical physics to address the main contradiction based on research needs.
Cross scale,  here referred to as cross Kn number, is a physical function inherited from the Boltzmann equation description.
These two different types of LBM work on complementary dimensions, with different goals and construction rules, each of which is reasonable.

DBM is often referred to as the Discrete Boltzmann Method, Discrete Boltzmann Modeling, and Discrete Boltzmann Model without causing ambiguity.
As mentioned earlier, the more complete term for the Discrete Boltzmann Method is Discrete Boltzmann modeling and analysis Method.
When compared with other methods, DBM's identity is ``physical model construction method+complex physical field analysis method'', or rather a physical model construction method with built-in complex physical field analysis function.
When compared with other physical models, DBM is a physical model with built-in complex physical field analysis capabilities.
Because the commonly used physical models currently do not provide complex physical field analysis capabilities for simulated data, when referring to the term ``discrete Boltzmann model or modeling'', people often do not realize that it also includes complex physical field analysis function.
This is why our research group has to refer to it as the ``physical model construction and complex physical field analysis method'' in some situations.

The physical model construction part of DBM includes determining specific control equations and providing the most necessary physical constraints to be followed  for selecting discrete velocities.
Just like in NS simulation, in DBM simulation, the specific discretization formats for spatial derivatives, time integrals, and particle velocities should be chosen reasonably based on the specific operating conditions and flow patterns.

According to Eq. (\ref{eq.18}), from a dimensional perspective and physical image, the physical meaning of non-equilibrium feature quantities $\bm{\Delta}_{n+1}^{*}$ is the non-organized flux of $\bm{\Delta}_{n}^{*}$.
It should be noted that in macroscopic description, a physical quantity (such as density $\rho$) multiplied by flow velocity $\mathbf{u}$, its physical meaning is the flux of this physical quantity, which is an organized flux.
The combination of organized flux and non-organized flux constitutes a more comprehensive description of complex flow kinetic behavior.
Among all the non-organized fluxes composed of non-conserved central moments of ($f-f^{(0)}$), only Non-Organized Momentum Flux (NOMF) $\bm{\Delta}_{2}^{*}$ and Non-Organized Energy Flux (NOEF) $\bm{\Delta}_{3,1}^{*}$ have counterparts in the NS description, corresponding to viscous stress and heat flux, respectively.
The remaining non-conserved central moments are not directly included in NS, but as mentioned earlier, their physical images are clear.
The absence of these physical quantities in the NS description is precisely the physical reason of NS's inability to describe well situations with higher degrees of discreteness/non-equilibrium.
As the degree of discreteness/non-equilibrium increases, these non-organized fluxes should receive more attention.

Since the traditional fluid mechanics theory is based on NS, we are most familiar with the NS equation.
Comparing with NS is an important way to understand new models.
Therefore, it is necessary to clarify the connection and difference between DBM and NS again: NS is just a physical model, and it does not inherently possess complex physical field analysis capabilities for data, it only considers first-order (in terms of the Kn number perspective) discrete/non-equilibrium effects, which is only applicable to quasi-continuous and near-equilibrium situations.
In terms of numerical simulation research, NS is only responsible for pre-simulation and not for post-simulation.
In contrast, DBM not only provides modeling capabilities equivalent to the corresponding level of Extended Hydrodynamic Equations (EHE) but also provides a set of complex physical field analysis methods for data.
In terms of numerical simulation research, DBM is not only responsible for pre-simulation, but also for post-simulation.

\begin{figure*}[htbp]
\center\includegraphics*
[width=0.8\textwidth]{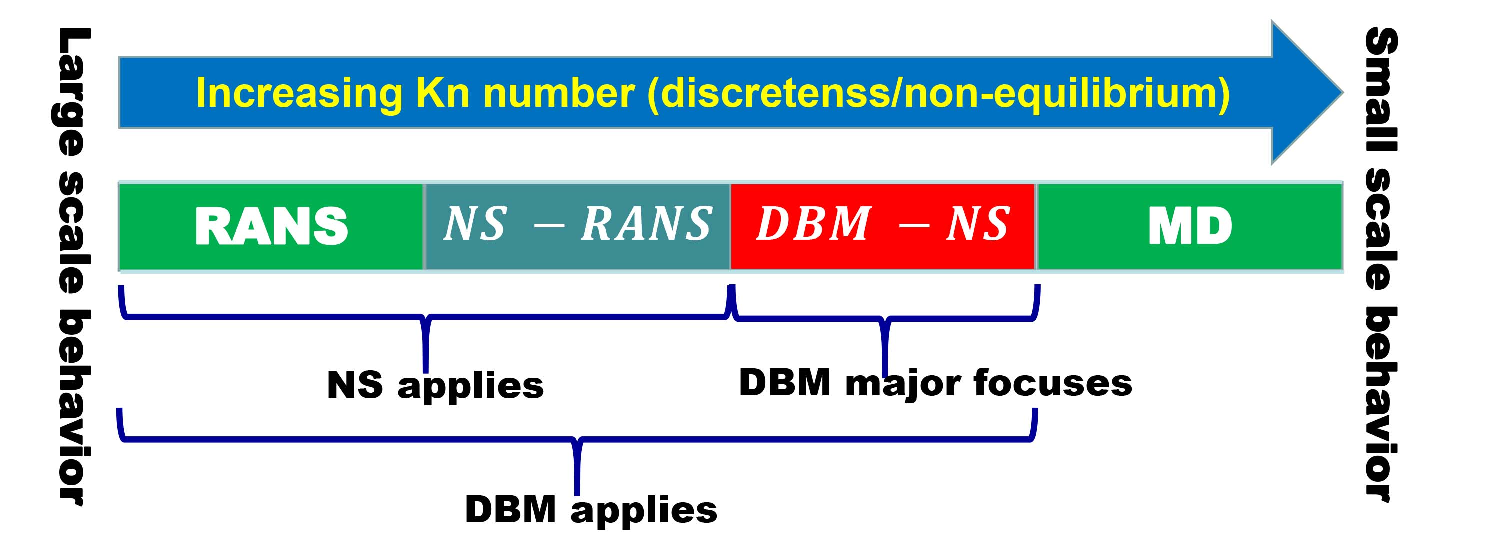}
\caption{ Schematic for application ranges of several physical models.
} \label{fig008}
\end{figure*}

DBM is beyond traditional fluid modeling in both its applicable range and physical capabilities, focusing more on the ``mesoscale'' dilemma where macroscopic continuous models lack physical functionality and microscopic molecular dynamics simulations are powerless due to limited scale applicability, as shown in Fig. \ref{fig008}.
In the schematic, RANS is the abbreviation for Reynolds Average Navier Stokes, which is a further coarse-grained physical modeling based on NS and describes large-scale behavior within the NS range.
The ``NS$-$RANS'' in the schematic represents the portion that RANS fails to well describe within the scope of NS application, where ``$-$'' is a minus sign, indicating subtraction from the scope of application.
The ``DBM$-$NS'' indicates the part within the scope of DBM that is not well described by NS and is the key focus of DBM.

The greater challenge in the study of mesoscale behavior lies in the modeling, simulation, and analysis of higher degrees of discreteness/non-equilibrium.
The necessity of the most closely related non-conserved moments increases as the degree of discreteness/non-equilibrium increases.
Although the goal is for situations with higher degrees of discreteness/non-equilibrium, almost everyone has to go through quasi continuous and near equilibrium situations because he or she requires cognitive and technical foundations.
This is why progress in the field of mesoscale behavior research can be relatively slow, especially for those just entering the field.

Compared with macroscopic behavior, the typical characteristics of ``mesoscale'' behavior are significant discrete effects and significant thermodynamic non-equilibrium effects. The description of mesoscale behavior requires more physical quantities.
The ``mesoscale'' characteristics of DBM include: (i) in terms of scale, between micro and macro, connecting micro and macro; (ii) in terms of functionality, surpassing NS in describing discrete/non-equilibrium effects, but weaker than MD.
In DBM, the corresponding solutions to the six basic scientific problems proposed in section \ref{sec:level2} are as follows: For problem (i), add physical quantities based on the definition of non-conserved kinetic moments according to the degree of discreteness/non-equilibrium.
In response to problem (ii), phase space description methods based on non-conserved moments of $(f-f^{(0)})$, morphological and other behavioral feature quantities are introduced.
In response to problem (iii), the model construction no longer relies on continuity and near equilibrium assumptions.
To address problem (iv), develop a description method that starts from macro continuous near equilibrium and gradually increases the degree of discreteness and non-equilibrium. For problem (v), develop mesoscale modeling based on the combination of kinetic theory and mean field theory.
For problem (vi), introduce the concept of non-equilibrium strength vector, where each component represents a non-equilibrium intensity, for cross localization descriptions.
Table \ref{Table1} summarizes the developmental milestones in the evolution of DBM.

\begin{center}
\begin{table*}
\begin{tabular}{p{2cm}<{\centering} p{15cm}<{\centering}}
\hline
\hline
Year    & Schemes for detecting and describing discrete/TNE behavior and effects \\
\hline
Before 2012    &There was no significant difference in physical function between the two types of LBMs. \\
2012 &    Proposed to use non-conserved moments of $(f-f^{(0)})$ to detect and describe discrete/TNE states and effects, which is the starting point of current DBM method~\cite{Xu2012FOP}.
 \\
2015 & Proposed to use the non-conserved moments of $(f-f^{(0)})$ as bases to open phase space, and use the distance from a state point to the origin to define the TNE strength of one perspective. This is the starting point of the phase space description method in DBM~\cite{Xu2015PRE}.
 \\
2018 & Proposed to use the distance between two points in the phase space to describe the difference between two discrete/TNE states, and use the mean distance between two points in a given time interval to describe the difference of two kinetic processes~\cite{Xu2018RGD}.   
\\
2021 & Extended the phase space method to describe any set of system features~\cite{Xu2021CJCP}.   
\\
2022 & Proposed the concept of non-equilibrium strength vector, each of whose components is one TNE strength of a perspective, to multi-perspective cross-locate the non-equilibrium strength of complex flow~\cite{Xu2022CMK,Zhang2022POF}.
\\
\hline
\hline
\end{tabular}
\caption{Milestones in the development of DBM.}
\label{Table1}
\end{table*}
\end{center}

\subsection{Modeling examples and flowchart}

According to the fundamental theory of non-equilibrium statistical physics, there are infinite ways in which the distribution function $f$ deviates from the equilibrium distribution function $f^{(0)}$.
There are infinitely many non-conserved moments of $(f-f^{(0)})$, and there are infinite possibilities for which moments have larger amplitudes (or intensities) and which moments have smaller amplitudes.
The approach of DBM involves breaking down complex problems, selecting a perspective, studying a set of kinetic properties of the system, and requiring the non-conserved moments that describe these properties to remain invariant during model simplification. Based on the independent components of the set of kinetic moments of $(f-f^{(0)})$, construct a phase space and use it and its subspaces to describe the discrete and non-equilibrium behavior characteristics of the system.
The research perspective and modeling accuracy adjust as the research progresses.
Therefore, in principle, DBM modeling itself can be independent of CE multiscale analysis. As long as it is possible to obtain which kinetic moments of $f^{(0)}$ are necessary for grasping system behavior in a certain way, it is only required that the corresponding kinetic moments remain unchanged when converted into summation for calculation.

Given the staged nature of the research, we will only provide the discussion on DBM modeling based on CE multi-scale analysis theory.
By utilizing CE multi-scale analysis and preserving the TNE effect of different orders, the Boltzmann equation can recover, which is actually reduced to the corresponding order of fluid dynamic equations..
In principle, it is not necessary, but most of the published DBM modeling is still within the framework of CE multi-scale analysis theory.
Unlike the LBM numerical solving scheme for some given control equations, in the DBM modeling process, the role of CE multiscale analysis is not to help derive the corresponding EHE functions, but only to quickly identify and confirm the kinetic moments of $f^{(0)}$ that need to be preserved based on the tensor order of particle velocity $\mathbf{v}$.
Whether  the specific form of the corresponding EHE is known or not does not affect DBM modeling and simulation.
The CE multi-scale analysis theory serves as  a reasonable theoretical basis for such kinetic models.

In fact, even in the case of a simple single-medium system, the theoretical derivation of 2nd-order or higher discrete and non-equilibrium effects in the KMM or EHE is already quite complex, let alone in the case of multi-medium scenarios.
Even if EHE, which is functionally equivalent to DBM, is ultimately derived, the feasibility of numerical simulation remains a huge challenge.
Because it involves stronger nonlinearity and higher-order spatiotemporal partial derivatives, and its number of terms increases sharply with the increase of discreteness and non-equilibrium degree, these pose substantial challenges to practical numerical simulation. Therefore, as the degree of discreteness or non-equilibrium increases, the KMM modeling and simulation approach quickly becomes infeasible.

Unlike KMM, DBM belongs to the kinetic direct modeling method.
Except for explaining and verifying the functions of the model, functionally equivalent EHE is not required in actual DBM simulations.
As the degree of discreteness or non-equilibrium increases, the complexity of DBM modeling and simulation approach also increases, but it is relatively slow, making it possible to go further.

Below, we illustrate the process of DBM modeling for an ideal monatomic gas in a single medium, showing the process of DBM modeling with different discrete or non-equilibrium degrees.
The general form of CE multi-scale expansion is
\begin{equation}
\begin{gathered}
  f = f^{(0)}  + {\text{Kn }}f^{(1)}  + {\text{Kn}}^2 f^{(2)}  +  \cdots {\text{            (a)}}, \hfill \\
  \frac{\partial }
{{\partial t}} = {\text{Kn}}\frac{\partial }
{{\partial t_1 }} + {\text{Kn}}^2 \frac{\partial }
{{\partial t_2 }} + \cdots {\text{                      (b)}}, \hfill \\
  {\text{ }}\frac{\partial }
{{\partial \mathbf{r}  }} = {\text{Kn}}\frac{\partial }
{{\partial \mathbf{r}_{1} }}+\cdots{\text{                                       (c)}}. \hfill \\
\end{gathered}
\end{equation}
To facilitate the interpretation of the compositions of distribution function, temporal rate of change, and spatial rate of change, we absorb the coefficients   $\text{Kn}^{j}$
into the corresponding non-equilibrium distribution function   $f^{(j)}$ , temporal rate of change $\partial/\partial t_{j}$  and spatial rate of change  $\partial/\partial \mathbf{r}_{j}$.
Thus, the CE multi-scale expansion becomes
\begin{equation}
\begin{gathered}
  f = f^{(0)}  + f^{(1)}  + f^{(2)}  + \cdots {\text{         (a)}} \hfill \\
  \frac{\partial }
{{\partial t}} = \frac{\partial }
{{\partial t_1 }} + \frac{\partial }
{{\partial t_2 }} + \cdots {\text{                      (b)}} \hfill \\
  {\text{ }}\frac{\partial }
{{\partial \mathbf{r}  }} = \frac{\partial }
{{\partial \mathbf{r}_{1}}}+\cdots{\text{                                (c)}} \hfill \\
\label{eq.27}
\end{gathered}
\end{equation}
In this way, the meaning of  $f^{(j+1)}$ is the newly added $(j+1)$th order more detailed description on basis of the previous $j$th order level coarse-grained description, $f = f^{(0)}  + f^{(1)}  + f^{(2)}  +  \cdot  \cdot  \cdot + f^{(j)} {\text{ }}$.
For the description of temporal change rate, the meaning of $\partial/\partial t_{j+1}$ is, based on the previous coarse-grained description $\partial t = \partial t_1 {+ }\partial t_2  +  \cdots  + \partial t_j $, as the temporal scale decreases from $t_{j}$ to $t_{j+1}$, the higher-frequency component that we observe.
For the description of spatial change rate, the meaning of $\partial /\partial {{\mathbf{r}}_{j+1}}$is the $(j+1)$th order more-detailed structure that we newly observe on the basis of the previous $j$th order coarse-grained description, $\partial \mathbf{r}=\partial {{\mathbf{r}}_{1}} + \partial {{\mathbf{r}}_{2}}+\cdots +\partial {{\mathbf{r}}_{j}}$, when the spatial scale decreases from ${{\mathbf{r}}_{j}}$ to ${{\mathbf{r}}_{(j+1)}}$. 
It should be noted that in the existing CE theory in the literature, the spatial change rate only retains the first order.
As mentioned earlier, the corresponding scenario of $f^{(0)}$ is continuous and equilibrium. So, what do $t_0$ and $\mathbf{r}_{0}$ correspond to?
Reference ~\cite{Xu2022CMK} points out that $t_0$ corresponds to the total time span of the system behavior that we aim to investigate.
When we use the time scale $t_0$ to examine the rate of change of system behavior over time, we cannot see the change, that is $\partial t_0 = 0$.
This is the reason why this term does not appear in CE expansion.
Similarly, $\mathbf{r}_{0}$ corresponds to the total space span of the system behavior that we aim to investigate.
When we use the space scale $\mathbf{r}_{0}$ to examine the rate of change of system behavior over space, we cannot see the change, that is $\partial \mathbf{r}_{0} = 0$.
This is the reason why this term does not appear in CE expansion.
In the following discussions on CE multiscale analysis, the spatial change rate $\partial \mathbf{r}$ is only discussed at one spatial scale $\mathbf{r}_{1}$.

Substituting CE expansion (\ref{eq.27}) into the BGK-Boltzmann equation,
\begin{equation}
\frac{{\partial f}}
{{\partial t}} + {\mathbf{v}} \cdot \frac{{\partial f}}
{{\partial {\mathbf{r}}}} =  - \frac{1}
{\tau }\left( {f - f^{(0)} } \right),
\end{equation}
gives
\begin{equation}
\begin{gathered}
  (\frac{\partial }
{{\partial t_1 }} + \frac{\partial }
{{\partial t_2 }} +  \cdots )(f^{(0)}  + f^{(1)}  + f^{(2)} {\text{ + }} \cdots ) \\
+ \mathbf{v} \cdot \frac{\partial }
{{\partial \mathbf{r}_{1} }}(f^{(0)}  + f^{(1)}  + f^{(2)} {\text{ + }} \cdots ) \hfill \\
  {\text{                                             }} =  - \frac{{\text{1}}}
{\tau }(f^{(1)}  + f^{(2)} {\text{ + }} \cdots ). \hfill \\
\label{eq.29}
\end{gathered}
\end{equation}
where $\tau$ is the relaxation time.
By equating the coefficients of the terms with the same order of Kn number on both sides of equation (\ref{eq.29}), we can obtain
\begin{equation}
f^{{\text{(1)}}}  = \tau [ - \frac{{\partial f^{(0)} }}
{{\partial t_1 }} - \frac{\partial }
{{\partial \mathbf{r}_{1} }}(f^{(0)} \mathbf{v} )] = \tau [ - \frac{{\partial f^{(0)} }}
{{\partial t_1 }} - \bm{\nabla}_1  \cdot (f^{(0)} {\mathbf{v}})] ,
\end{equation}
\begin{equation}f^{{\text{(2)}}}  =  - \tau [\frac{{\partial f^{(0)} }}
{{\partial t_2 }} + \frac{{\partial f^{(1)} }}
{{\partial t_1 }} + \bm{\nabla}_1  \cdot (f^{(1)} {\mathbf{v}})] ,
\end{equation}
\begin{equation}
f^{(3)}  =  \cdots .
\end{equation}

Assume we want to propose a DBM that only includes first-order discrete and non-equilibrium effects. It is easy to find
\begin{equation}
\begin{gathered}
  {\bm{\Delta }}_2^{*(1)}  = {\bm{\Delta }}_2^{(1)}  = {\mathbf{M}}_2 (f_{}^{(1)} ) \hfill \\
   = \tau {\bm{M}}_2 ( - \frac{{\partial f_{}^{(0)} }}
{{\partial t_1 }} - \bm{\nabla} _1  \cdot (f_{}^{(0)} {\mathbf{v}}_{} )) \hfill \\
   \Rightarrow {\mathbf{M}}_2 \left( {f^{\left( 0 \right)} } \right),{\mathbf{M}}_3 \left( {f^{\left( 0 \right)} } \right), \hfill \\
\end{gathered}
\end{equation}
\begin{equation}
\begin{gathered}
  {\bm{\Delta }}_{3,1}^{(1)}  = {\mathbf{M}}_{3,1} (f_{}^{(1)} ) \hfill \\
   = \tau {\mathbf{M}}_{3,1} ( - \frac{{\partial f_{}^{(0)} }}
{{\partial t_1 }} - \bm{\nabla} _1  \cdot (f_{}^{(0)} {\mathbf{v}}_{} )) \hfill \\
   \Rightarrow {\mathbf{M}}_{3,1} \left( {f^{\left( 0 \right)} } \right),{\mathbf{M}}_{4,2} \left( {f^{\left( 0 \right)} } \right), \hfill \\
\end{gathered}
\end{equation}
\begin{equation}
\begin{gathered}
  {\bm{\Delta }}_{3,1}^{{\text{*}}(1)}  = {\bm{\Delta }}_{3,1}^{(1)}  - {\mathbf{u}} \cdot {\bm{\Delta }}_2^{*(1)}  \hfill \\
   =  \cdots  \hfill \\
   \Rightarrow {\mathbf{M}}_{3,1} \left( {f^{\left( 0 \right)} } \right),{\mathbf{M}}_{4,2} \left( {f^{\left( 0 \right)} } \right). \hfill \\
\end{gathered}
\end{equation}
Among them, the meaning of the symbol ``$ \Rightarrow $'' is "involving".
Therefore, the set of kinetic moments involved in the viscous stress and heat flux calculation formulas are the following seven kinetic moments, $\left\{ {{\mathbf{M}}_0 ,{\mathbf{M}}_1 ,{\mathbf{M}}_{2,0} ,{\mathbf{M}}_2 ,{\mathbf{M}}_{3,1} ,{\mathbf{M}}_3 ,{\mathbf{M}}_{4,2} } \right\}$.
Consequently, DBM requires that these seven kinetic moments be strictly preserved during the process of converting from integrals to summations:	
\begin{equation}
\mathbf{M}_{0}=\overset{N}{\underset{i=1}{\sum }}f_{i}^{eq}=\rho \text{,}
\label{eq.36}
\end{equation}%
\begin{equation}
\mathbf{M}_{1}=\overset{N}{\underset{i=1}{\sum }}f_{i}^{eq}\mathbf{v}%
_{i}=\rho \mathbf{u}\text{,}  \label{M1}
\end{equation}%
\begin{equation}
\mathbf{M}_{2,0}=\overset{N}{\underset{i=1}{\sum }}\frac{1}{2}%
f_{i}^{eq}(v_{i}^{2}+\eta _{i}^{2})=\frac{1}{2}\rho \lbrack (D+n)T+u^{2}]%
\text{,}  \label{M20}
\end{equation}%
\begin{equation}
\mathbf{M}_{2}=\overset{N}{\underset{i=1}{\sum }}f_{i}^{eq}\mathbf{v}_{i}%
\mathbf{v}_{i}=\rho (T\mathbf{I}+\mathbf{uu})\text{,}  \label{M2}
\end{equation}%
\begin{equation}
\mathbf{M}_{3,1}=\overset{N}{\underset{i=1}{\sum }}\frac{1}{2}%
f_{i}^{eq}(v_{i}^{2}+\eta _{i}^{2})\mathbf{v}_{i}=\frac{1}{2}\rho \mathbf{u}%
[(D+n+2)T+u^{2}]\text{,}  \label{M31}
\end{equation}%
\begin{equation}
{\mathbf{M}}_{3}=\overset{N}{\underset{i=1}{\sum }}{f_{i}^{eq}}\mathbf{v}%
_{i}\mathbf{v}_{i}\mathbf{v}_{i}=\rho (T\bm{\Theta}+\mathbf{uuu})\text{,}
\end{equation}%
\begin{equation}
\begin{gathered}
{\mathbf{M}}_{4,2}=\overset{N}{\underset{i=1}{\sum }}{f_{i}^{eq}}\frac{%
v_{i}^{2}+\eta _{i}^{2}}{2}\mathbf{v}_{i}\mathbf{v}_{i}=\rho \lbrack (\frac{%
D+n+2}{2}T+\frac{u^{2}}{2})T\mathbf{I}\\
+(\frac{D+n+4}{2}T+\frac{u^{2}}{2})%
\mathbf{uu}]\text{,}
\label{eq.42}
\end{gathered}
\end{equation}%
with $\bm{\Theta}=(u_{\alpha }\delta _{\beta \gamma }+u_{\beta }\delta
_{\alpha \gamma }+u_{\gamma }\delta _{\alpha \beta })\widehat{\mathbf{e}}%
_{\alpha }\widehat{\mathbf{e}}_{\beta }\widehat{\mathbf{e}}_{\gamma }$, $(%
\widehat{\mathbf{e}}_{\alpha },\widehat{\mathbf{e}}_{\beta },\widehat{%
\mathbf{e}}_{\gamma })$ denote unit vectors along the $\alpha $, $\beta $,
and $\gamma $ axes of a fixed coordinate system.
In equations (\ref{eq.36})-(\ref{eq.42}), the left side represents the analytical solution for the kinetic moment of the equilibrium distribution function $f^{eq}=f^{(0)}$.
These equations (\ref{eq.36})-(\ref{eq.42}) can be written in matrix form as,
\begin{equation}
\hat f_k ^{eq}  = m_{ki} f_i ^{eq},
\label{eq.43}
\end{equation}
or
\begin{equation}
{\mathbf{\hat f}}^{eq} = {\mathbf{m\cdot f}}^{eq} .
\label{eq.44}
\end{equation}
Equations (\ref{eq.36})-(\ref{eq.42}), or the matrix equation (\ref{eq.43}) or (\ref{eq.44}) represent the physical constraint that need to be satisfied for the discrete velocity selection in the DBM evolution equation.

If we want to achieve a DBM that includes second-order discrete/non-equilibrium effects, we only need to consider: which additional kinetic moments of $f^{(0)}$ should be preserved after adding the second-order non-equilibrium contributions to viscous stress and heat flux?
Because
\begin{equation}
\begin{gathered}
  {\bm{\Delta }}^{(2)}  = {\mathbf{M}}_2 (f_{}^{(2)} ) \Rightarrow {\mathbf{M}}_2 (f_{}^{(1)} ),{\mathbf{M}}_3 (f_{}^{(1)} ), \hfill \\
  {\text{where }}{\mathbf{M}}_3 (f_{}^{(1)} ) \Rightarrow {\mathbf{M}}_4 (f_{}^{(0)} ), \hfill \\
\end{gathered}
\end{equation}
\begin{equation}
\begin{gathered}
  {\mathbf{M}}_{3,1} (f_{}^{(2)} ) \Rightarrow {\mathbf{M}}_{3,1} (f_{}^{(1)} ),{\mathbf{M}}_{4,2} (f_{}^{(1)} ), \hfill \\
  {\text{where }}{\mathbf{M}}_{4,2} (f_{}^{(1)} ) \Rightarrow {\mathbf{M}}_{5,3} (f_{}^{(0)} ). \hfill \\
\end{gathered}
\end{equation}
It can be seen that there are two more kinetic moments, $\bm{\Delta}_4$ and $\bm{\Delta}_{5,3}$, that need to be conserved.
 It is clear that, in a DBM containing only up to the second-order discrete/non-equilibrium effects, the set of kinetic moments that need to be preserved are $\left\{ {{\mathbf{M}}_0 ,{\mathbf{M}}_1 ,{\mathbf{M}}_{2,0} ,{\mathbf{M}}_2 ,{\mathbf{M}}_{3,1} ,{\mathbf{M}}_3 ,{\mathbf{M}}_{4,2} ,{\mathbf{M}}_4 ,{\mathbf{M}}_{5,3} } \right\}$.

\begin{figure*}[htbp]
\center\includegraphics*
[width=0.6\textwidth]{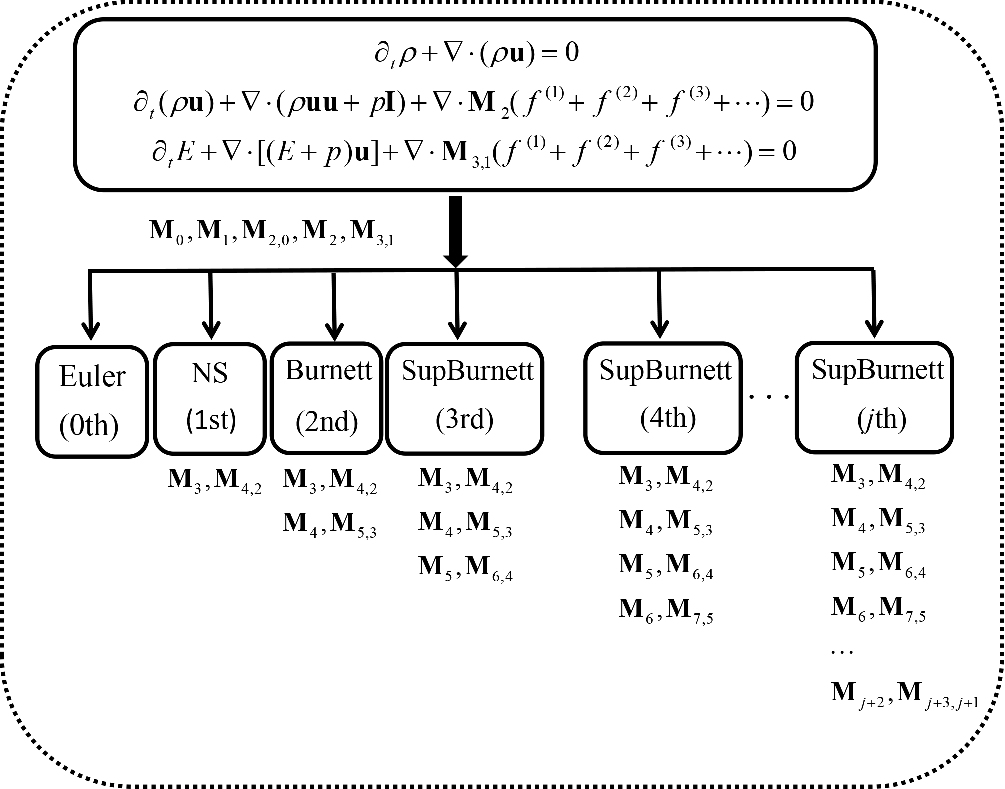}
\caption{  Kinetic moments that should keep value in various levels of DBM.
} \label{fig009}
\end{figure*}

\begin{figure*}[htbp]
\center\includegraphics*
[width=0.5\textwidth]{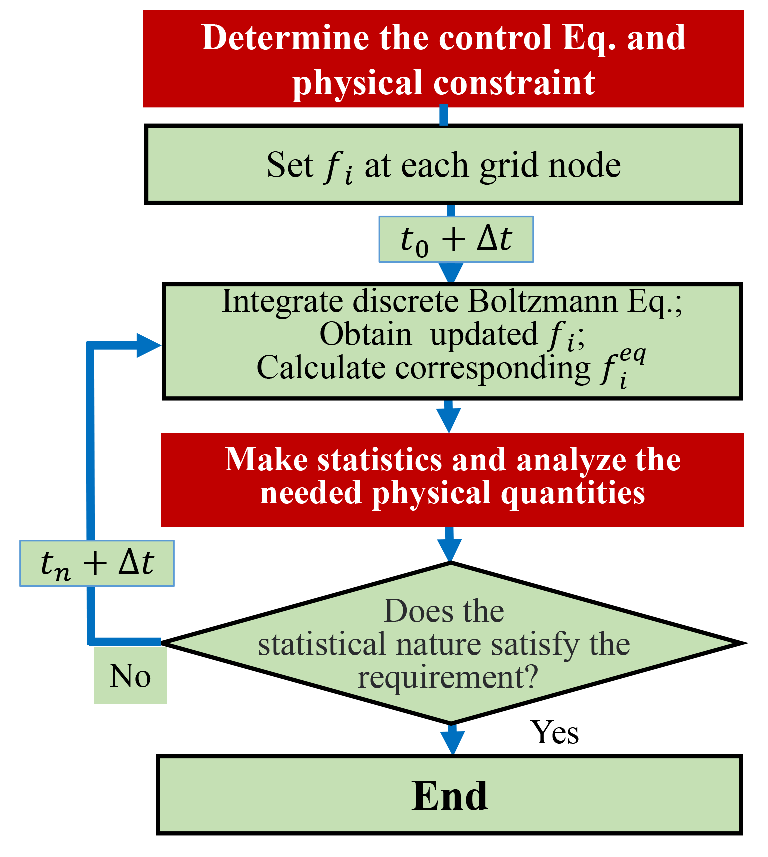}
\caption{  Flowchart for DBM simulation.
} \label{fig0010}
\end{figure*}

Figure \ref{fig009} shows the kinetic moments that need to be preserved in the step-by-step DBM construction as the degree of discreteness/non-equilibrium represented by the Kn number increases and the corresponding hydrodynamic equations.
It is easy to find that when considering up to the $j$-th order discreteness/non-equilibrium, the number of kinetic moments that need to be preserved is $3+2 (j+1)$.
For more details, please refer to references~\cite{Zhang2019Dr, Gan2018PRE, Gan2022JFM, Zhang2022POF}.
It is emphasized again that the ability to recover the corresponding level of hydrodynamic equations is only a part of the physical functionality of DBM.
Roughly equivalent to the physical functions of DBM is the extended hydrodynamic equation systems, which should include some of the most closely related non-conserved moment evolution equations in addition to the usual density, momentum, and energy evolution equations.
Among them, the extended part is referred to the set of evolution equations of corresponding non-conserved moments, and its necessity increases with the increase of system discreteness/non-equilibrium degree.
We provide a DBM simulation flowchart in Fig. \ref{fig0010} corresponding to the work positioning shown in Fig. \ref{fig003}.

\subsection{ Brief review of steady DBM }

\begin{figure*}[htbp]
\center\includegraphics*
[width=0.5\textwidth]{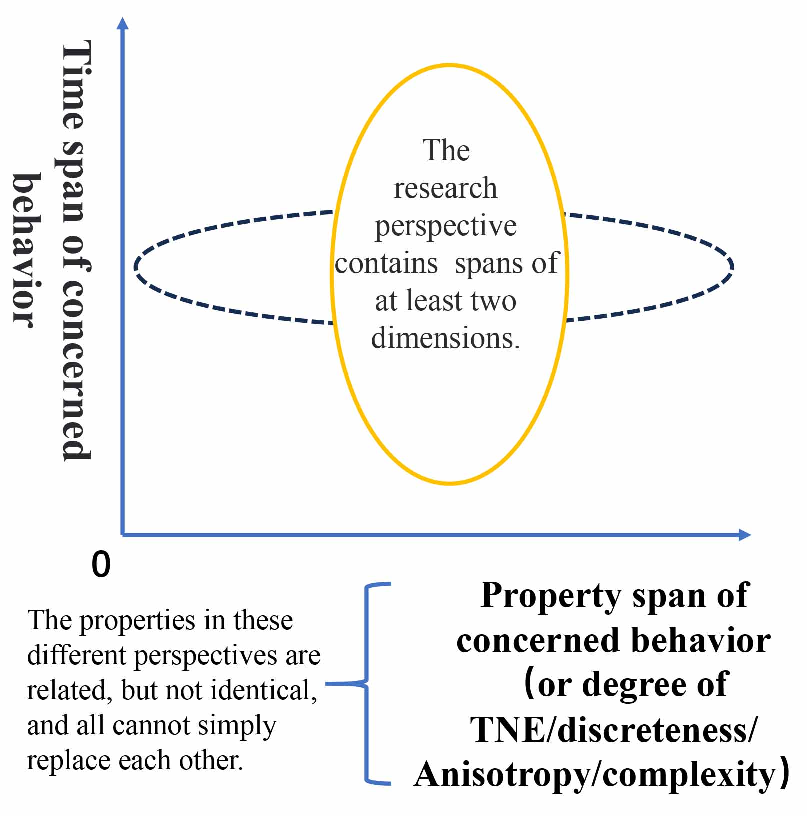}
\caption{ Two dimensions of the perspective of complex flow behavior research.
} \label{fig0011}
\end{figure*}

When conducting research on complex flow behaviors, the research perspective includes at least two dimensions of span, as shown in Fig. \ref{fig0011}: the span of characteristics that make up this set of behaviors and the time span of these behaviors.
Among them, the span of kinetic behavior often correlates with the discrete/non-equilibrium degree of the system being described.
In addition to the kinetic characteristics that evolve over time, people often need to conduct detailed research on some steady-state behavior of the system.
The steady DBM is designed for this requirement.

Formally, in the discrete Boltzmann equation (\ref{eq.15}), setting the time derivative term to zero yields the control equation for steady-state DBM,
\begin{equation}
\mathbf{v}_{i} \cdot \frac{\partial f_{i}}{\partial \mathbf{x}} + (\text{force} \  \text{term})_{i} = (\text{collision}\   \text{term})_{i}
\label{eq.47}
\end{equation}
However, the selection rules for discrete velocities here are completely different from the case of unsteady DBM, and the physical meaning of the discrete distribution function $f_{i}$ is the actual distribution function of corresponding discrete velocity $\mathbf{v}_{i}$.

\begin{figure*}[htbp]
\center\includegraphics*
[width=0.6\textwidth]{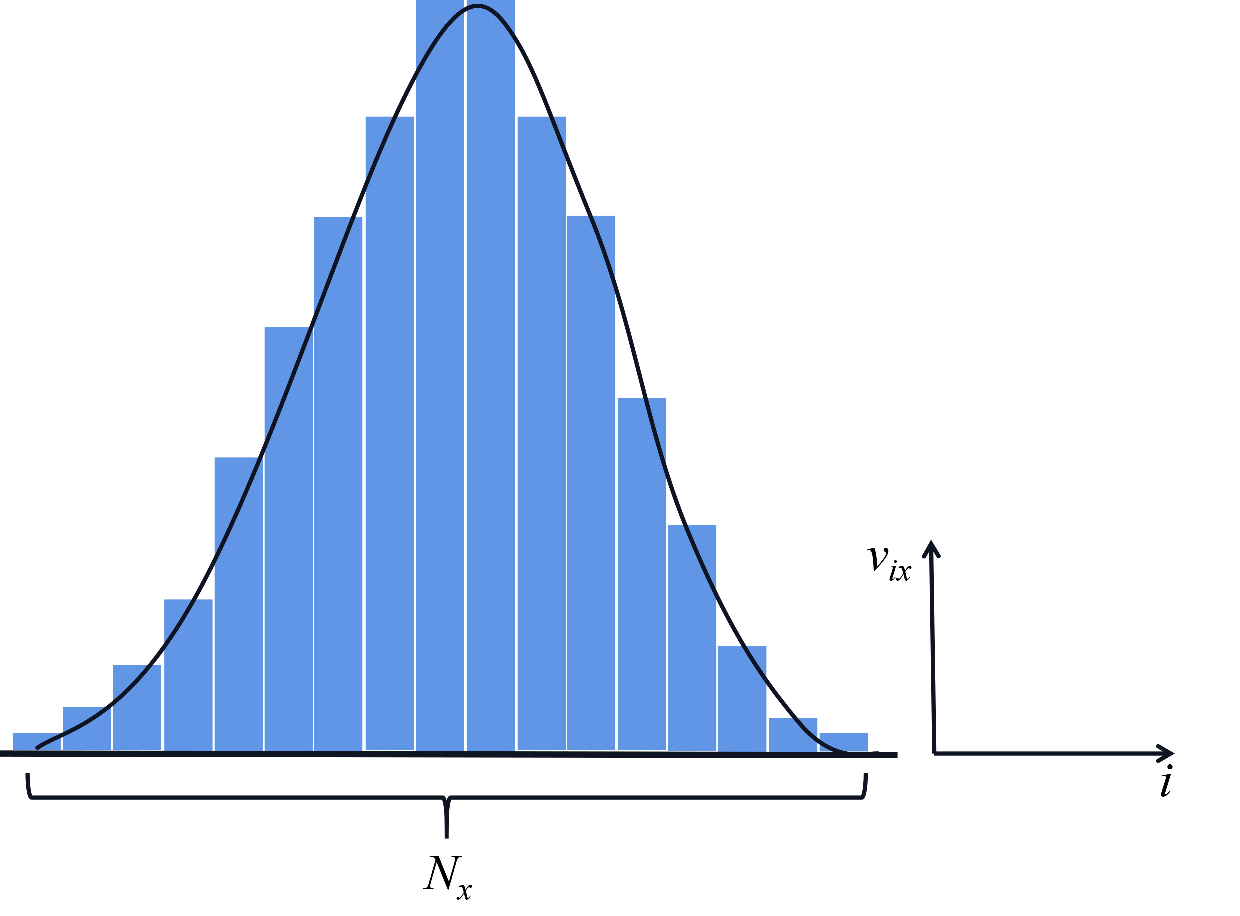}
\caption{  Schematic of discretization of particle velocity space.
The velocity discretization in $x$ direction is taken as example.
} \label{fig0012}
\end{figure*}

Considering that the distribution function $f$ has a small amplitude when
$\left| {{\mathbf{v}} - {\mathbf{u}}} \right|$ is large, the simplest way in steady DBM is to uniformly discretize the particle velocity space in a finite region centered on the local flow velocity $\mathbf{u}$.
In the case of low particle number density, a more reasonable approach is to use non-uniform grids
 with local refinement around the local velocity field $\mathbf{u}$~\cite{Wu2014JFM}.
In reference~\cite{Zhang2023POF}, the discretization of particle velocity space introduces a Lagrangian grid that moves with local flow velocity $\mathbf{u}$.
In this way, as shown in Fig. \ref{fig0012}, in the three-dimensional case, the direct uniform discretization of particle velocity can be written as:

\begin{equation}
\begin{gathered}
  v_{ix}  = \left( {i - \frac{{N_{vx}  + 1}}
{2}} \right)/\left( {\frac{{N_{vx}  - 1}}
{2}} \right)v_{x,\max } , \hfill \\
  v_{jy}  = \left( {j - \frac{{N_{vy}  + 1}}
{2}} \right)/\left( {\frac{{N_{vy}  - 1}}
{2}} \right)v_{y,\max } , \hfill \\
  v_{kz}  = \left( {k - \frac{{N_{vz}  + 1}}
{2}} \right)/\left( {\frac{{N_{vz}  - 1}}
{2}} \right)v_{y,\max } . \hfill \\
\end{gathered}
\end{equation}
where $(N_{vx} ,N_{vy} ,N_{yz} )$ are the total numbers of grid in the particle velocity space at three degrees of freedom, $(v_{x,\max } ,v_{y,\max } ,v_{z,\max } )$ are the maximum values of particle velocity in the three dimensions.
Because it is a uniform grid, the weight coefficients of each grid point are the same.
In many cases, it is reasonable to take $v_{x,\max }  = v_{y,\max }  = v_{z,\max }  = v_{\max}$.
But in the case of local refinement, the weight coefficients of grid points are no longer the same.
Taking direction of $\mathbf{v}_{y}$ as an example, we provide an example of using a set of non-uniform grid $v_{jy}$ and weight coefficient $w_j$:
\begin{equation}
\begin{gathered}
  v_{jy}  = \left( {j - \frac{{N_{vy}  + 1}}
{2}} \right)^\lambda  /\left( {\frac{{N_{vy}  - 1}}
{2}} \right)^\lambda  v_{y,\max } , \hfill \\
  w_j  = \lambda \left( {j - \frac{{N_{vy}  + 1}}
{2}} \right)^{\lambda  - 1} /\left( {\frac{{N_{vy}  - 1}}
{2}} \right)^\lambda  v_{y,\max } . \hfill \\
\end{gathered}
\end{equation}
The discretization of particle velocity space in steady DBM given here is only an example,
The purpose is to highlight the differences from non-steady-state DBM.
For more details, please refer to more professional literature.

A brief methodological review on the capabilities and characteristics of the two types of DBMs, time-dependent DBM and steady-state DBM, is provided as follows:

(i) Due to limitations in simulation capabilities, we often need to compromise and trade-offs between the two spans, time span and behavior span, of the descriptive perspective.
The kinetic behavior span is closely related to the degree of discreteness/non-equilibrium.
If we compress the time span of our attention, it is possible to exchange it for broadening the behavioral span of our attention.
If the time span of the focus is compressed to a single point, the span of the behavior of the focus can be expanded to all the kinetic moments of the distribution function $f$, i.e., all the kinetic features of the complete and real $f$.
This is the limit extension of the degree of discreteness/non-equilibrium that can be described under the effective within the framework of BGK-like model and Chapman-Enskog multi-scale expansion theory.

(ii) At this limit, the number of kinetic moments to be preserved becomes infinite.
In the ideal case, $f_i$ becomes the real distribution function corresponding to $\mathbf{v}_{i}$, and particle velocity $\mathbf{v}_{i}$ becomes the real particle velocity.
At this limit,  the linear equations about $f^{(0)}$ that were originally designed to preserve only a finite number of kinetic moments, such as equation (\ref{eq.43}), become infinitely high-dimensional.
Therefore, the original method of selecting discrete velocities is no longer applicable.
When complexity and difficulty reach a certain level, they often turn around and become simpler.
At this point, the discretization of particle velocity space can return to a direct discretization scheme similar to spatial and temporal discretization.

(iii) In time-dependent DBM, only a finite number of kinetic moments of $f$ are preserved, and what has a clear physical meaning is the kinetic moment of $f_i$, but $f_i$ itself does not have a clear physical meaning.
The recovered distribution function is only approximately accurate, and is only accurate in terms of the features of interest.
When it comes to the steady DBM,
the focus is on the actual distribution function, which means that essentially all kinetic moments need to be preserved.
It's not just the kinetic moments of that have clear physical interpretations, but also the distribution function itself, which represents the real distribution function at that particular particle velocity.

(iv) Both time-dependent DBM and time-independent DBM can be used together, achieving the ultimate or maximum descriptive power in two dimensions: the maximum in terms of time span (sacrificing behavior span or degree of discreteness/non-equilibrium, etc.) and the maximum in terms of degree of discreteness/non-equilibrium (sacrificing time span, etc.). The time-dependent DBM and steady-state DBM, similar to the time-dependent Schrödinger equation and the steady-state Schrödinger equation, also show significant differences in numerical simulation methods.
Of course, it also contains the same or similar parts. It is noted that the Schrödinger equation has a clear physical image correspondence and is a physical model description equation.
It does not contain numerical schemes for simulation.

(v) Compared to steady-state DBM, time-dependent DBM compresses the system's behavior description ability from all (infinite) kinetic moments of $f$ to a finite set.
The gain in the system's behavior description ability is that the time span of the behavior is amplified infinitely from 0.
However, there's a trade-off involved in this.
The gain is ``mesoscopic'' and the cost is also ``mesoscopic''.
Both types of DBMs are typical representatives of ``choosing something to do'' or ``doing something and abandoning the others''.

(vi) The transformation of coordinate systems from Eulerian to Lagrangian during the construction process of steady-state DBM in Ref.~\cite{Zhang2023POF} has not yet involved discrete formats and belongs to the theoretical modeling part.
Please note that MD simulation is Lagrangian, and the coordinate system follows the molecule.
The reason why the coordinate system can run with molecule is because in MD, the position of molecule at the next moment is deterministic!
In kinetic theory, the direction and speed $\mathbf{v}$ of molecule at the next moment are uncertain and are probability events.
So, in Lagrangian DBM, the coordinate system cannot follow the (uncertain and probabilistic) molecule, but can and should follow the fluid element with a macroscopic flow velocity $\mathbf{u}$.
This is the technical key in the construction of Lagrangian DBM.

\section{\label{sec:level4} Verification and validation}

In this section we illustrate some verification and validation of the method. 
Figures \ref{fig0012-1} to \ref{fig0027} are for unsteady DBM, and Fig. \ref{fig0028} is for steady DBM.
It should be mentioned that the model verification and validation are based on known results, aims to verify and validate the chosen discrete format, program implementation, parameter settings, etc.
This is a basic process but does not have any contribution to the new physics cognition.

\subsection{ Combustion, detonation and shock wave }

The DBM study of complex flows started from combustion and detonation. 
The first work was completed by Bo Yan \cite{Yan2013FOP}, a postdoctoral fellow of the first author's research group, and then Chuandong Lin \cite{Lin2016CNF,Lin2018CNF,Lin2014CTP,Lin2017SR,Lin2018CAF,Lin2019PRE,
Lin2020CESW,Lin2020Entropy}, Yudong Zhang \cite{Zhang2016CNF}, Yiming Shan \cite{Shan2022JMES}, and some other graduate students of the research group participated in this direction. 
Currently, there is a relatively substantial body of literature on this subject from various research groups.

\begin{figure*}[htbp]
\center\includegraphics*
[width=0.5\textwidth]{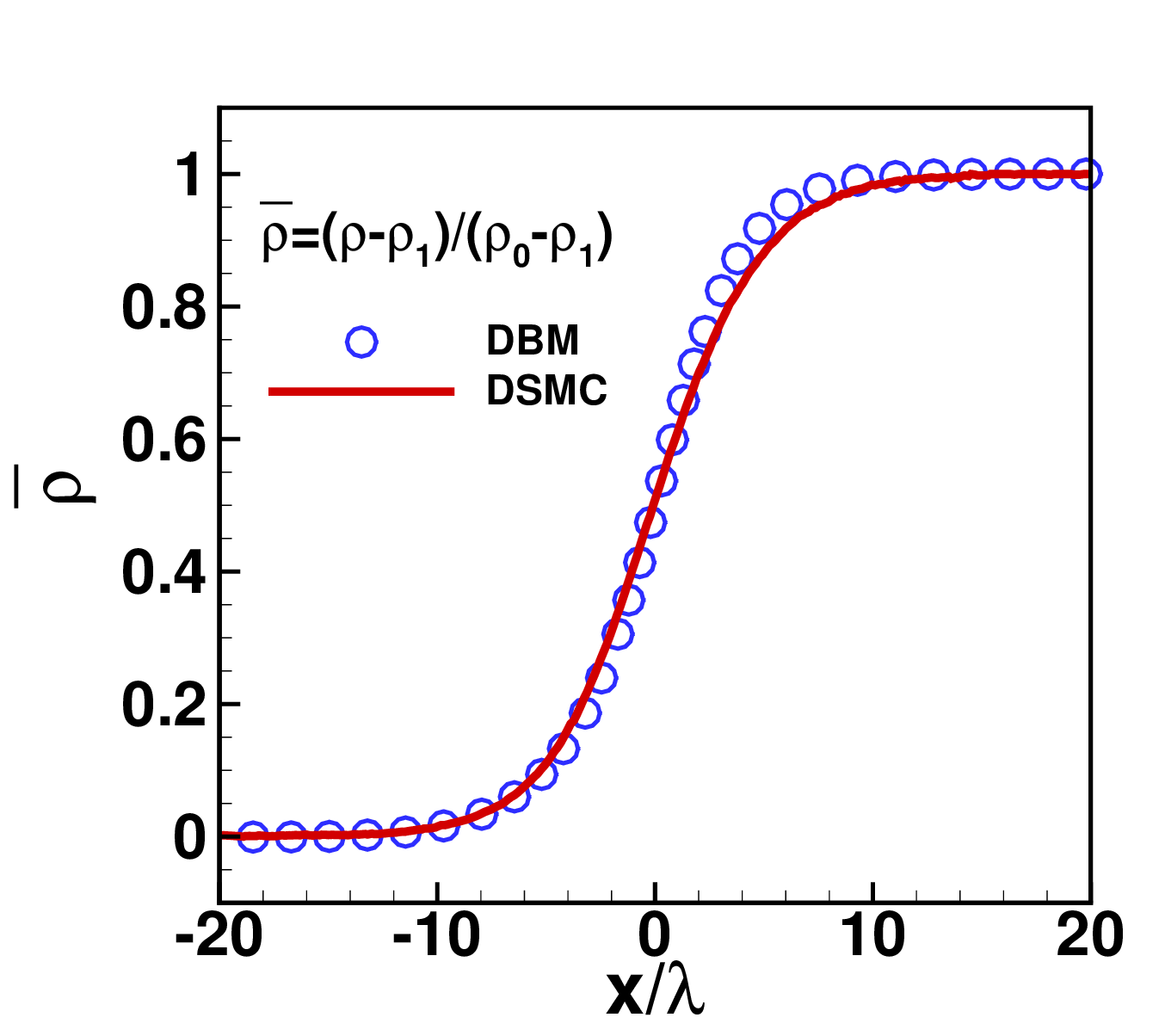}
\caption{ (From Ref. ~\cite{Zhang2022POF} Fig. 7 with permission. 
Comparison of DBM simulation and DSMC simulation of a shock structure. 
} \label{fig0012-1}
\end{figure*}

Detonation wave can be regarded as shock wave sustained by chemical reaction. 
The DBM in published literature is only concerned with the non-equilibrium induced by the small structure and/or fast flow behavior, so the description of chemical reaction description uses the traditional phenomenological model. 
From the perspective of constructing new model, what needs to be verified is the capability to describe the structure and behavior of shock waves. 
Figure \ref{fig0012-1} shows the comparison of the shock wave structures obtained by DBM and DSMC. 
It should be pointed out that for the simulations of steady-state structures such as shock waves, the use of steady-state DBM is more effective.

\subsection{Microscale flow}

\begin{figure*}[htbp]
\center\includegraphics*
[width=0.6\textwidth]{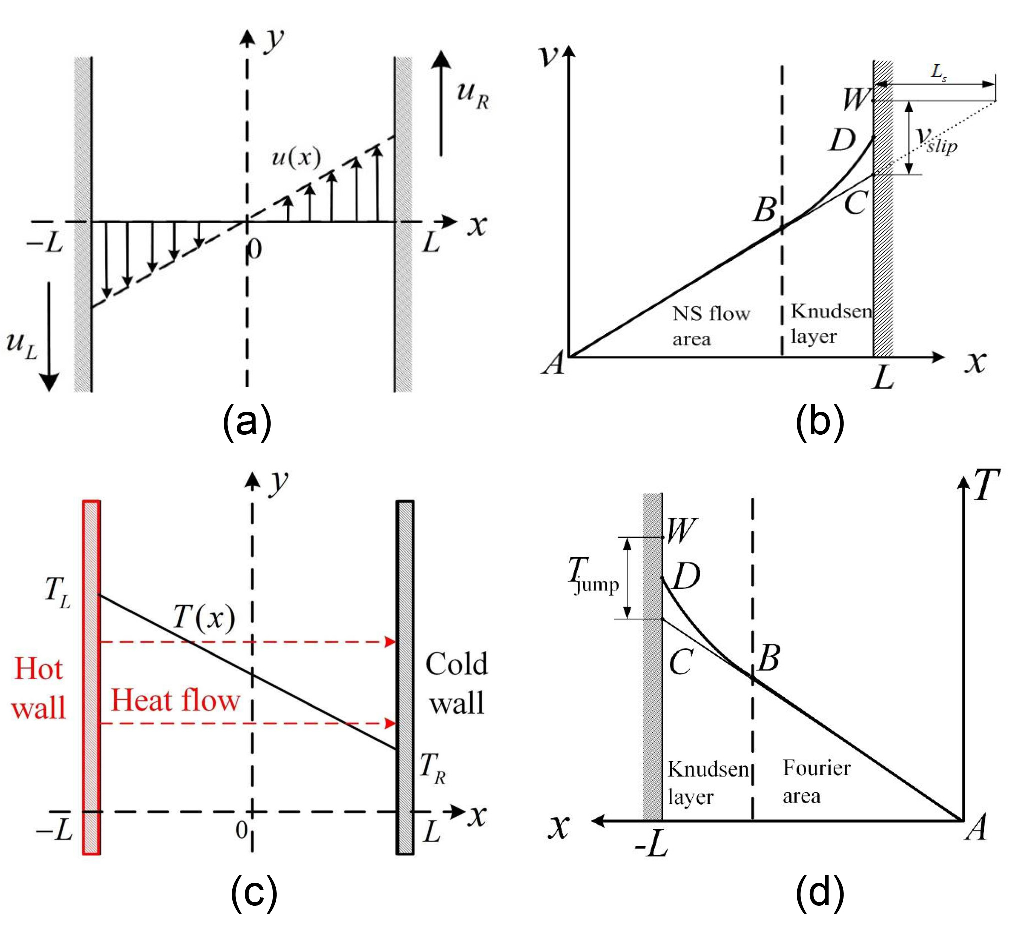}
\caption{ Schematic of velocity slip and temperature jump in Knudsen layer.
} \label{fig0013}
\end{figure*}

Micro-nano flow is a common phenomenon in micro-nano electromechanical systems.
It is well-known that mass and heat transfer in micro-nano flows exhibit many unique behaviors, such as the appearance of Knudsen layer near the wall.
Non-linear mass transport such as velocity slip, and non-Fourier heat transfer behavior such as temperature jump would come out within the Knudsen layer (as shown in Fig. \ref{fig0013}).
There are two main reasons for these phenomena: (i) With the decrease in the transverse size of the channel, the average molecular distance is no longer negligible compared to the transverse characteristic scale of the flow behavior.
(ii) The non-equilibrium driving mechanism in the fluid close to the wall is very different from that in internal fluid away from the wall, because the former carries the energy exchange (if no material exchange) between the system and the outside world.
Therefore, in addition to considering the order of Kn numbers to be retained in the internal flow description, it is also necessary to consider the design of suitable kinetic boundary conditions~\cite{Xu2022CMK,Zhang2018CTP,Zhang2019CPC,Zhang2022AIP}.

\begin{figure*}[htbp]
\center\includegraphics*
[width=0.8\textwidth]{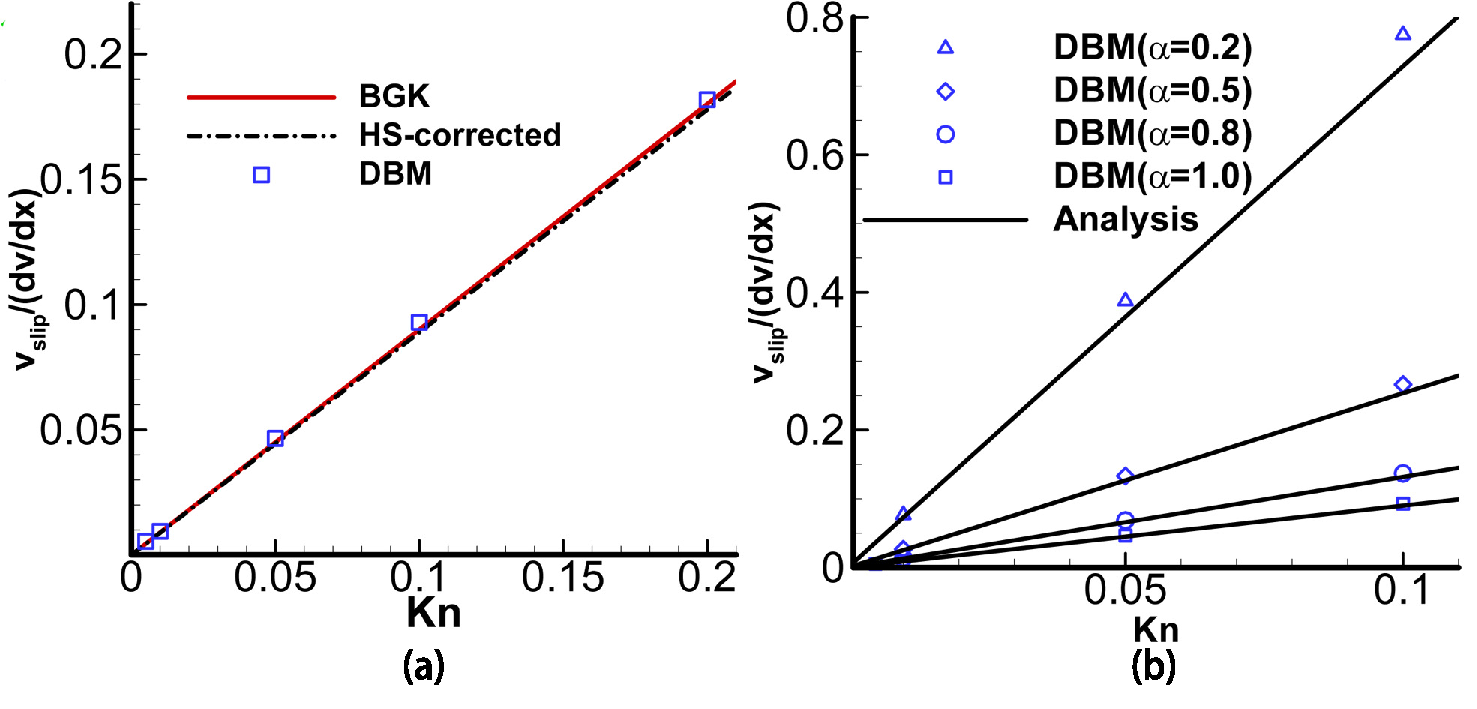}
\caption{ (From Ref. ~\cite{Zhang2018CTP} Figs. 4 and 6 with permission.) Comparison between DBM simulation results and analytical solutions: slip velocity.
} \label{fig0014}
\end{figure*}

\begin{figure*}[htbp]
\center\includegraphics*
[width=0.5\textwidth]{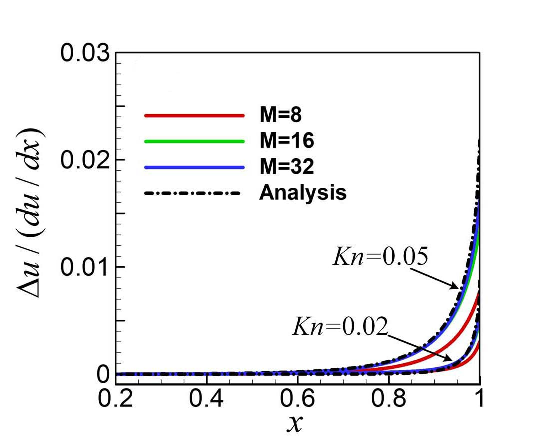}
\caption{  (From Ref. ~\cite{Zhang2022AIP} Fig. 2(b) with permission.) Comparison between DBM simulation results and analytical solutions: velocity difference profile.
} \label{fig0015}
\end{figure*}

Couette flow, Poiseuille flow, and so on are classic problems in the study of slip phenomenon.
For the convenience of discussion, the velocity Knudsen layer shown in Figs.  \ref{fig0013}(a)-\ref{fig0013}(b) and the temperature Knudsen layer shown in Figs. \ref{fig0013}(c)-\ref{fig0013}(d) are discussed separately.
The schematic diagram for the velocity Knudsen layer is shown in Fig. \ref{fig0013}(a).
It depicts a scenario where two long parallel plates are filled with gas.
The two plates move in the $y$ direction with velocities $u_{L}=-v$ and $u_{R}=v$, respectively, while velocities in the $x$ direction are zero.
The temperature of the two plates is fixed at $T_0$.
The following two cases would make the Kn number increase and the velocity slip phenomenon becomes more significant: (i) the distance between the plates decreases, (ii) the gas density between the two plates decreases.
After entering the steady state, although nonlinear velocity distribution [BD segment in Fig. \ref{fig0013}(b)] and velocity slip appear in the Knudsen layer, the velocity distribution in the region away from the wall still changes linearly [AB segment in Fig. \ref{fig0013}(b)].
Extend the line segment AB to the wall, the intersection point is C, and the corresponding flow rate is $v_{\rm{c}}$.
At this time, the difference between the plate speed $v_{w}$ and $v_{\rm{c}}$ is:
\begin{equation}
v_{slip}  = v_w  - v_{\rm{c}} ,
\end{equation}
which is defined as slip velocity.
The distribution along $x$ direction of velocity difference $\Delta u(x)$ between the curve BD and BC is called velocity Knudsen profile, which is an important parameter to describe the velocity Knudsen layer.
The analytical solutions of $v_{slip}$ and  $\Delta u(x)$ can be found in Refs. ~\cite{Sone2007book,Onishi1974}.
Considering the Maxwell-type boundary condition, the expression of slip velocity is $v_{slip}=k_{s}\lambda \sqrt{\pi}/2 \cdot \rm{d}\emph{v}/d\emph{x}$, where $k_{s}$ is a coefficient related to the tangential momentum accommodation coefficients (TMAC, $\alpha$, which is used to measure the proportion of complete diffuse reflection)~\cite{Onishi1974}.
$\lambda$ represents the mean free path of molecules.
In BGK collision model, $\lambda_{BGK}=\mu/p \sqrt{8RT/ \pi}$, and in hard sphere collision model, $\lambda_{HS}=4/5 \lambda_{BGK}$.
Figure \ref{fig0014}(a) shows the comparison of slip velocity normalized by $\rm{d}\emph{v}/d\emph{x}$ between DBM results and two types of analytical solutions at different Kn numbers~\cite{Zhang2018CTP}.
In Fig. \ref{fig0014}(b), DBM simulation results for various TMAC ($\alpha$) are compared with analytical solutions.
Figure \ref{fig0015} shows the curves of velocity difference profile $\Delta u(x)$ for different Kn numbers, where $M$ is the number of discrete velocity directions~\cite{Zhang2022AIP} .
The analytical solution of $\Delta u(x)$ is $\Delta u(x)=Y_{0}(\eta)\sqrt{2} \rm{Kn}\cdot \rm{d}\emph{u}_{\emph{y}}/ \rm{d} \emph{x}$, where $Y_{0}$ is the velocity Knudsen-layer function~\cite{Sone2007book} and $\eta=(1-x)/(\sqrt{2}\rm{Kn})$.
The DBM results are in good agreement with the analytical solutions, and it can be seen that DBM can accurately capture the velocity slip and the flow velocity behavior in the Knudsen layer.

\begin{figure*}[htbp]
\center\includegraphics*
[width=0.5\textwidth]{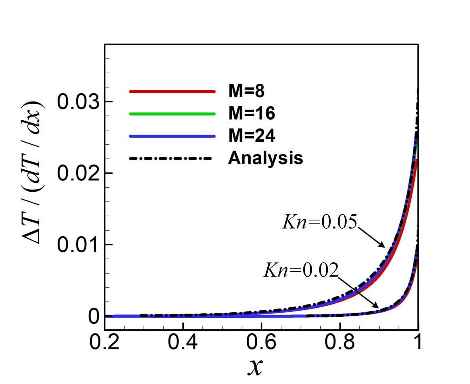}
\caption{  (From Ref. ~\cite{Zhang2022AIP} Fig. 6(b) with permission.) Comparison between DBM simulation results and analytical solutions: temperature difference profile.
} \label{fig0016}
\end{figure*}

\begin{figure*}[htbp]
\center\includegraphics*
[width=0.8\textwidth]{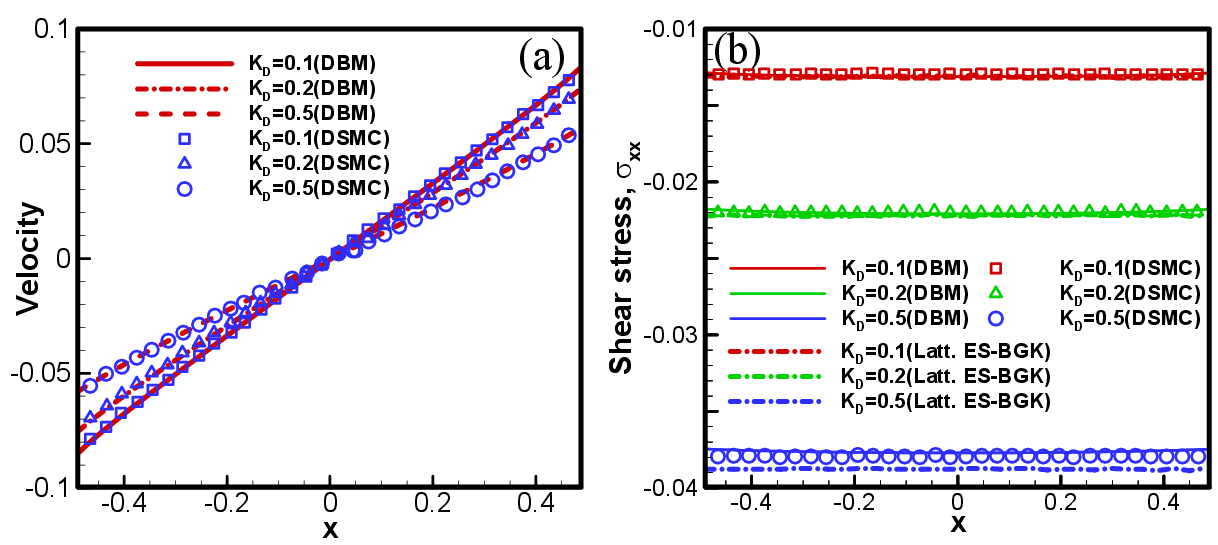}
\caption{  (From Ref. ~\cite{Zhang2019CPC} Fig. 5 with permission.) 
 The results of Couette flow at steady state: (a) velocity profiles, the symbols are DSMC data and the lines denote DBM results; (b) viscous shear stress profiles, the symbols denote DSMC data, the dashed line represents the Lattice ES-BGK results, and the solid line represents the DBM results.
} \label{fig0017}
\end{figure*}

The schematic diagram for the temperature Knudsen layer is depicted in Fig. \ref{fig0013}(c).
Similar to that of plate shear flow,  it involves two long parallel plates filled with gas, with zero velocity between the plates.
The temperatures of the two plates are maintained at $T_L$ and $T_R$, respectively.
After entering the steady state, although the nonlinear distribution of temperature [section BD in Figure \ref{fig0013}(d)] and the temperature jump occurs in the Knudsen layer, the temperature distribution in the area away from the wall still changes linearly [section AB in Figure \ref{fig0013}(d)].
Extending line segment AB to the wall, the intersection point is denoted as C, and the corresponding temperature is represented as $T_{\rm{c}}$.
At this point, the difference between the plate temperature $T_w$ and $T_{\rm{c}}$ is:
\begin{equation}
T_{slip}  = T_w  - T_{\rm{c}} ,
\end{equation}
which is defined as the slip temperature.
The temperature difference between the curves BD and BC, i.e., $\Delta T(x)$, as a function of position, is known as the temperature Knudsen profile,
which is an important parameter to describe the temperature Knudsen layer.
The analytical solutions of $T_{slip}$ and  $\Delta T(x)$ can be found in Ref. ~\cite{Sone2007book}.
Figure \ref{fig0016} displays the agreement of the temperature difference profiles $\Delta T(x)$ between DBM results and analytical solutions, for cases with two different Kn numbers~\cite{Zhang2022AIP}.
The analytical solution of temperature difference profiles is $\Delta T(x) = \Theta_{1}(\eta)\sqrt{2}\rm{Kn} \cdot \rm{d}\emph{T}/\rm{d}\emph{x}$, where $\Theta_{1}(\eta)$ represents the temperature Knudsen-layer function.
Figure \ref{fig0017}(a) shows the comparison of velocity distribution between DBM results and DSMC results when Couette flow develops to a steady state, with two different Kn numbers~\cite{Zhang2019CPC}.
Figure \ref{fig0017}(b) shows the profile of the $xx$ component of shear stress, which contains the results of DBM, DSMC, and lattice ES-BGK under three Kn numbers.
It can be seen that the DBM results are consistent with the DSMC results at several Kn numbers.
However, at a high Kn number (Kn $=$ 0.5), the DBM results are better than those of the lattice ES-BGK model.

\begin{figure*}[htbp]
\center\includegraphics*
[width=0.8\textwidth]{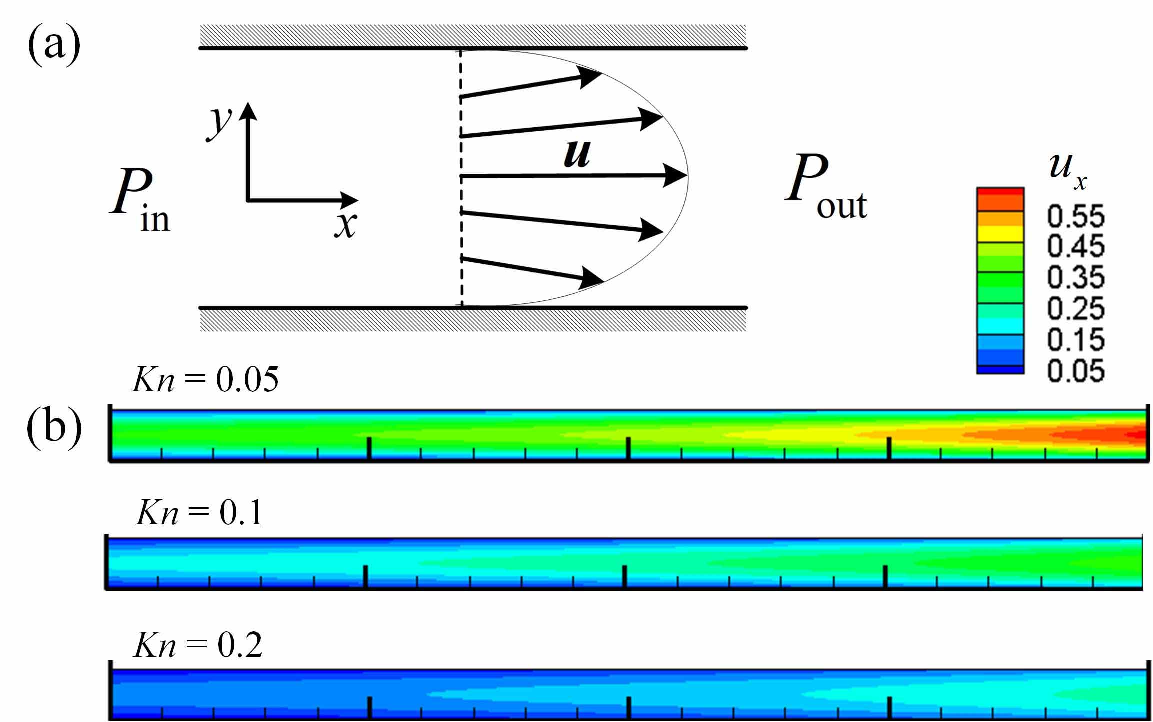}
\caption{   (From Ref. ~\cite{Zhang2022AIP} Fig. 10 with permission.) Pressure-driven flow. (a) Schematic, and (b) DBM simulation results of velocity contour under different Kn numbers.
} \label{fig0018}
\end{figure*}

\begin{figure*}[htbp]
\center\includegraphics*
[width=0.8\textwidth]{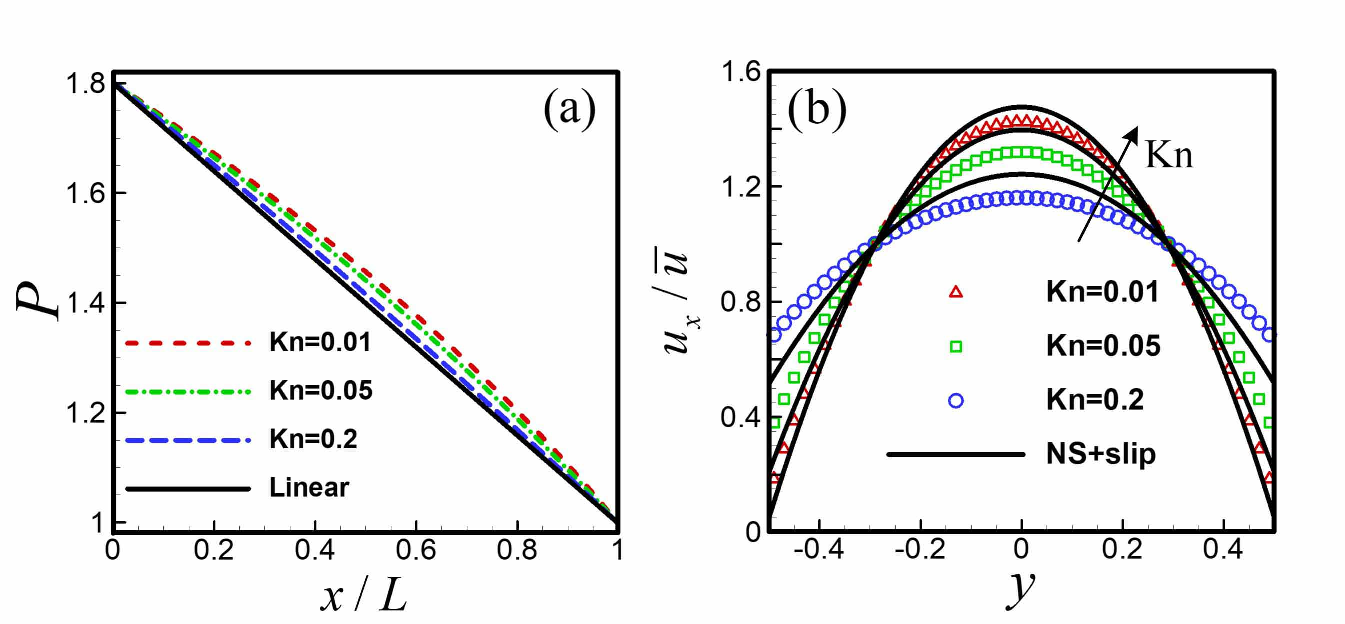}
\caption{  (From Ref. ~\cite{Zhang2022AIP} Fig. 11 with permission.) (a) Profiles of pressure $p(x)$ along the central line under different Kn numbers.
(b) Profiles of $u_x$ along the $y$ direction under different Kn numbers.
} \label{fig0019}
\end{figure*}

\begin{figure*}[htbp]
\center\includegraphics*
[width=0.5\textwidth]{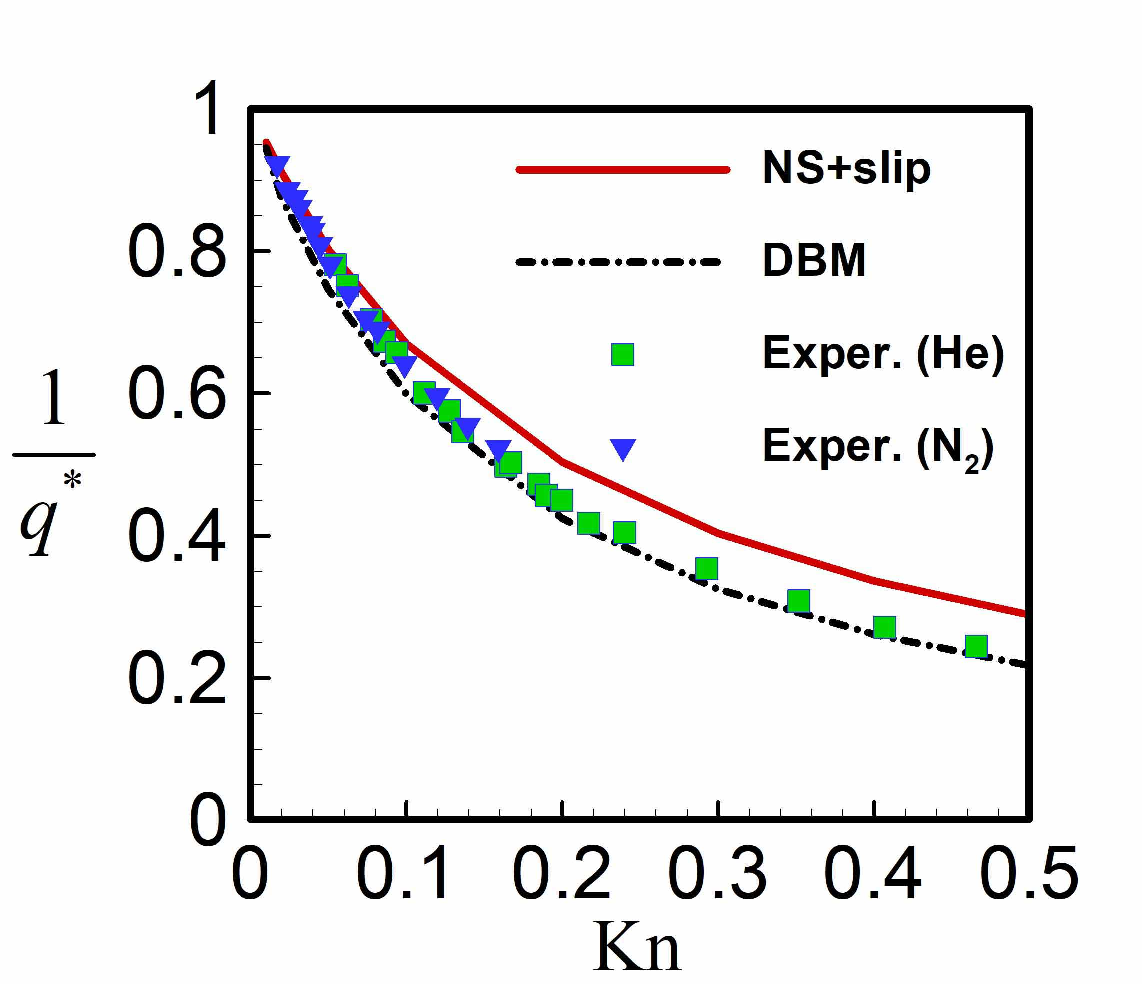}
\caption{  (From Ref. ~\cite{Zhang2022AIP} Fig. 12 with permission.) Inverse reduced mass flow rate under various Knudsen numbers, in which including the results of DBM simulation, NS simulation and experiment.
} \label{fig0020}
\end{figure*}

In a pressure-driven Poiseuille flow, the wall is fixed and there is a pressure difference between the two ends of the pipe.
The flow of fluid is driven by the pressure difference.
Due to the action of the wall surface, the velocity of the fluid in the center of the pipeline is larger, gradually decreasing from the center to the wall, as shown in Fig. \ref{fig0018}(a).
Figure \ref{fig0018}(b) shows the $u_x$ contour for three different Kn numbers.
Figure \ref{fig0019}(a) illustrates the distribution of pressure $p$ along the $x$ direction at the  center line.
It can be seen that the pressure distribution is nonlinear and has negative curvature, which is consistent with the previous experimental results and analytical theory~\cite{Zhang2022AIP}. 
Figure \ref{fig0019}(a) explains the phenomenon that appeared in Fig. \ref{fig0018}(b), showing that the more linear the pressure distribution, the smaller the difference in velocity in the cross section plane.
Figure \ref{fig0019}(b) displays the velocity distribution $u_x(y)$ along the $y$ direction under different Kn numbers.
It can be seen that as the Kn number increases, the deviation between the analytical solution based on the NS equations and DBM simulation results becomes more pronounced.
This is physically reasonable because as the Kn increases, slip boundary analysis gradually becomes less applicable.
Figure \ref{fig0020} shows the variation of inverse reduced mass flow rate with Kn number, including DBM simulation results, NS combined with slip boundary simulation results, and experimental results of two different gases ($\rm{He}$ and $\rm{N_2}$).
It can be seen that when the Kn number is less than 0.1, the results of NS are in good agreement with the experimental results.
With the increase of the Kn number, the difference between the results of NS and the experimental results becomes larger.
However, the DBM result is still in good agreement with the experimental result when the Kn number approaches 0.5~\cite{Zhang2022AIP}.

\subsection{Cavity flow}

\begin{figure*}[htbp]
\center\includegraphics*
[width=0.5\textwidth]{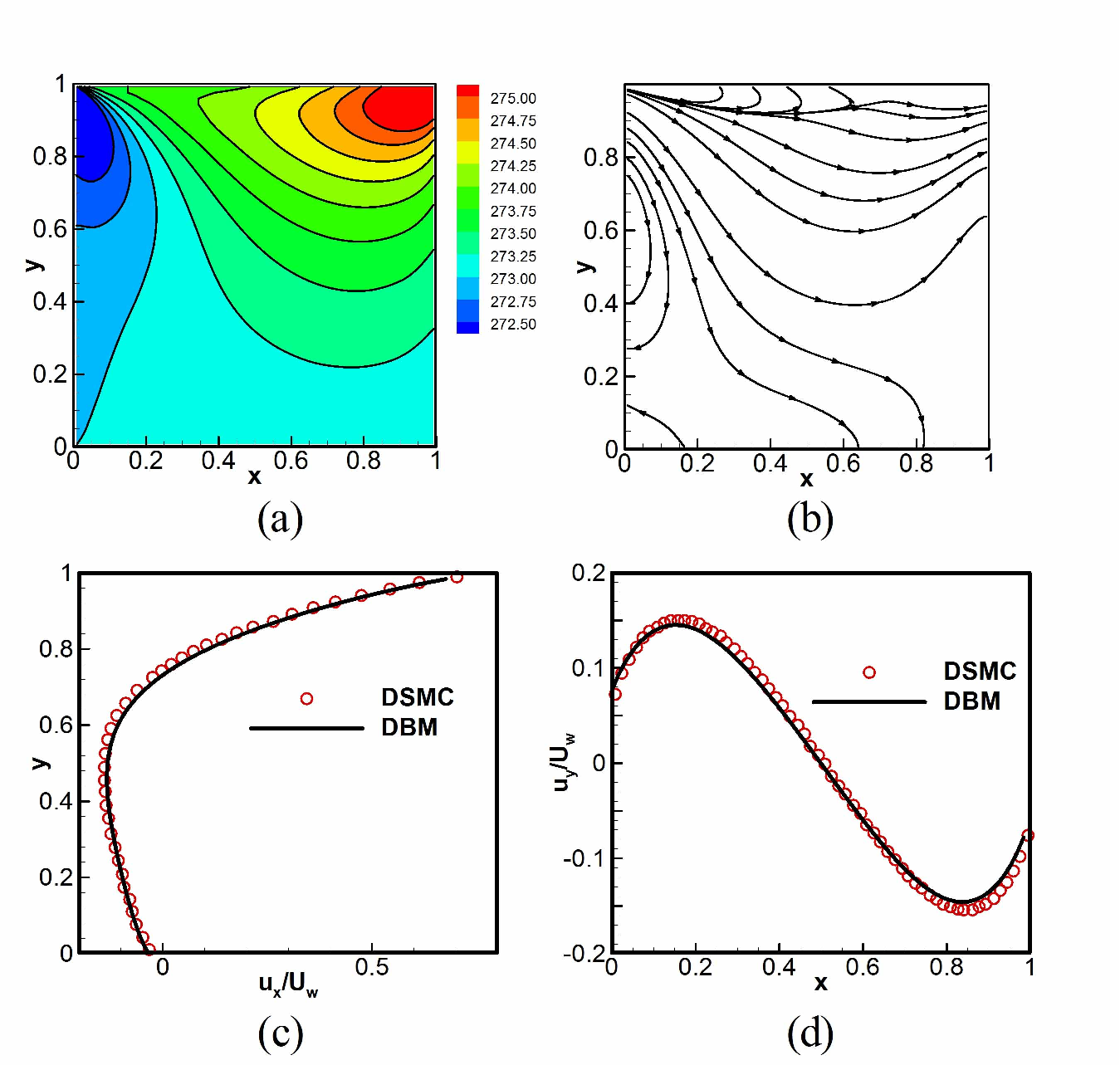}
\caption{  (From Ref. ~\cite{Zhang2019CPC} Fig. 12 with permission.) Simulation results of non-equilibrium cavity flow in steady state.
(a) Temperature contour. (b) Heat flow streamline. (c) Horizontal velocity distribution on the vertical centerline. (d) Vertical velocity distribution on the vertical centerline.
} \label{fig0021}
\end{figure*}

Two-dimensional square cavity flow is a common boundary-driven flow, which is also frequently encountered in non-equilibrium flow problems~\cite{Zhang2019CPC}.
In a cavity with size $L \times L$ filled with gas, the upper wall of the cavity moves to the right at a constant horizontal velocity $U_{w}$, while the other three walls remain stationary.
The temperature on the four walls is fixed at $T_0$.
In a specific practical scenario, we choose argon as the gaseous medium in the square cavity.
The mass of the argon atom is $\rm{m} = 6.63 \times 10^{-26}kg$, the temperature of the wall is $T_0 = 273 \rm{K}$, and the motion velocity of the upper wall is $U_w=50m/s$.
DBM simulation uses dimensionless quantities.
Figure \ref{fig0021}(a) shows the temperature contour of the flow field when the square cavity flow develops to a steady state, and Figure \ref{fig0021}(b) is the corresponding heat flux streamline.
From Fig. \ref{fig0021}(a), it can be seen that the left side of the flow field is the low-temperature region and the right side is the high-temperature region.
However, it is shown in Fig. \ref{fig0021}(b) that the heat flux is from the left side to the right side.
This is inconsistent with Fourier's thermal conductivity law but is reasonable. 
This is a typical strong thermodynamic non-equilibrium effect. According to the Chapman-Enskog multi-scale analysis, when the Knudsen number is small, only the first order term in Knudsen number is significant. 
In other words, when the degree of TNE is low, the linear response theory works. 
In this case, the heat flux is positively proportional to the negative temperature gradient. 
But with increasing the Knudsen number, the second and even higher order terms in the Knudsen number can no longer be ignored. 
In other words, with increasing the degree of TNE, the linear response theory gradually fails and the nonlinear responses begin to become more significant. 
The heat flux is no longer only related to the temperature gradient, but
also influenced by velocity gradient and/or density gradient.
When the negative contribution from density gradient and/or velocity gradient surpasses the contribution of the temperature gradient, the heat flux will flow from the low-temperature region to the high-temperature region. 
The DBM results in this case are confirmed by the DSMC predictions.
Figures \ref{fig0021}(c) and \ref{fig0021}(d) demonstrate the comparison between DBM results and DSMC results of the velocity distribution, showing good agreement between the two.

\subsection{Fluid collision  \label{Fluid-collision}}

Collision of two fluids will lead to the increase of gradients of density, temperature, flow velocity, and pressure at the collision interface, resulting in strong non-equilibrium effects at the interface~\cite{Gan2018PRE,Zhang2022POF}.
Therefore, fluid collisions are often used to assess the ability of models to describe non-equilibrium effects.
According to CE multi-scale analysis theory, the degree of nonlinearity of constitutive relation varies with the degree of non-equilibrium.
Even for the same flow behavior, using different physical quantities to investigate,  it may require different levels of accuracy in the DBM physics.

\begin{figure*}[htbp]
\center\includegraphics*
[width=0.9\textwidth]{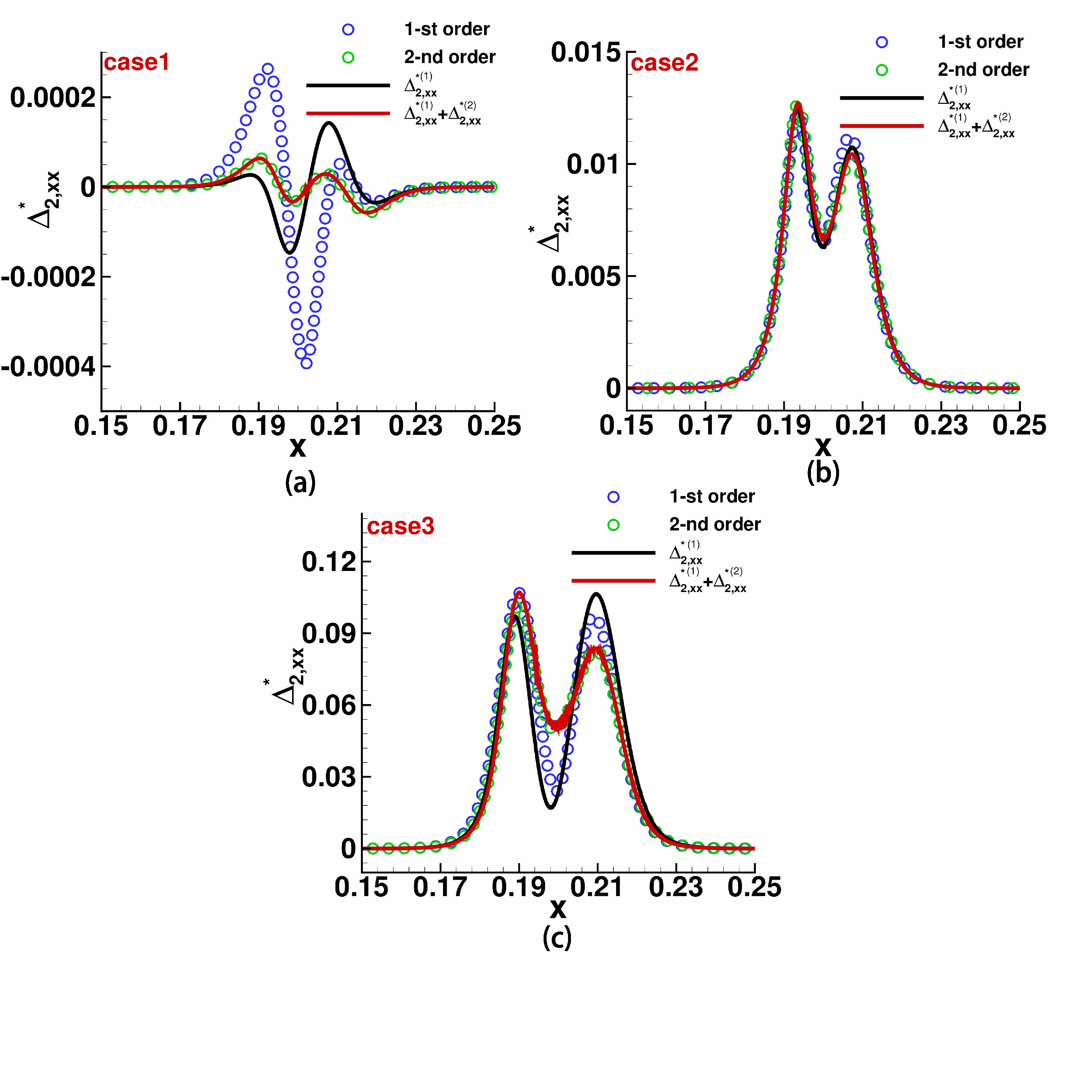}
\caption{  (From Ref. ~\cite{Zhang2022POF} Fig. 9 with permission.) Comparison of $ \Delta_{2,xx}^{*}$ profile between DBM simulation results and analytical results. The three figures represent the cases of weak, medium, and strong TNE strength, respectively. The two symbols indicate the results of first-order and second-order DBM, respectively, with lines representing the analytical results at two TNE levels.
} \label{fig0022}
\end{figure*}

\begin{figure*}[htbp]
\center\includegraphics*
[width=0.9\textwidth]{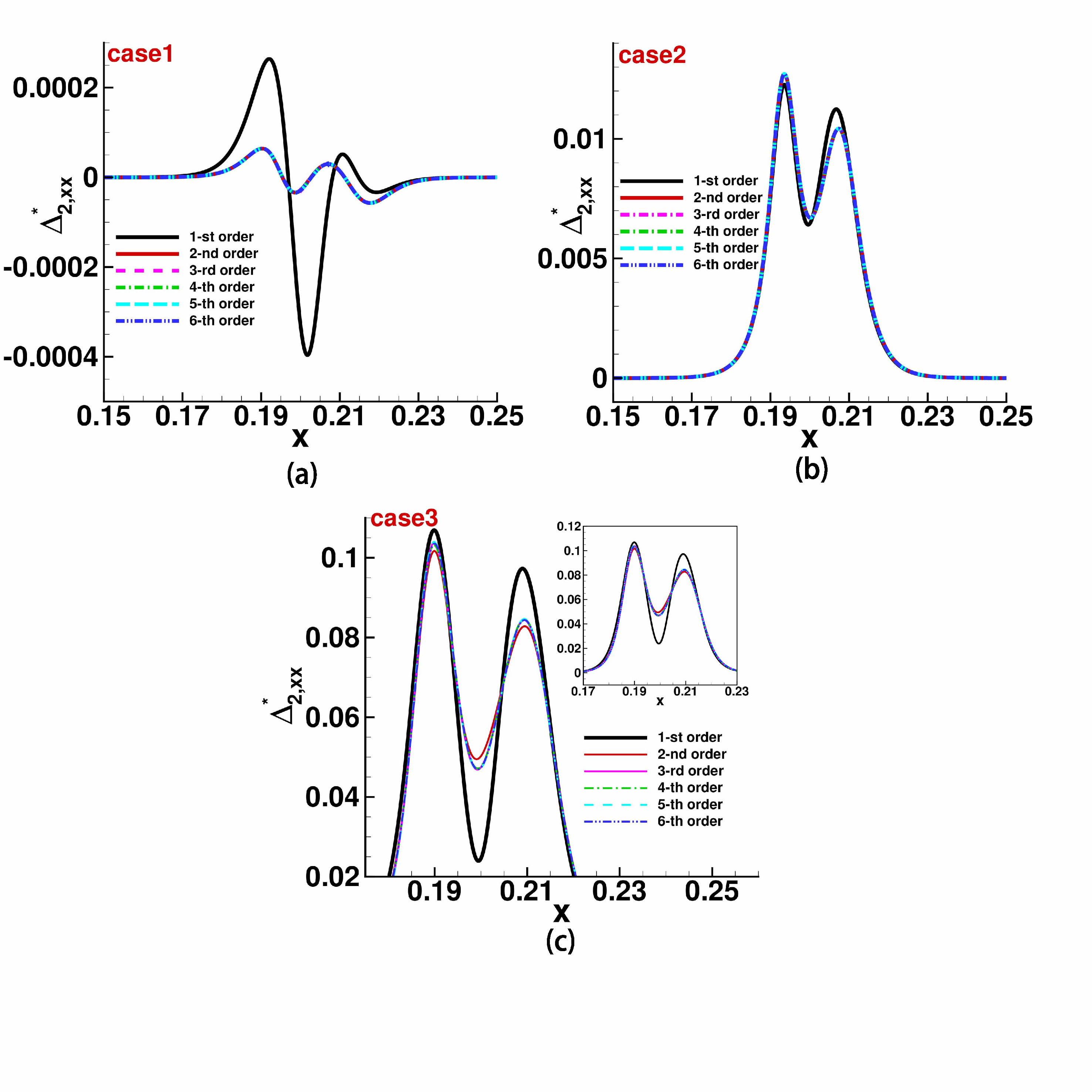}
\caption{   (From Ref. ~\cite{Zhang2022POF} Fig. 11 with permission.) The DBM simulation results from different orders of DBM.
} \label{fig0023}
\end{figure*}

Figures \ref{fig0022} and \ref{fig0023} show the distribution of the $xx$ component of $\bm{\Delta}_{2}^{*}$ (corresponding to the viscous stress tensor in the macroscopic fluid dynamics equations) for detecting the system's non-equilibrium behavior~\cite{Zhang2022POF}.
The higher the degree of TNE (from the perspective of Kn number) considered in the  modeling process, the corresponding constitutive relationship of $\Delta_{2,xx}^{*}$ in EHE is more accurate.
The expressions for first-order and second-order non-equilibrium stresses can refer to the Appendix \ref{sec:AppendixesB}.
It is clear that relaxation time $\tau$ and macroscopic gradients are the driving forces of non-equilibrium effects.
For $\Delta_{2,xx}^{*(1)}$, the main driving factors are relaxation time $\tau$ and velocity gradient.
But for $\Delta_{2,xx}^{*(2)}$, it is also related to density gradient and temperature gradient.
Quantities such as Kn number, relaxation time $\tau$, the spatial gradient and temporal change rate of macroscopic physical quantities, are all measures of TNE intensity/degree/strength from a certain perspective.
The intensity of $\Delta_{2,xx}^{*}$ is also a measure of TNE strength from a particular perspective.
Therefore, it is again stated that for complex flow studies, using a single perspective to define the non-equilibrium strength is easy to draw one-sided conclusions.
This is the reason why DBM introduces a non-equilibrium intensity vector (with each component being a measure of TNE strength from a specific perspective) to carry out multi-perspectives cross-positioning for the non-equilibrium intensity.

Based on the above considerations, in order to demonstrate the ability of the DBM model to describe different degrees of non-equilibrium, three different cases are designed.
If only see from the view of $\Delta_{2,xx}^{*}$, cases 1, 2, and 3 correspond to weak, moderate, and strong non-equilibrium strengths, respectively.
Among these cases, case 1 has an initial velocity of 0, so its strength of $\Delta_{2,xx}^{*}$ is the weakest.
In case 2, the initial collision velocity is increased,  thereby incorporating first-order non-equilibrium effects, resulting in a greater $\Delta_{2,xx}^{*}$ strength compared to case 1.
In case 3, we increase the relaxation time $\tau$ ten times compared with the $\tau$ in case 2, so that it produces a strength approximately ten times greater than that of case 2.
However, the non-equilibrium strength of the system cannot be solely measured with a single indicator.
For example, in case 1, its velocity gradient is small, so the density gradient and temperature gradient play a significant role, resulting in the relative value of $\Delta_{2,xx}^{*(2)}$ to $\Delta_{2,xx}^{*(1)}$ can not be negligible.
Therefore, DBM further introduces the relative non-equilibrium strength quantity $\Delta_{2,xx}^{*(2)}/\Delta_{2,xx}^{*(1)}$ to more accurately describe the relative importance of different levels of non-equilibrium effects.
From the perspective of relative non-equilibrium strength, the non-equilibrium strength in case 2 is smaller than that in cases 1 and 3.

In Figure \ref{fig0022}, the physical accuracies of the used DBM are first order and second order (represented by circles with different colors), respectively.
The analytic solution results correspond to first order [$\Delta_{2,xx}^{*(1)}$] and second order [$\Delta_{2,xx}^{*(1)}+\Delta_{2,xx}^{*(2)}$] (represented by lines with different colors), respectively.
It can be seen that in case 1, although the non-equilibrium strength from the view of $\Delta_{2,xx}^{*}$ is weak, the results of the first order DBM show significant deviations from both first-order and second-order analytical results.
This is because in case 1, $\Delta_{2,xx}^{*(2)}$  is non-negligible relative to $\Delta_{2,xx}^{*(1)}$ [i.e., the relative non-equilibrium strength $\Delta_{2,xx}^{*(2)}/\Delta_{2,xx}^{*(1)}$ is relatively large].
The physical accuracy of second order non-equilibrium (from the perspective of Kn number) is not guaranteed in the first order DBM during the model construction process, so the first order DBM does not guarantee the accuracy of description of $\Delta_{2,xx}^{*(2)}$.
However, the results of second-order DBM match perfectly with the second-order analytical results.
This is because in the construction process of second-order DBM, the accuracy of the second-order non-equilibrium description (from the Knudsen number perspective) is ensured.
In case 2, although the non-equilibrium strength from the view of $\Delta_{2,xx}^{*}$ is increased, $\Delta_{2,xx}^{*(2)}$ becomes negligible compared to $\Delta_{2,xx}^{*(1)}$ [$\Delta_{2,xx}^{*(2)}/\Delta_{2,xx}^{*(1)}$ is small].
As a result, the results of first-order DBM, second-order DBM, and first-order and second-order analytical results are all in agreement.
In case 3, the non-equilibrium strength from the view of $\Delta_{2,xx}^{*}$ is further increased, $\Delta_{2,xx}^{*(2)}$ becomes non-negligible compared to $\Delta_{2,xx}^{*(1)}$.
In this case, the results of first-order DBM show deviations from the second-order analytical results, while the results of second-order DBM align well with the second-order analytical results.
Figure \ref{fig0022} demonstrates one specific aspect of the complexity of non-equilibrium flow behavior.

The higher the physical accuracy of the DBM model, the more kinetic moments need to be retained in the modeling process, which in turn increases the computational cost of simulations.
So, in the actual simulation study, how many orders of physical accuracy of DBM need to be considered?
Meeting the needs, the simplest, the least cost, is the first choice.
The required level of physical accuracy can be determined through the convergence of simulation results.
The specific examples are shown in Figure \ref{fig0023}, in which the above three working cases are simulated with first order to sixth order DBM (from the view of Kn number), respectively.
It can be observed that, except for the results of the first order DBM, the rest of the DBM results are consistent.
For cases 1 and 2, to obtain accurate viscous stresses, at least a second order model is required.
However, for case 3, the second-order DBM results differ from the higher-order DBM results and the results from the third-order model and above overlap.
Therefore, for case 3, to accurately describe the non-equilibrium behavior from the view of $\Delta_{2,xx}^{*}$, at least a third order DBM is required.

\subsection{Multiphase flow  \label{Multi-phase-flow}}

\begin{figure*}[htbp]
\center\includegraphics*
[width=0.7\textwidth]{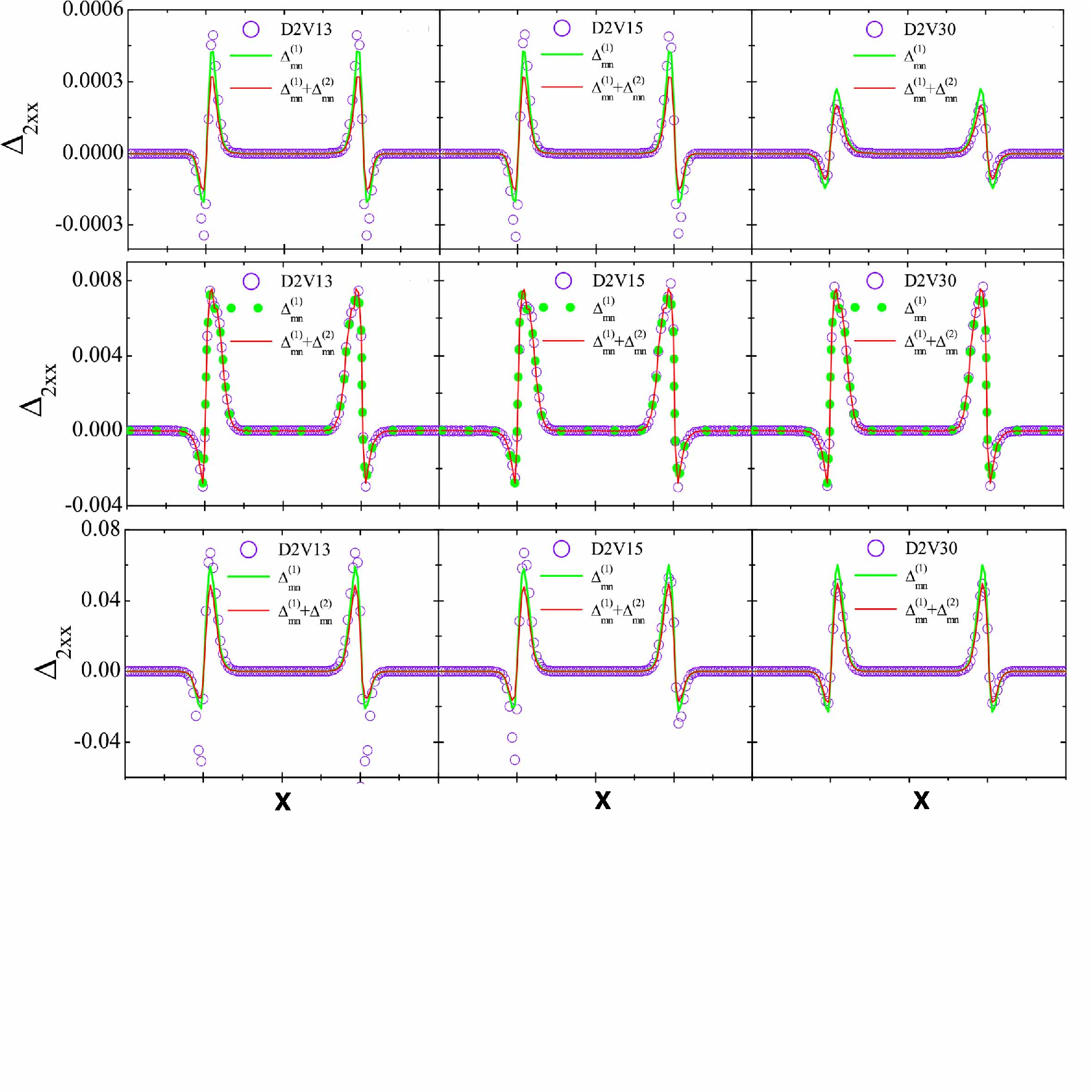}
\caption{   (From Ref. ~\cite{Gan2022JFM} Figs. 7, 8 and 9, with permission.) Comparison of $\Delta_{2,xx}$ profile between DBM simulation results and analytical results.
Each row of the pictures show corresponds to cases with weak, moderate, and strong TNE strength, and each column corresponds to the results of D2V13, D2V15, and D2V30, respectively.
In the figures, the circles represent results from DBM at various levels, the green lines and green points are results calculated from the first theoretical analysis, and red lines denote results obtained from the second theoretical analysis.
} \label{fig0024}
\end{figure*}

Multiphase flows and phase change heat transfer phenomena are widely
exist in the natural world and engineering applications~\cite{Chen1996PRL,Wagner1998PRL,Michael1995PRL}.
Reference~\cite{Gan2022JFM} further explores the ability of different orders of DBM (from the perspective of Kn number) to describe the non-equilibrium effect in liquid-vapor phase transition systems.
Reference~\cite{Gan2022JFM} analyzed the ability of the DBM model to describe different degrees of non-equilibrium from four basic perspectives, i,e., $\Delta_{2,xx}$, $\Delta_{3,xxx}$, $\Delta_{3,1x}$, and $\Delta_{4,2xx}$.
Figure \ref{fig0024} displays the spatial distribution of typical thermo-dynamic non-equilibrium effects, i.e., $\Delta_{2,xx}$, (the combination of hydrodynamic non-equilibrium and thermodynamic non-equilibrium) near the gas-liquid interface.
If only look at the view from $\Delta_{2,xx}$, each row of Fig. \ref{fig0024} corresponds to weak, moderate, and strong non-equilibrium scenarios.
Each column in the figure corresponds to three cases where the selection of discrete velocity is D2V13, D2V15, and D2V30, respectively.
In terms of the retained order of the Knudsen number, the first two correspond to the NS level (retaining up to the first-order terms of the Knudsen number).
However, D2V15 retains $\rm{M}_{5,1}$ moment relations more than the D2V13 model, making the constitutive relations in the EHE more complete.
The D2V30 case corresponds to the super-Burnett level (retaining up to the third-order terms of the Knudsen number).
For comparison, the figure also provides analytical results for the first-order ($\Delta_{mn}^{(1)}$) and second-order ($\Delta_{mn}^{(2)}$) non-equilibrium effects.
For the case of weak non-equilibrium (from the perspective of $\Delta_{2xx}$), as can be seen in the first row of Fig. \ref{fig0024}, there exists a deviation between the first-order analytical solution and the second-order analytical solution.
It indicates that the second order non-equilibrium ($\Delta_{2xx}^{(2)}$) cannot be ignored compared with the first order non-equilibrium ($\Delta_{2xx}^{(1)}$), that is, the relative non-equilibrium strength $\Delta_{2xx}^{(2)}/\Delta_{2xx}^{(1)}$ is larger.
Because the first order DBM does not guarantee the accuracy of the description of $\Delta_{2xx}^{(2)}$ in model construction, the results of D2V13 and D2V15 are both deviated from the analytical solutions.
However, the D2V30 model which considers up to the third order non-equilibrium, provides numerical results that match well with the analytical results.
For the case where the non-equilibrium strength is moderate (from the perspective of $\Delta_{2xx}$) and the relative non-equilibrium strength $\Delta_{2xx}^{(2)}/\Delta_{2xx}^{(1)}$ is small, as shown in the second row of Fig. \ref{fig0024}, all three models  can effectively describe non-equilibrium.
In the case of strong non-equilibrium (from the perspective of $\Delta_{2xx}$), as shown in the third row of Fig. \ref{fig0024}, the results from the D2V30 model which considers the third order non-equilibrium agree well with the analytic solutions, while the other two models deviate significantly.
But relatively speaking, the results from D2V15 are better than that from D2V13.
These results indicate that higher-order DBM models have a stronger ability to describe non-equilibrium effects.
As discussed in Sections \ref{Fluid-collision} and \ref{Multi-phase-flow}, the Kn number, relaxation time $\tau$, macroscopic quantity gradient, TNE quantity, and relative TNE quantity can all be used to describe the non-equilibrium strength, but the results obtained by relying on a single quantity and from a single perspective are obviously incomplete.
It is more reasonable to introduce the non-equilibrium strength vector and describe the non-equilibrium strength from multiple perspectives.

\subsection{Hydrodynamic instability  \label{Hydrodynamic-instability}}

\begin{figure*}[htbp]
\center\includegraphics*
[width=0.5\textwidth]{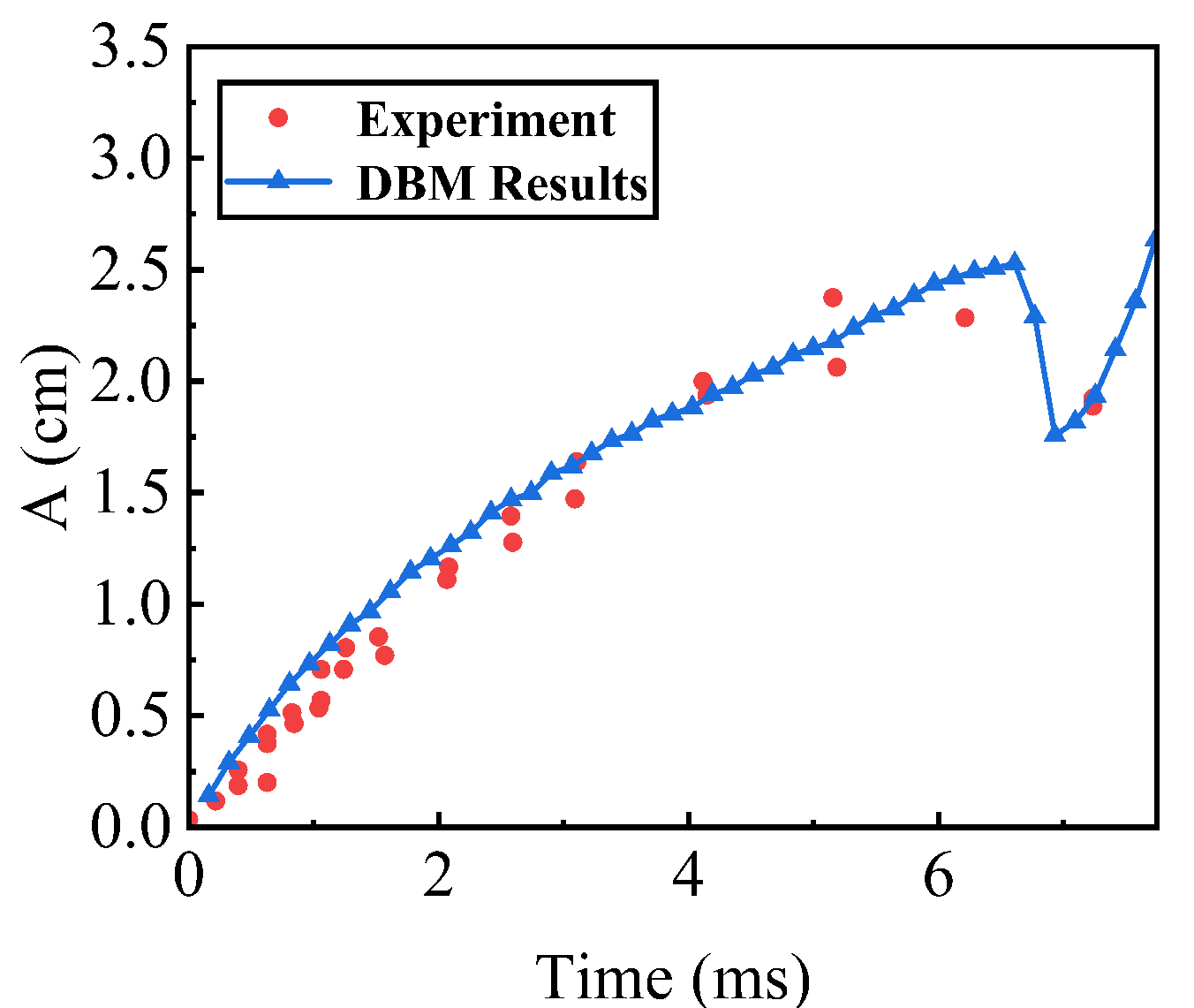}
\caption{  (From Ref. ~\cite{Shan2023CTP} Fig. 3, with permission.) Evolution of spike amplitude.
The red points in the picture are experimental results from Ref. ~\cite{Collins2002JFM}, 
and blue line represent DBM results.
} \label{fig0025}
\end{figure*}

RM instability under re-shock condition is an important research topic in ICF field.
In Ref. ~\cite{Shan2023CTP}, the re-shock process of $\rm{SF}_6$ and air was studied by using two-fluid DBM, and the corresponding amplitude evolution curve was shown in Fig. \ref{fig0025}.
It can be seen that DBM can capture the amplitude evolution accurately. The RMI under same conditions were also simulated in Ref.~\cite{Latini2007POF}.
It should be noted that in the literature ~\cite{Latini2007POF}, the numerical simulation uses Euler equation, so artificial viscosity has to be added, and in order to better conform to the experiment, the diffusion effect was also artificially corrected.
In DBM simulations, such artificial interventions are not used.

\subsection{Shock-bubble interaction  \label{Shock-bubble interaction}}

\begin{figure*}[htbp]
\center\includegraphics*
[width=1.0\textwidth]{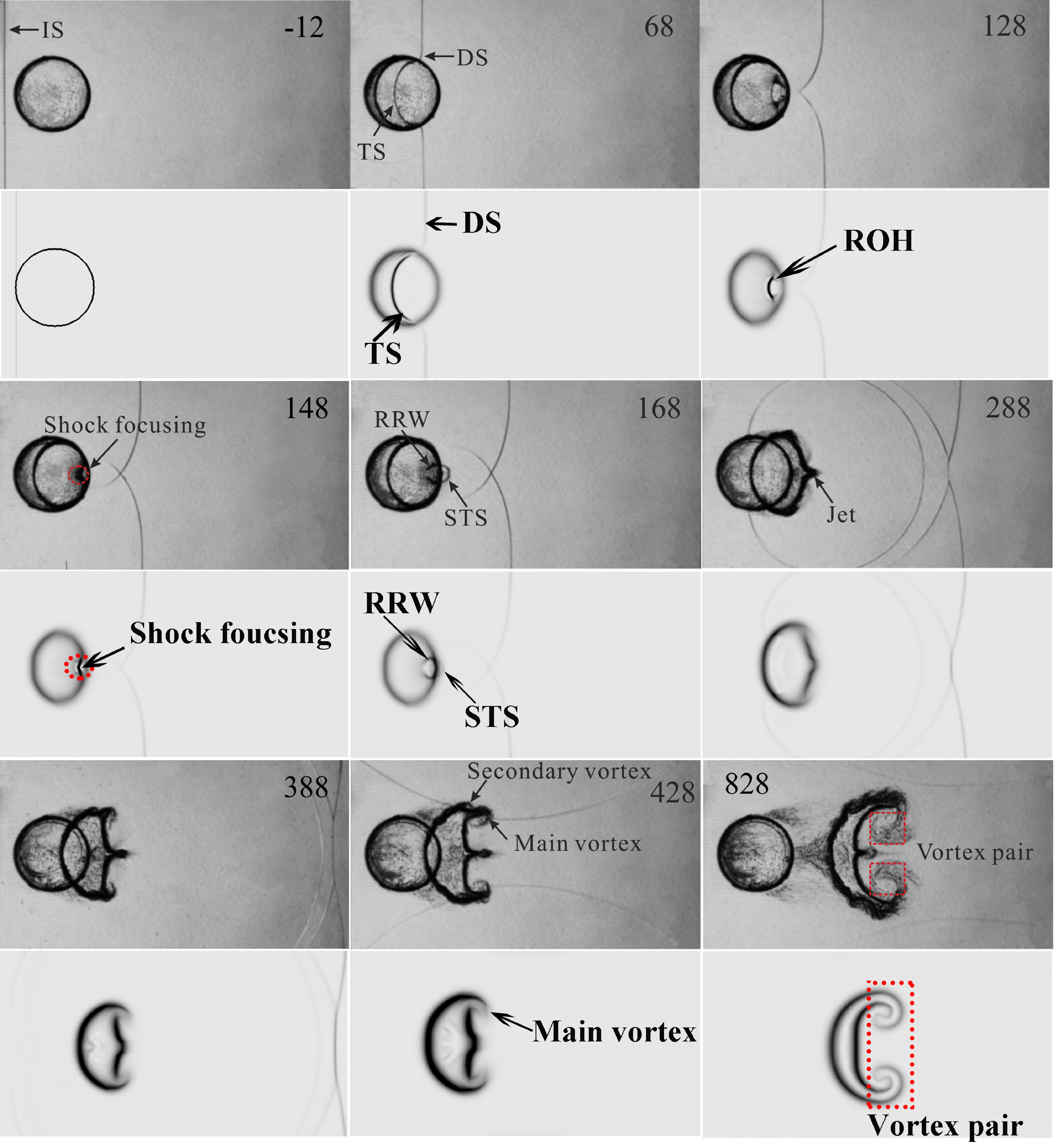}
\caption{   (From Ref. ~\cite{Zhang2023CAF} Fig. 4, with permission.) Snapshots of schlieren images of the interaction between a shock wave and a heavy-cylindrical bubble. 
The odd rows represent experimental results from Ref.~\cite{Ding2018POF} with permission, and the even rows are DBM simulation results. 
Numbers in the picture represent the time in $\mu s$.
} \label{fig0025-1}
\end{figure*}

\begin{figure*}[htbp]
\center\includegraphics*
[width=0.5\textwidth]{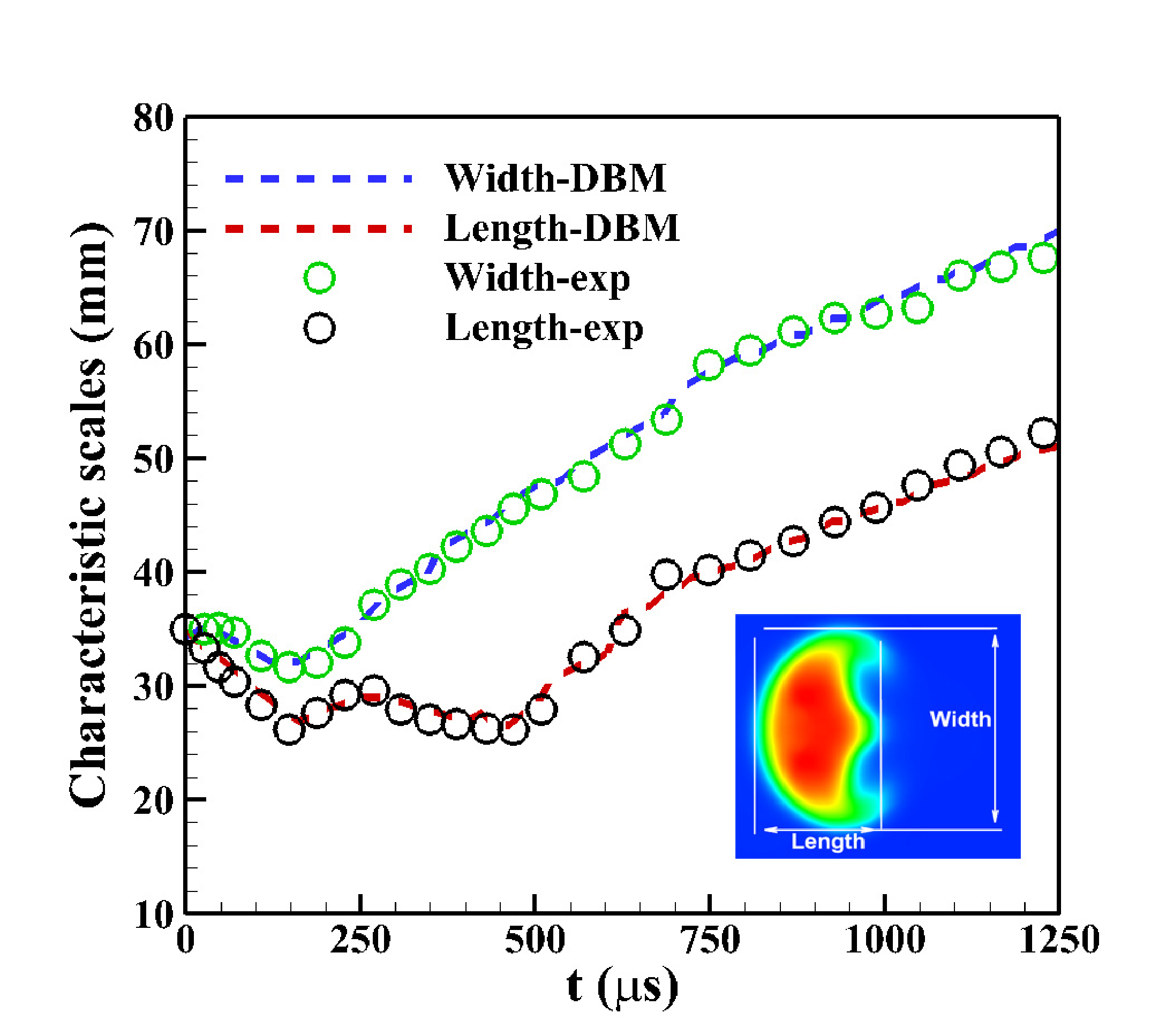}
\caption{   (From Ref. ~\cite{Zhang2023CAF} Fig. 5, with permission.) Evolution curves of the characteristic scale of shock-bubble interaction.
The lines indicate DBM simulation results and the symbols are experimental results extracted from Ref. ~\cite{Ding2018POF}.
} \label{fig0026}
\end{figure*}

The interaction between a shock wave and a heavy-cylindrical bubble (shock-bubble interaction, SBI) is studied in Ref. ~\cite{Zhang2023CAF}.
Figure \ref{fig0025-1} shows the comparison of snapshots of schlieren images between experimental results and DBM simulation results.
In Fig. \ref{fig0025-1}, the odd rows represent experimental results from Ref.~\cite{Ding2018POF}, and the even rows are DBM simulation results. 
The schlieren images of DBM simulation are calculated from the density gradient formula, i.e., $|  \nabla \rho | / |  \nabla \rho |_{max}  $, with $ |\nabla \rho| = \sqrt{(\partial \rho/\partial x)^2+(\partial \rho/\partial y)^2}  $.
The typical wave patterns and the bubble's main characteristic structure are marked out in the figures. 
For example, ``TS'' represents the transmitted shock propagating downstream inside the bubble, which is generated after the incident shock impacts the bubble.
The shock focusing occurs when two high-pressure regions meet, which in turn produces secondary transmitted shock waves (STS), reflected rarefaction waves (RRW), and jet.
These typical wave patterns during the SBI can all be captured by DBM.
The evolution of bubble characteristic scales during the SBI process are demonstrated in Fig. \ref{fig0026}.
The lines in the figure are DBM simulation results, and the symbols are the experimental results extracted from Ref. ~\cite{Ding2018POF}.
It can be seen that DBM can accurately describe the bubble deformation process in the SBI process.

\subsection{Plasma system  \label{Plasma-system}}

\begin{figure*}[htbp]
\center\includegraphics*
[width=0.4\textwidth]{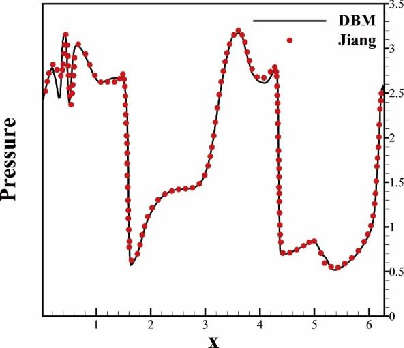}
\caption{   (From Ref. ~\cite{Song2023POF} Fig. 2, with permission.)
Comparison between DBM results and NS results of pressure profile at the line  $y=0.625\pi$ in simulation of Orsazg-Tang magnetic turbulence.
} \label{fig0027}
\end{figure*}

The compressible Orsazg-Tang (OT) magnetic turbulence problem was first proposed by Orsazg and Tang.
It is widely used to validate the effectiveness of magnetohydrodynamics models due to the complex wave structures that emerge during the evolution of the flow field.
Figure \ref{fig0027} shows the pressure distribution at the line $y=0.625\pi$ when the OT magnetic turbulence problem develops to steady state~\cite{Song2023POF} .
The black lines represent the DBM simulation results, and the red dots are the simulation results from Ref. ~\cite{Jiang1999JCP}.
DBM can accurately capture the pressure distribution in this problem.

\subsection{ Steady flow  \label{Steady-results}}
\begin{figure*}[htbp]
\center\includegraphics*
[width=0.8\textwidth]{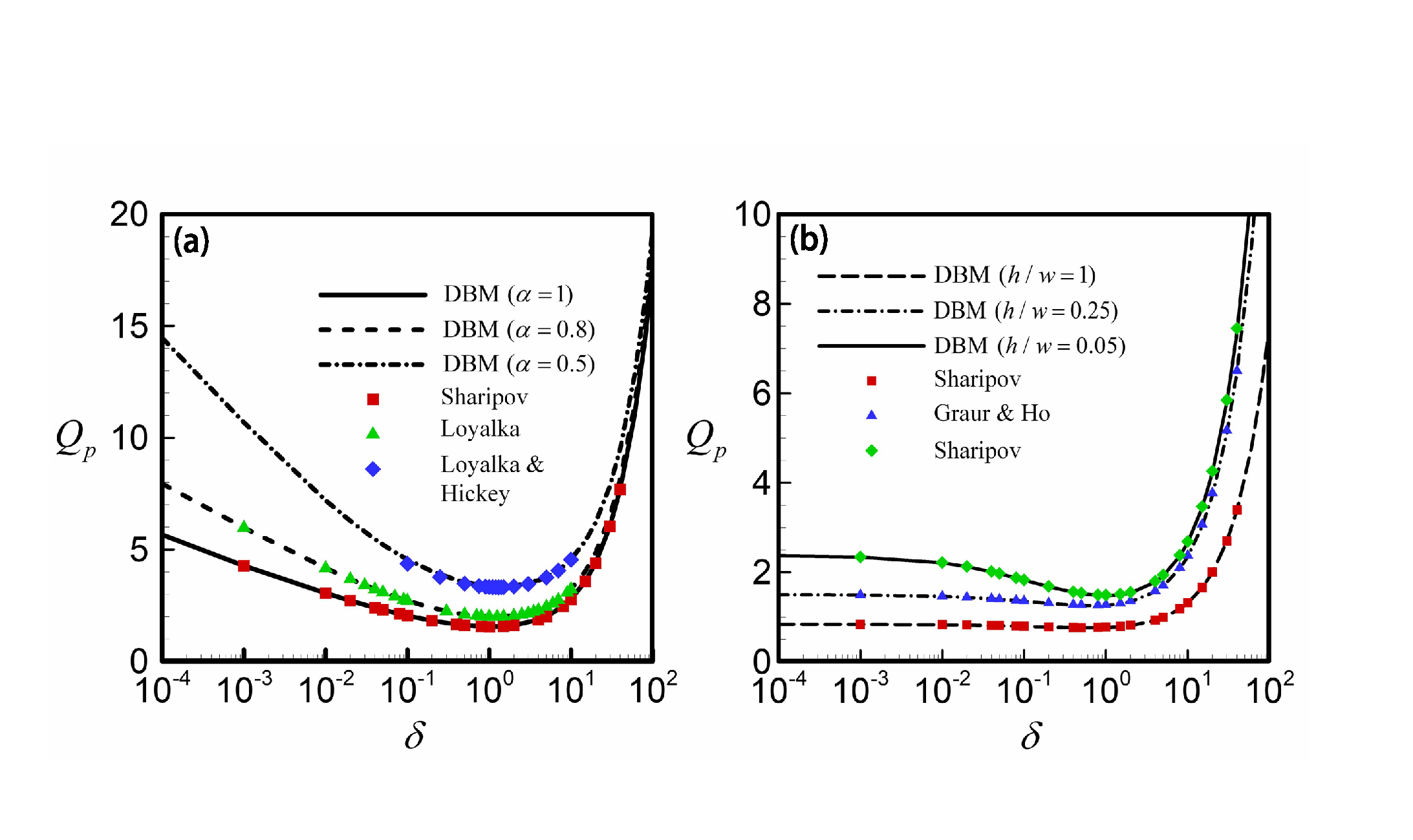}
\caption{   (From Ref. ~\cite{Zhang2023POF} Figs. 1 and 3, with permission.) (a) Profiles of the dimensionless mass flow rate with various rarefaction  parameters under three different values of TMAC for the one-dimensional pressure driven flow through a microchannel. (b) Profiles of the dimensionless mass flow rate with rarefaction parameters for three different aspect ratios in the context of the two-dimensional pressure-driven flow through a microchannel.
} \label{fig0028}
\end{figure*}

Figure \ref{fig0028} presents a set of validation results for steady-state DBM simulations~\cite{Zhang2023POF}.
Specifically, Figures \ref{fig0028}(a) and \ref{fig0028}(b) show the curves of mass flow rate $Q_{p}$ versus the rarefaction coefficient $\delta$ in one- and two-dimensional pressure-driven flow problems, respectively.
Figure \ref{fig0028}(a) displays results for different TMAC $\alpha$ and Fig. \ref{fig0028}(b) shows the results for different aspect ratios $h/w$ of pipelines.
 The lines represent DBM simulation results, and the symbols represent data from previous literature.
The results show that DBM can accurately describe the mass flow rate of micropipe flow with different TMAC and different aspect ratios.

\section{\label{sec:level5} Applied research}

To study the unknown is the original intention and purpose of the new model.
New models are often built because existing models do not meet requirements.
A new model, if it does not exceed any of the physical functions of the earlier model, the necessity of its existence is greatly reduced!
In addition to the introduction of some specific research works, the author also wants to convey a more important idea: as the degrees of discreteness and thermodynamic non-equilibrium increase, the complexity of system behavior rises sharply;
it is incomplete to rely only on the macroscopic quantities in the NS model, and more physical quantities need to be introduced to describe the state and behavior of the system to ensure that the actual control ability of system behavior does not decline.
Therefore, this paper presented below includes results obtained from multiple analysis methods and approaches.
From the perspective of the ways of extracting complex physical field information, the following results are mainly obtained by three methods: macroscopic quantitative analysis, morphological feature analysis, and thermodynamic non-equilibrium effect/behavior analysis.
Among these, the thermodynamic non-equilibrium effect/behavior analysis means the description method based on the non-conserved moment of $(f-f^{(0)})$, as shown in Sections \ref{Brief-review} and \ref{Description}.
It should be noted that the definition of non-equilibrium strength in the following results may not be the same.
However they are all selected according to the specific system characteristics, and all describe the characteristics of the system behavior from the corresponding perspective.
Morphological feature analysis refers to the morphological analysis method based on Minkowski measure~\cite{Xu2022CMK}.
In a $D$-dimensional space, a set of convex sets satisfying motion invariance and additivity can be completely described by $D+1$ Minkowski measures.
For the two-dimensional case, the three Minkowski measures are the proportion $A$ of the area of the high-$\Theta$ region, the length $L$ of the boundary between the high-$\Theta$ and low-$\Theta$ regions, and the Euler characteristic coefficient $\chi$ describing the degree of connectivity of the regions.
Here $\Theta$ can be any physical quantity, such as density, temperature, non-equilibrium quantity, etc.
Different methods and different angles of the same method constitute a more complete description of the behavior of fluid systems.

Similarly, we first introduce the application of unsteady DBM and then introduce the application of steady DBM.
In terms of physical results, the simulation results shown below are divided into two categories, one that can also be obtained by NS, and another that NS cannot provide.
The former, such as the evolution characteristics of interface spikes in RT instability systems and the deformation process of bubbles during the shock-bubble interaction, will not be introduced here.
The latter represents where DBM surpasses NS.
These previously poorly understood discrete effects and non-equilibrium behavior characteristics contain a large number of physical functions to be developed.
The physical problems that DBM is concerned with include, but are not limited to, hydrodynamic instability, phase separation/non-equilibrium phase change, bubble/droplet collision, fusion and fragmentation, shock and detonation, and plasma kinetic theory, etc.

\subsection{ Detonation and shock wave }

\begin{figure*}[htbp]
\center\includegraphics*
[width=0.5\textwidth]{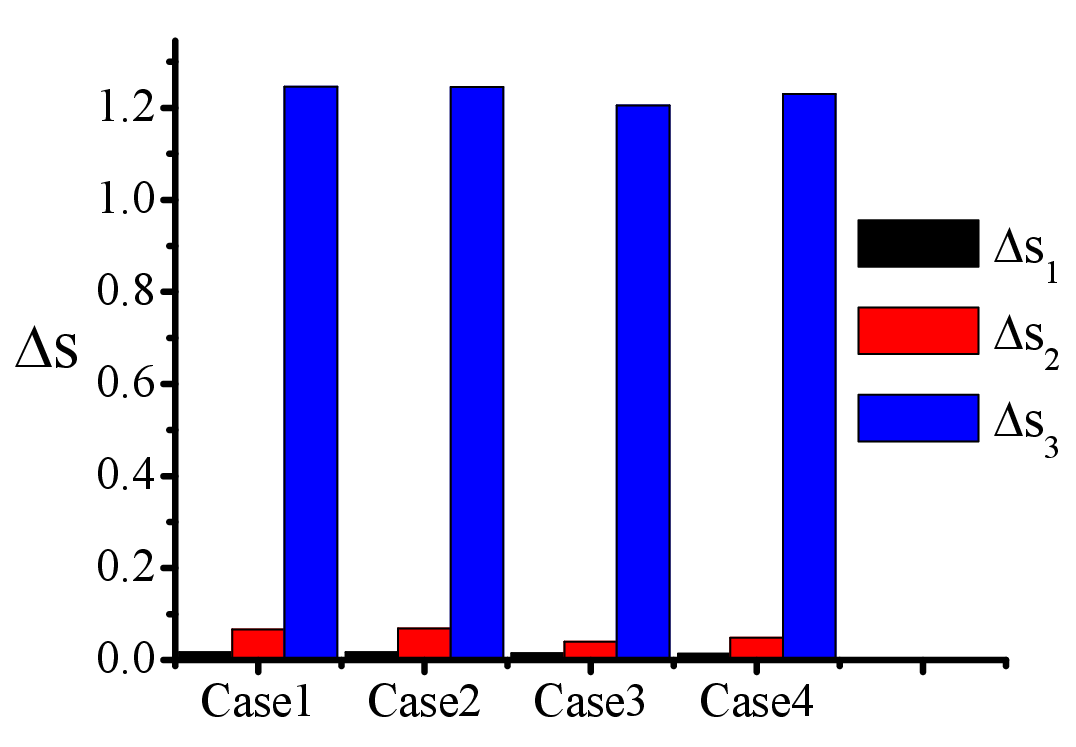}
\caption{   (From Ref. ~\cite{Zhang2016CNF} Fig. 12, with permission.) Comparison of three entropy production rates in fluid systems of detonation problems with chemical reactions.
} \label{fig0029}
\end{figure*}

Entropy production rate is an important concern in many fields related to compression science, such as shock wave and detonation physics, ICF, and aerospace~\cite{Song2023AAAS}.
By detecting TNE features from different perspectives, we can study the main mechanisms causing entropy production in the system and their relative importance.
In detonation problems involving chemical reactions, the entropy production rate contributes in three ways: chemical reactions, NOMF, and NOEF.
As shown in Fig. \ref{fig0029}, the entropy production rate for detonation problems involving chemical reactions is investigated~\cite{Zhang2016CNF}.
The results show that the entropy production rate ($\Delta S_3$) caused by the chemical reaction is much larger than the other two ($\Delta S_1$ and $\Delta S_2$). 

In many cases, the initial shock waveform is not flat. In the case where the initial waveform can be approximated as a plane plus disturbance, such shock waves are often referred to as disturbance shock waves for convenience of description. 
In general, the waveform of such a disturbed shock gradually returns to a plane as it propagates through the material. Such a process often referred to as shock shaping.
The RMI caused by the interaction between a plane shock wave and a disturbed interface has been extensively studied. 
In recent years, more attention has been paid to the RMI caused by interaction between a disturbed shock wave and plane interface\cite{Liao2019PRE,Zou2019JFM,Zou2020SSPMA}. At present, all the studies on shock shaping and RMI induced by the interaction between disturbed shock wave and plane interface in the literature are based on traditional fluid modeling. 
Shan \emph{et al.} are using DBM modeling and analysis method to investigate the kinetic behaviors, particularly those ignored by traditional fluid modeling \cite{Shan2024-in-preparation}.

\subsection{ Shock-bubble interaction  }

\begin{figure*}[htbp]
\center\includegraphics*
[width=0.8\textwidth]{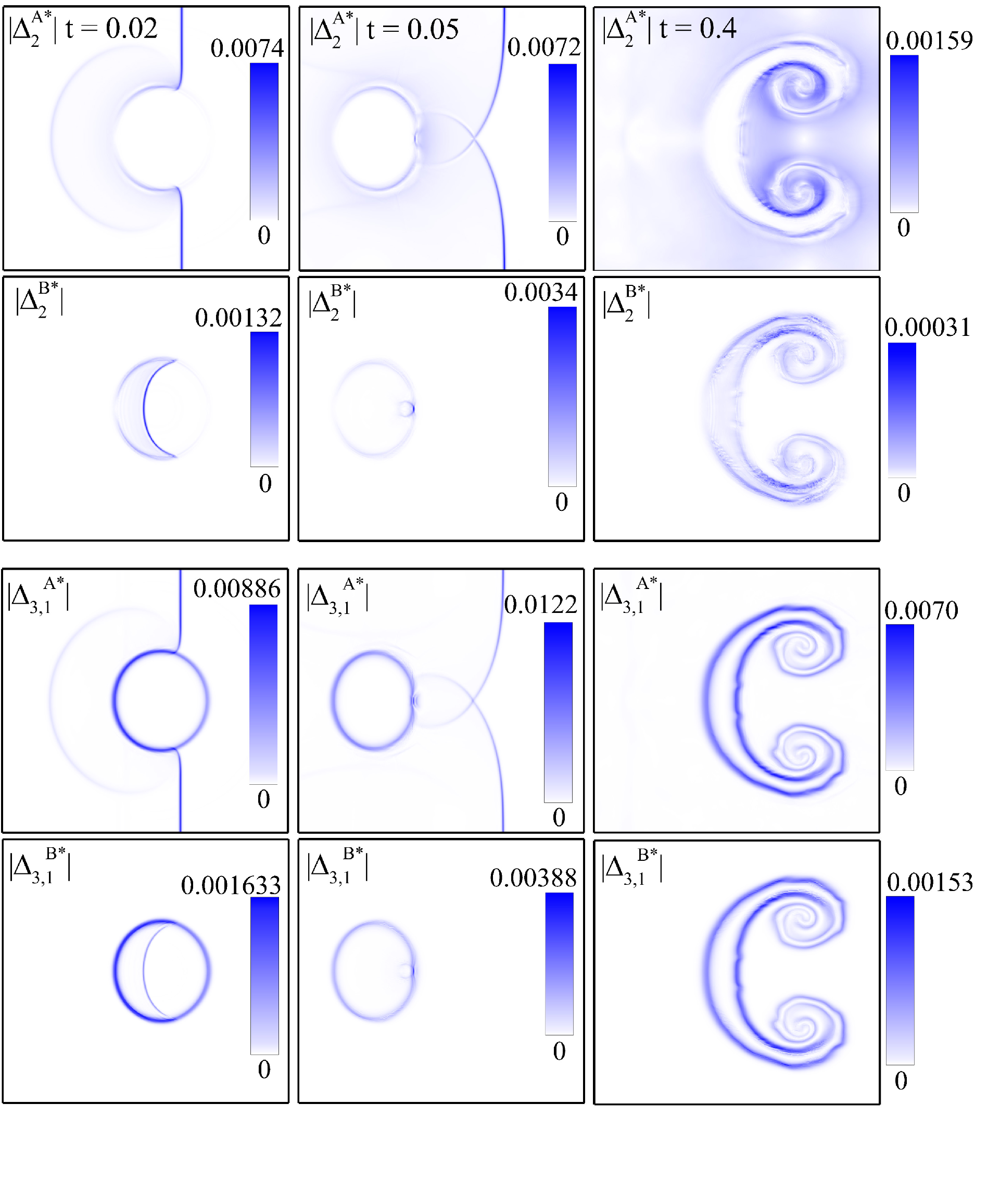}
\caption{   (From Ref. ~\cite{ZhangDJ2023POF} Fig. 9, with permission.)
Contours of the two typical TNE quantities $| \bm{\Delta}_{2}^{\sigma *} |$ and $| \bm{\Delta}_{3,1}^{\sigma *} |$ at three different moments.
} \label{fig0032}
\end{figure*}

\begin{figure*}[htbp]
\center\includegraphics*
[width=0.6\textwidth]{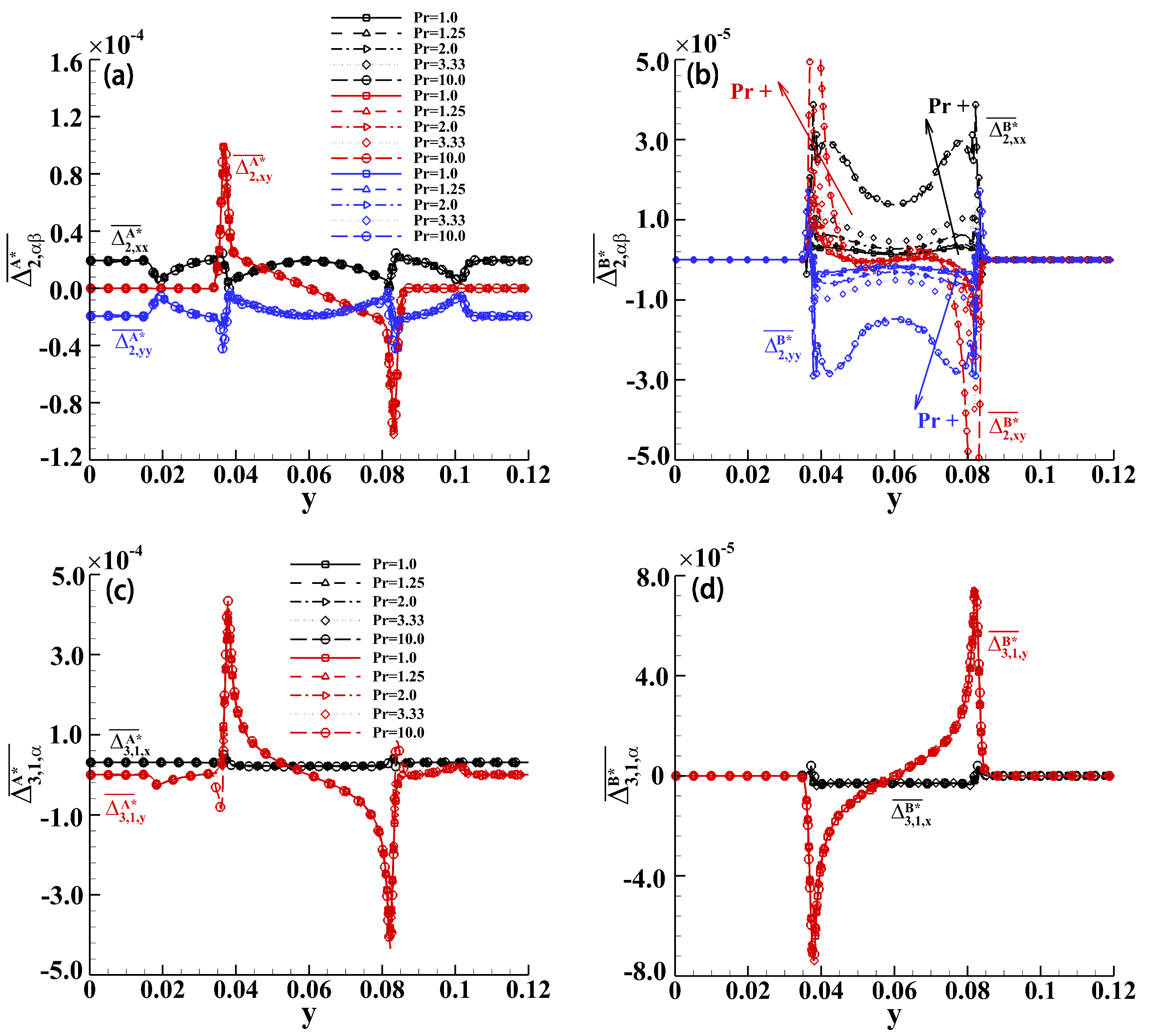}
\caption{   (From Ref. ~\cite{ZhangDJ2023POF} Fig. 10, with permission.)
Profiles of average TNE strength along the $y$ direction.
} \label{fig0033}
\end{figure*}

\begin{figure*}[htbp]
\center\includegraphics*
[width=0.45\textwidth]{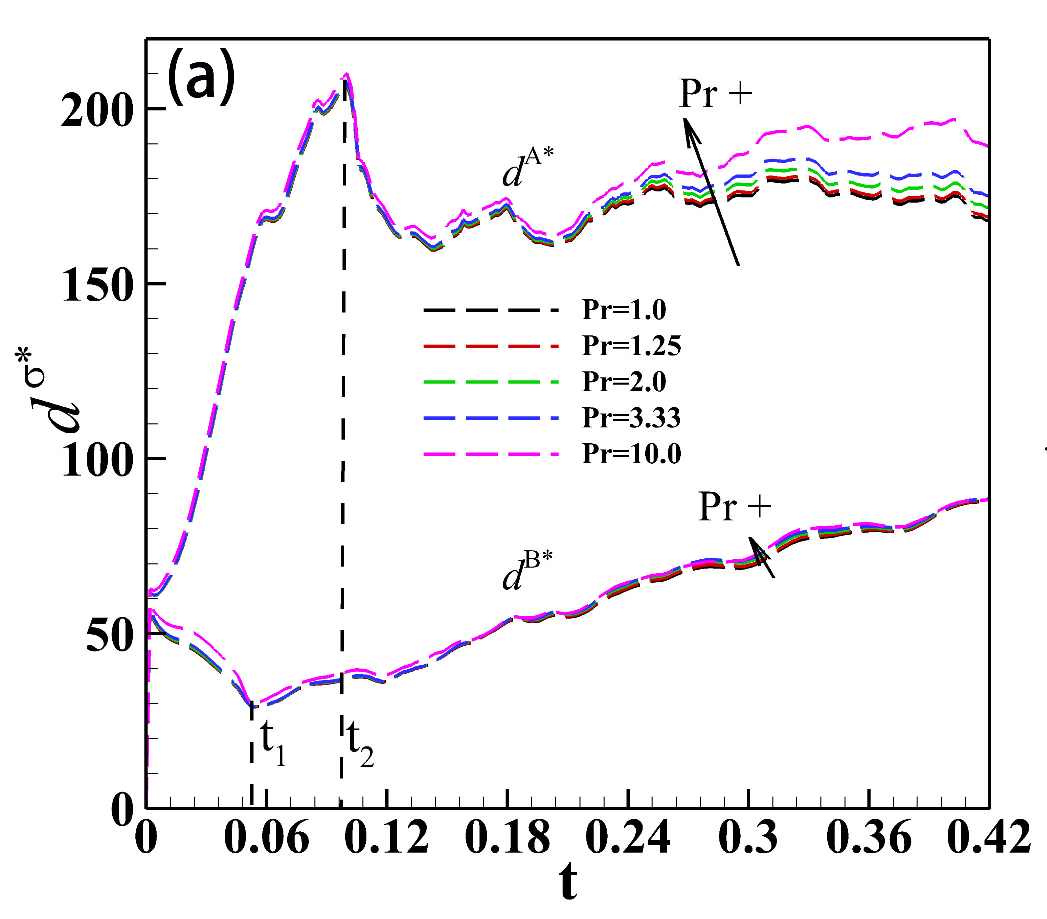}
\caption{   (From Ref. ~\cite{ZhangDJ2023POF} Fig. 11, with permission.) Evolution curves of the quantity of global TNE strength.
} \label{fig0034}
\end{figure*}

Below, we present examples from references Refs. ~\cite{Zhang2023CAF} and ~\cite{ZhangDJ2023POF} to illustrate the specific applications of TNE quantities.
These two works introduce a non-equilibrium intensity vector to study the non-equilibrium effect and behavior of the SBI process from multiple perspectives.
Figure \ref{fig0032} shows the contour for two typical non-equilibrium quantities of two different fluid components during the SBI process~\cite{ZhangDJ2023POF}.
In this figure, the first two rows are the spatial distribution of $| \bm{\Delta}_{2}^{*} |$, and the last two rows correspond to the spatial distribution of $| \bm{\Delta}_{3,1}^{*} |$.
The odd rows are the distribution of component A and the even rows are the distribution of component B.
Different columns in the figure correspond to different typical moments.
The location and characteristics of shock waves (incident shock wave, transmitted shock wave, etc.) and rarefaction wave are clearly captured by the non-equilibrium quantity.
Their spatial distribution is highly related to interfaces with strong macroscopic gradients, so they can be used to identify interface features.
At the same time, these non-equilibrium quantities also have rich physical connotations in phase space.
The positive and negative values of non-equilibrium quantity represent the direction that deviates from the equilibrium state, and the magnitude of a non-equilibrium quantity indicates the extent of deviation from equilibrium.
The non-equilibrium strength of the flow field is analyzed from another perspective.
Figure \ref{fig0033} shows the spatial profile of the average non-equilibrium strength along the $y$ direction obtained by summing and averaging the non-equilibrium quantity along the $x$ direction.
Figure \ref{fig0033} also shows the viscous effects on the two types of average non-equilibrium strengths.
It can be seen that the spatial profile of these non-equilibrium quantities shows interesting symmetries, in which the rich and complex non-equilibrium effects and mechanisms are far from being fully understood.
Taking $\overline{\Delta_{2,\alpha \beta}^{A*}}$ as an example, the profiles of $\overline{\Delta_{2,xx}^{A*}}$ is symmetry about the line $y=0.6$, indicating that the upper and lower parts of the flow field deviate from equilibrium in the same way.
From the perspective of $\overline{\Delta_{2,xy}^{A*}}$,  the profile is symmetric about the origin, suggesting that the upper and lower portions of the flow field deviate from equilibrium in opposite directions but with consistent magnitudes.
The profiles of $\overline{\Delta_{2,xx}^{A*}}$ and $\overline{\Delta_{2,yy}^{A*}}$ are both symmetric about the line $\overline{\Delta_{2,\alpha \beta}^{A*}} = 0$, indicating that when looking at the non-equilibrium strength of the fluid field from these two different perspectives, yields results with consistent magnitudes but opposite directions.
In other words, looking at the non-equilibrium strength of the system from different perspectives, the result may be the same or opposite.
Figure \ref{fig0033} only displays results for a particular time step, while results for other time steps are equally important.
It is because the non-equilibrium characteristics of the flow field are very complex and far from being fully understood, so multi-level and multi-perspective research is needed. Understanding the time evolution of non-equilibrium quantities is helpful in understanding the kinetics of SBI processes.
Figure \ref{fig0034} shows the global TNE strength of the two fluid components, where $t_1$ corresponds to the moment when the incident shock wave just sweeps the bubble, and $t_2$ is the moment when the incident shock wave exits the flow field, and these two moments correspond to the extremum of the global TNE strength curve.
After $t_2$, it can be seen that the global TNE strength curve is oscillatory, which is attributed to the influence of reflected shock waves.
Hence, global TNE strength can provide a comprehensive reflection of certain flow field characteristics, serving as another perspective for understanding system properties.
Different perspectives and different levels of description constitute a relatively complete understanding of the fluid system.

\subsection{  Phase transition and phase separation }

\begin{figure*}[htbp]
\center\includegraphics*
[width=0.5\textwidth]{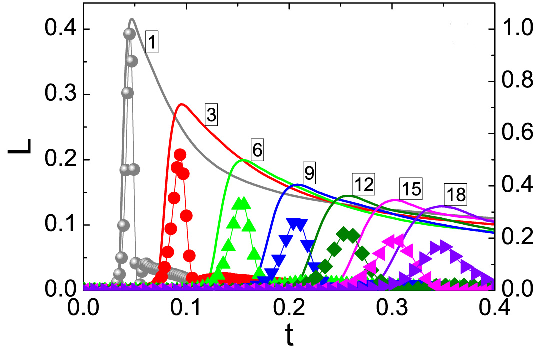}
\caption{ (From Ref. ~\cite{Gan2015Soft} Fig. 10, with permission.) Evolution curves of boundary length $L$ and non-equilibrium strength $D$ in non-isothermal phase separation process.
} \label{fig0038}
\end{figure*}

\begin{figure*}[htbp]
\center\includegraphics*
[width=0.6\textwidth]{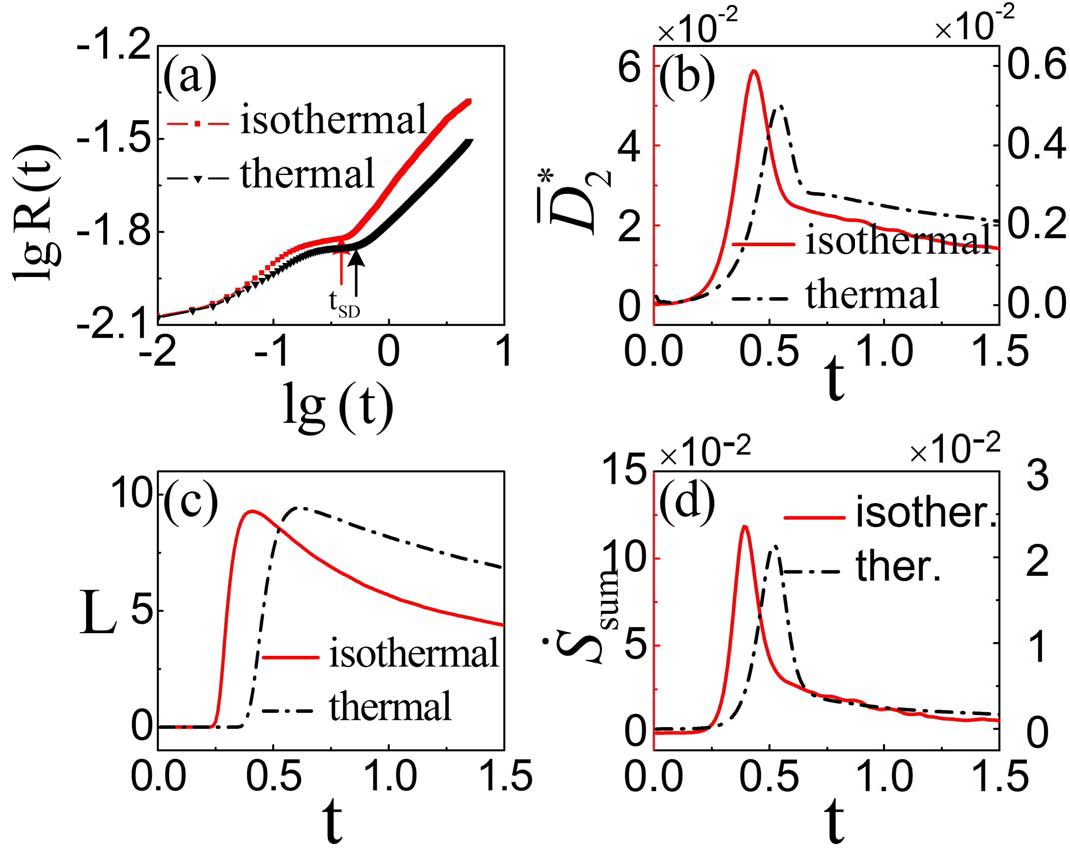}
\caption{ (From Ref. ~\cite{Zhang2019Soft} Fig. 5, with permission.) Evolution curves of several characteristic quantities in the phase separation process.
} \label{fig0039}
\end{figure*}

It is found that in the case of non-equilibrium phase transition and phase separation, as shown in Fig. \ref{fig0038}, the TNE strength $D$ gradually increases in the first stage of phase separation (Spin Decomposition stage, SD stage), and gradually decreases in the second stage (Domain Growth stage, DG stage).
Therefore, its maximum point can be used as the first physical criterion to divide the two stages~\cite{Gan2015Soft}.
Similar characteristics exist in the evolution of morphological characteristic quantity $L$, so it can also serve as a physical criterion to divide the two stages.
Figure \ref{fig0039} shows the evolution of logarithm of the characteristic length $\lg R$, non-equilibrium strength $\overline{D}_{2}^{*}$, boundary length $L$ and entropy production rate $\dot{S}_{sum}$ during phase transition.
The changing trends of these quantities can be used to identify the two stages~\cite{Zhang2019Soft}.
Specifically, the entropy production rate $\dot{S}_{sum}$ gradually increases in the first phase of phase separation (SD stage), and gradually decreases in the second phase (DG stage), so its maximum point can be used as the second physical criterion to divide the two stages.

\subsection{  Phase transition and bubble coalescence }

\begin{figure*}[htbp]
\center\includegraphics*
[width=0.5\textwidth]{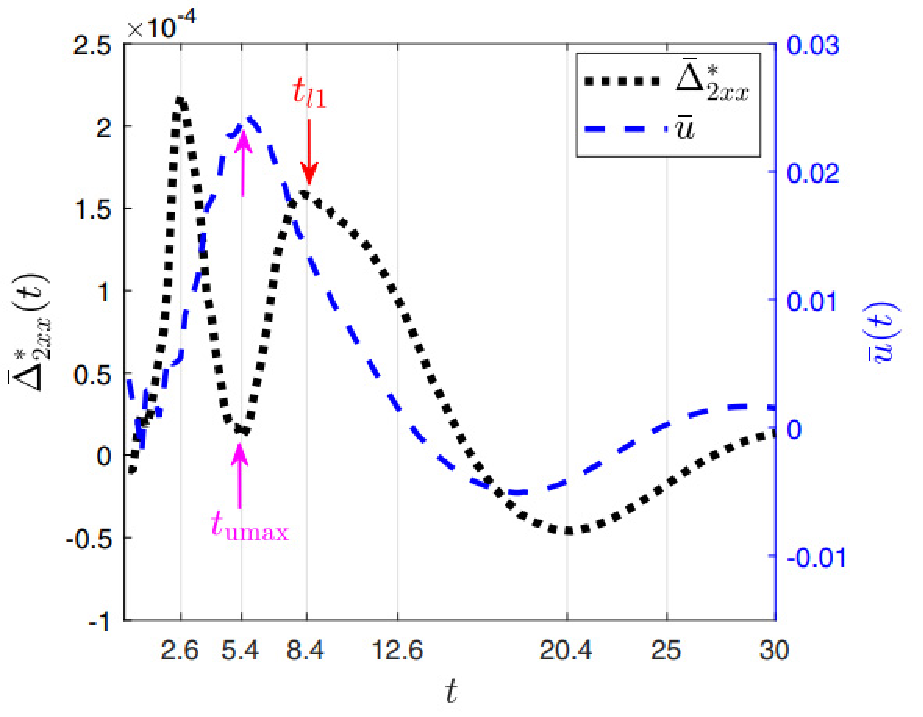}
\caption{ (From Ref. ~\cite{Sun2022PRE} Fig. 6, with permission.)
Evolution curves of average coalescence velocity $\bar{u}$ and $\Delta_{2xx}^{*}$.
} \label{fig0040}
\end{figure*}

\begin{figure*}[htbp]
\center\includegraphics*
[width=0.5\textwidth]{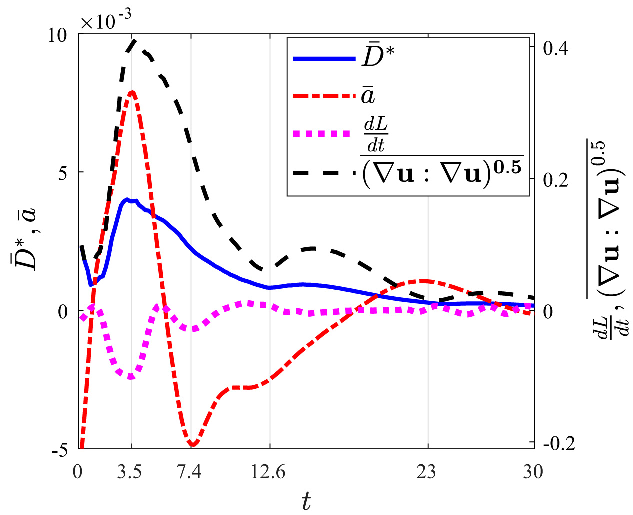}
\caption{ (From Ref. ~\cite{Sun2022PRE} Fig. 8, with permission.) Evolution curves of the average strength $\overline{D}^{*}$, average coalescence velocity $\overline{a}$, the slope of boundary length $dL/dt$, and $\overline{(\nabla \mathbf{u}:\mathbf{u})^{0.5}}$.
} \label{fig0041}
\end{figure*}

During the non-equilibrium phase transition and bubble coalescence, the characteristics of each component of average stress exhibit distinct stages~\cite{Sun2022PRE}.
Therefore, the quantity of $\bar{\Delta}_{2,xx}^{*}$ inside the bubble can be used to calibrate the bubble merging features and divide the merging stages.
As shown in Fig. \ref{fig0040}, the evolution of $\bar{\Delta}_{2,xx}^{*}$ shows two characteristic instants: the moment $t_{umax}$ when the mean coalescence speed $\bar{u}$ gets the maximum (at this time the ratio of minor and major axes is about 0.5) and the moment when the ratio of minor and major axes becomes 1 for the first time.
These two characteristic instants divide the process of bubble coalescence into three stages.
As also shown in Fig. \ref{fig0041}, the average TNE strength $\overline{D}^{*}$, coalescence acceleration $\overline{a}$ and the slope of boundary length $dL/dt$ are highly correlated in the early stage, and their amplitudes reach their maximum values almost simultaneously.
Therefore, their maximum point can be regarded as the end of the first stage of bubble coalescence and the beginning of the second stage.
In another work, Ref. \cite{Sun2023arXiv}, it is shown that the temporal evolutions of the total TNE strength and the total entropy production rate can both provide physical criteria to distinguish the stages of droplet coalescence.

\subsection{  Droplet collision }

\begin{figure*}[htbp]
\center\includegraphics*
[width=0.9\textwidth]{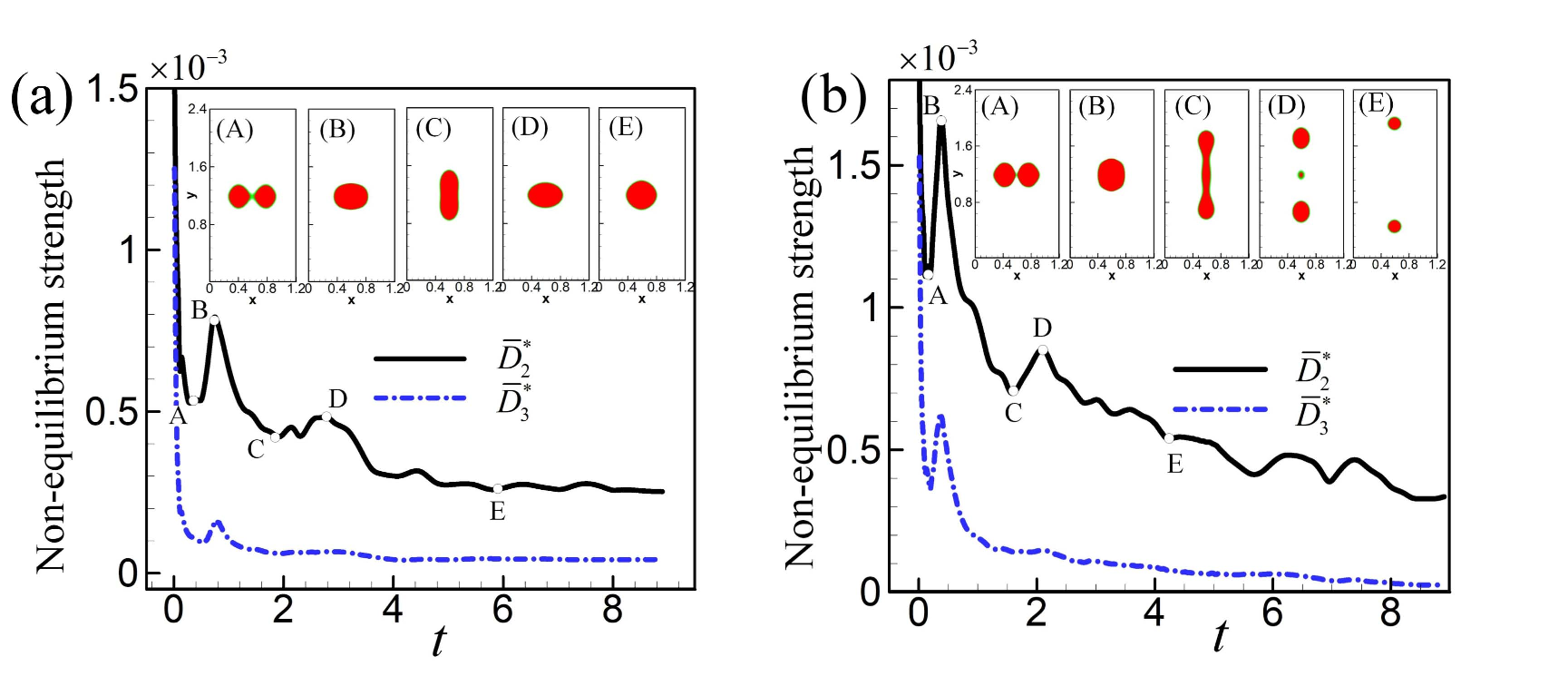}
\caption{  (From Ref. ~\cite{Zhang2020FOP} Fig. 10, with permission.) Evolution curves of two TNE quantities ($\overline{D}_{2}^{*}$ and $\overline{D}_{3}^{*}$) during the droplet collisions, where Figure (a) represents the two droplets are fused together after the head-on collision, the Figure (b) indicates the two droplets are separated after collision.
} \label{fig0042}
\end{figure*}

In droplet collision kinetics, the characteristic of the average non-organized momentum flow (NOMF) strength ($\overline{D_{2}^{*}}$) can be used to distinguish the type of droplet collision and divide the collision stage~\cite{Zhang2020FOP}.
As shown in Fig. \ref{fig0042}, in the collision-fusion case, in the late period of collision (after point D in the figure), $\overline{D_{2}^{*}}$ rapidly decreases and then  oscillates towards  to a steady state.
However, in collision-separation case, $\overline{D_{2}^{*}}$ continues to exhibit significant oscillations during the descent.

\subsection{  RTI and anti-RTI: effects of intermolecular potential and high compressibility }

\begin{figure*}[htbp]
\center\includegraphics*
[width=0.5\textwidth]{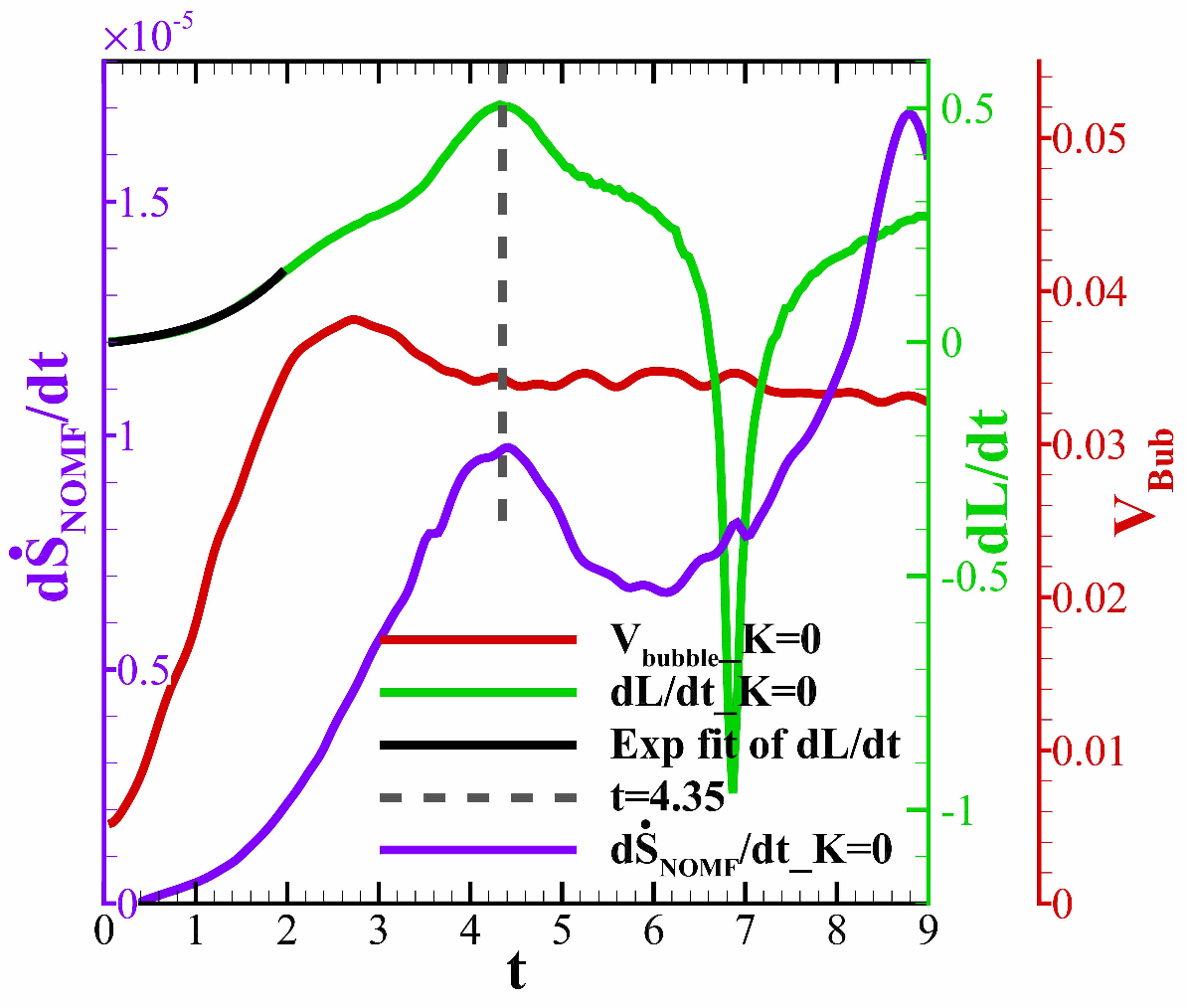}
\caption{  (From Ref. ~\cite{Chen2022PRE} Fig. 11, with permission.)
Evolution curves of change rate of entropy production rate $d\dot{S}_{\rm{NOMF}}/dt$, change rate of boundary length $dL/dt$, and bubble velocity $V_{\rm{bubble}}$.
} \label{fig0043}
\end{figure*}

In the study of RTI with consideration of intermolecular potential effects, the surface tension has a stage-wise impact on perturbation amplitudes, bubble velocity, and two entropy generation rates~\cite{Chen2022PRE}.
The interface tension inhibits the evolution of RTI during the bubble acceleration stage.
In the stage of bubble asymptotic velocity, the interface tension first promotes and then inhibits the evolution of RTI.
In addition, as shown in Fig. \ref{fig0043}, the first maximum point of $dL/dt$ (green line in the figure) and the first maximum point of $d\dot{S}_{\rm{NOMF}}/dt$ induced by NOMF (purple line in the figure) can both be used as  physical criteria for the bubble velocity entering the asymptotic stage in RTI.
Interface length $L$ is also one of the perspectives for the study/characterization of TNE behavior, and the exponential growth stage of $L$ corresponds to the bubble acceleration stage. 

RTI under constant acceleration has been studied extensively and a relatively mature physical understanding has been obtained, but the acceleration of the RTI occurring in ICF is variable. The research on RTI under variable acceleration is relatively few, and the physical understanding is insufficient. The reverse of acceleration will cause the anti-RTI mechanism and behavior, and Jia \emph{et al.} \cite{Jia2024-in-preparation} are conducting research on this issue. It has been shown that the magnitude of acceleration and the time of acceleration reversal have significant influences on the evolution of RTI, such as the location of the interface, the location and development trend of the spike bubble, the evolution of local and global TNE characteristics and effects, etc.. In the strongly compressible case, some new behaviors and mechanisms may also emerge in the evolution of the RTI system. For example, the density of the fluid from the upper layer which is originally higher may become lower dense than that of the surrounding originally lighter fluid, resulting in the exchange of identities of the light and heavy fluids, leading to anti-RTI mechanism and behavior. The emergence of these new mechanisms and behaviors invalidate earlier methods of determining the source of matter particles based on density. Therefore, Chen \emph{et al.} \cite{Chen2024-in-preparation} are introducing tracer particles into the DBM model of van der Waals fluids to study the influence of flow compressibility on the evolution of single-mode RTI systems of van der Waals fluids from the perspective of statistical physics combined with morphological analysis.

\subsection{  KHI and coupled RT-KHI system }

In order to solve the problem of analyzing various complex physical fields during the evolution of KHI, in the literature \cite {Gan2019FOP}, Gan \emph{et al.} combined tracking non-equilibrium behavior characteristics and morphological analysis techniques to carry out physical identification and tracking technology design of feature structures or patterns, and quantitatively characterize the width and development rate of KHI mixed layer. It is found that the width of the mixed layer, the non-equilibrium strength and the length of the interface boundary show a high spatiotemporal correlation. The viscosity, which is highly correlated with the NOMF, inhibits the development rate of KHI, prolongates the duration of linear phase, reduces the maximum disturbance kinetic energy, improves the overall and local thermal non-equilibrium strength, and expands the range of non-equilibrium. The thermal conduction effect, generally consistent with the viscous effect, inhibits the growth of structures with small wavelengths during the evolution of KHI, thus making it easier for the system to form large-scale structures. Different from the consistent inhibition of viscosity on KHI, the thermal conduction effect is first suppressed and then increased in phases when the heat conduction is strong. The inhibition is due to the heat conduction extending the interface width, reducing the macroscopic quantity gradient and the non-equilibrium driving force strength, and the enhancement is due to the broadened density transition layer making the KHI more easily absorb energy from the fluid on both sides, thus enhancing the KHI. The competition of these two effects accompanied the whole process of KHI evolution, the inhibitory effect is dominant in the early stage, and the enhancement effect is dominant in the late stage.

\begin{figure*}[htbp]
\center\includegraphics*
[width=1.0\textwidth]{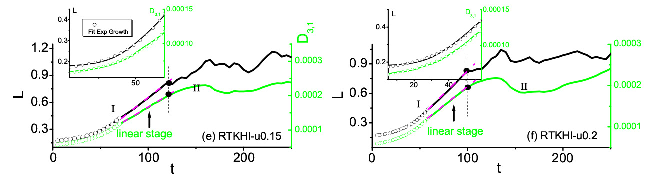}
\caption{  (From Ref. ~\cite{Chen2020POF} Fig. 13, with permission.)
Evolution curves of boundary length $L$ and strength of NOEF $D_{3,1}$ in coupled RT-KHI system.
These two figures correspond to the cases with different initial shear velocity.
} \label{fig0044}
\end{figure*}

In the coupled RT-KHI system, the TNE quantities can serve as criteria for identifying the transition of dominant instability mechanisms~\cite{Chen2020POF,Chen2022FOP}.
As shown in Fig. \ref{fig0044}, in the case where KHI dominates at an earlier time and the RTI dominates at a later time (corresponding to shear dominance and buoyancy dominance, respectively), the boundary length $L$ and NOEF strength show similar linear growth stages.
The end of the linear growth stage of boundary length $L$ and NOEF strength $D_{3,1}$ can be used as a physical criterion for the transition from KHI dominance to RTI dominance~\cite{Chen2020POF}.

\subsection{  Shock wave/boundary layer interaction}

There exist complex shock wave/boundary layer interaction phenomena in supersonic flow. 
The main sources of aerodynamic drag in scramjet engines are ``friction drag'' caused by boundary layer and ``wave drag'' caused by shock wave, both of which are directly related to entropy increase, so entropy increase is the key parameter to evaluate the aerodynamic drag. 
Song \emph{et al.}\cite{Song2023AAAS} recently investigated this problem using the DBM modeling and analysis method. 
Their results show that for regular reflection, the non-equilibrium intensity of the reflected shock wave is stronger than that of the incident shock wave. 
For shock wave/laminar boundary layer interaction, entropy production rate caused by viscosity is dominant in shock wave, and entropy production rate caused by heat conduction is dominant in boundary layer. 
The intensity of the two entropy production rates increases with the increase of Mach number. 
The research results can provide theoretical guidance for evaluating the flow quality in the inlet.

\subsection{  Plasma shock wave }

\begin{figure*}[htbp]
\center\includegraphics*
[width=0.8\textwidth]{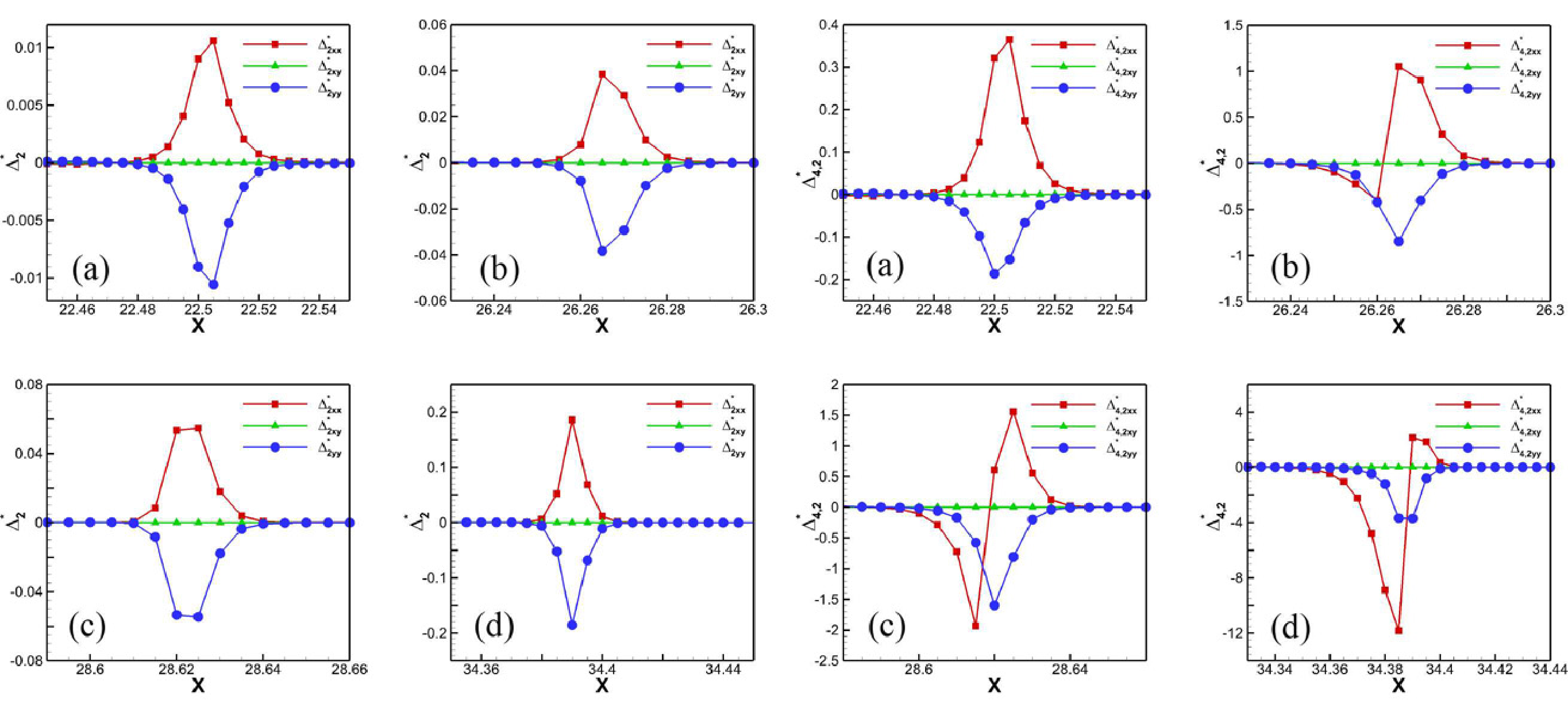}
\caption{  (From Ref. ~\cite{Liu2023JMES} Figs. 12 and 15, with permission.)
Profiles of  $\bm{\Delta}_{2}^{*}$ and $\bm{\Delta}_{4,2}^{*}$ of plasma shock wave with different Ma numbers.
} \label{fig0045}
\end{figure*}

In plasma research, TNE characteristics can be used to physically distinguish plasma shock waves from neutral fluid shock waves.
This distinction can help in the design of tracking techniques for the respective interfaces~\cite{Liu2023JMES}.
Figure \ref{fig0045} shows profiles of TNE quantity near the shock front in plasma shock wave under different Ma numbers.
The left four figures are for $\bm{\Delta}_{2}^{*}$ and the right four figures are for $\bm{\Delta}_{4,2}^{*}$.
In a neutral fluid shock wave, the TNE quantities $\bm{\Delta}_{2}^{*}$ and $\bm{\Delta}_{4,2}^{*}$ exhibit symmetry with their $xx$ and $yy$ components deviating from equilibrium in opposite directions (results for neutral fluids are not shown here).
For plasma shock waves, the profiles of non-equilibrium quantity $\bm{\Delta}_{2}^{*}$ behave the same as that of neutral fluid.
However, the profile of $\bm{\Delta}_{4,2}^{*}$ does not exhibit the symmetric deviation.
Thus, these differences can be used to distinguish plasma shock waves from neutral fluid shock waves.

\subsection{  Plasma Orszag-Tang vortex }

\begin{figure*}[htbp]
\center\includegraphics*
[width=0.8\textwidth]{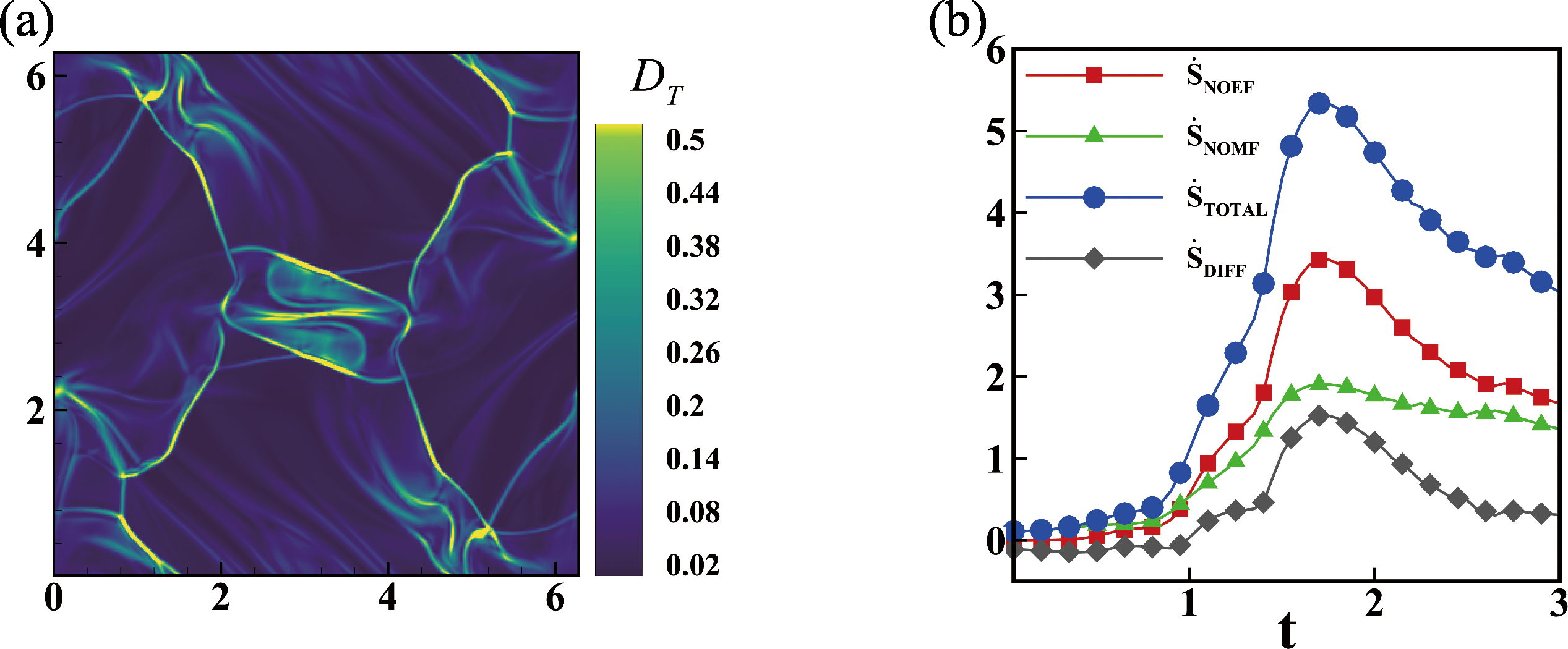}
\caption{  (From Ref. ~\cite{Song2023POF} Fig. 3, with permission.)
TNE effects of Orszag-Tang vortex problem. 
(a) Contour of total TNE strength at $t=3$. 
(b) Evolution of four kinds of entropy production rates with time.
} \label{fig046}
\end{figure*}

Through a DBM for plasma system, the kinetic behaviors of the Orszag-Tang vortex problem are investigated for the first time~\cite{Song2023POF}.
Figure \ref{fig046}(a) shows the contour of total TNE strength at $t=3$. 
It is observed that the total TNE strength is mainly distributed near the shock front. Figure \ref{fig046}(b) shows the evolution of four kinds of entropy production rates with time. 
In general, the entropy production rate is related to the difficulty of compression. 
From Fig. \ref{fig046}(b), it is found that the compression difficulty is divided into two stages. 
Before $t=1.7$, the entropy production rate and compression difficulty increase with time. 
After $t=1.7$, the entropy production rate and compression difficulty decrease with time.

\subsection{  Plasma RMI }

\begin{figure*}[htbp]
\center\includegraphics*
[width=0.8\textwidth]{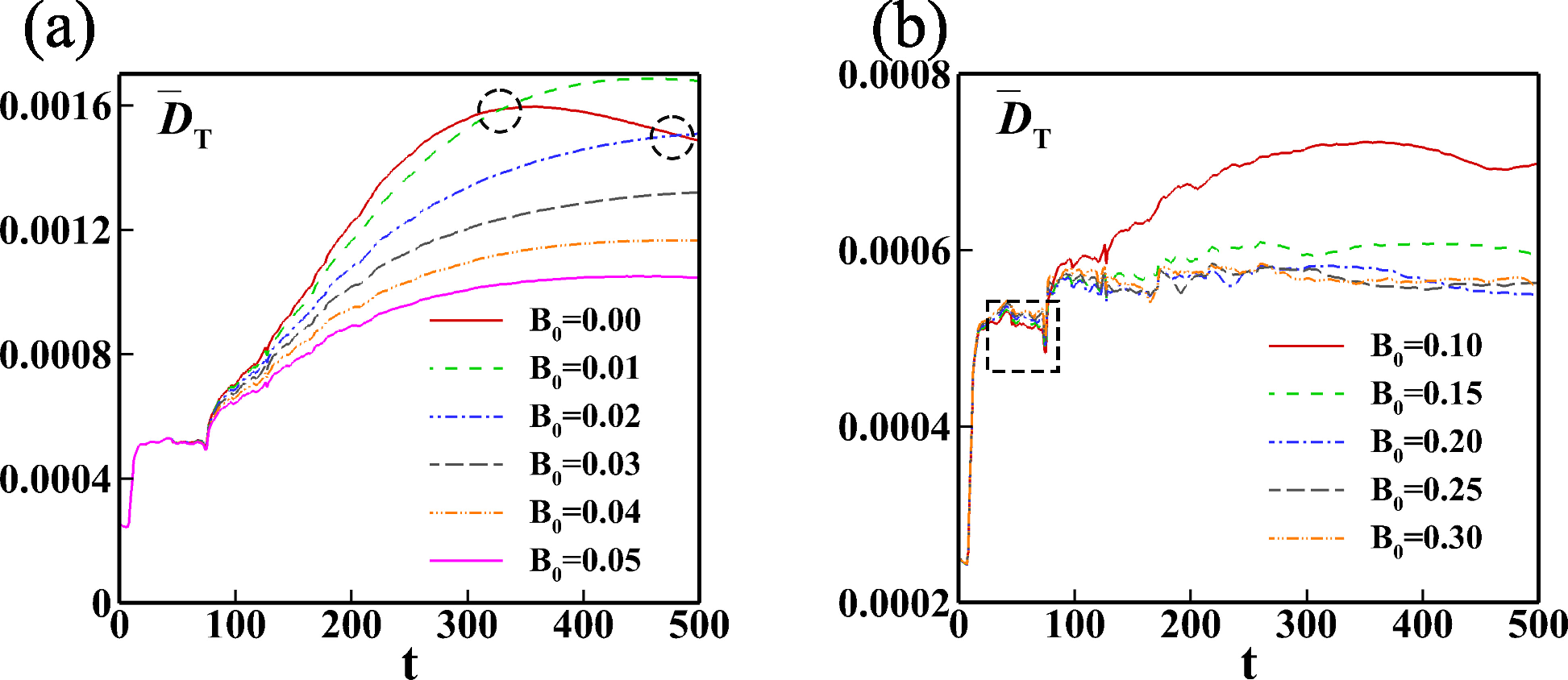}
\caption{  (From Ref. ~\cite{Song2023POF} Fig. 13, with permission.)
Evolution of global average TNE effects with different initial applied magnetic fields from $t = 0$ to 500.
} \label{fig030}
\end{figure*}

\begin{figure*}[htbp]
\center\includegraphics*
[width=0.8\textwidth]{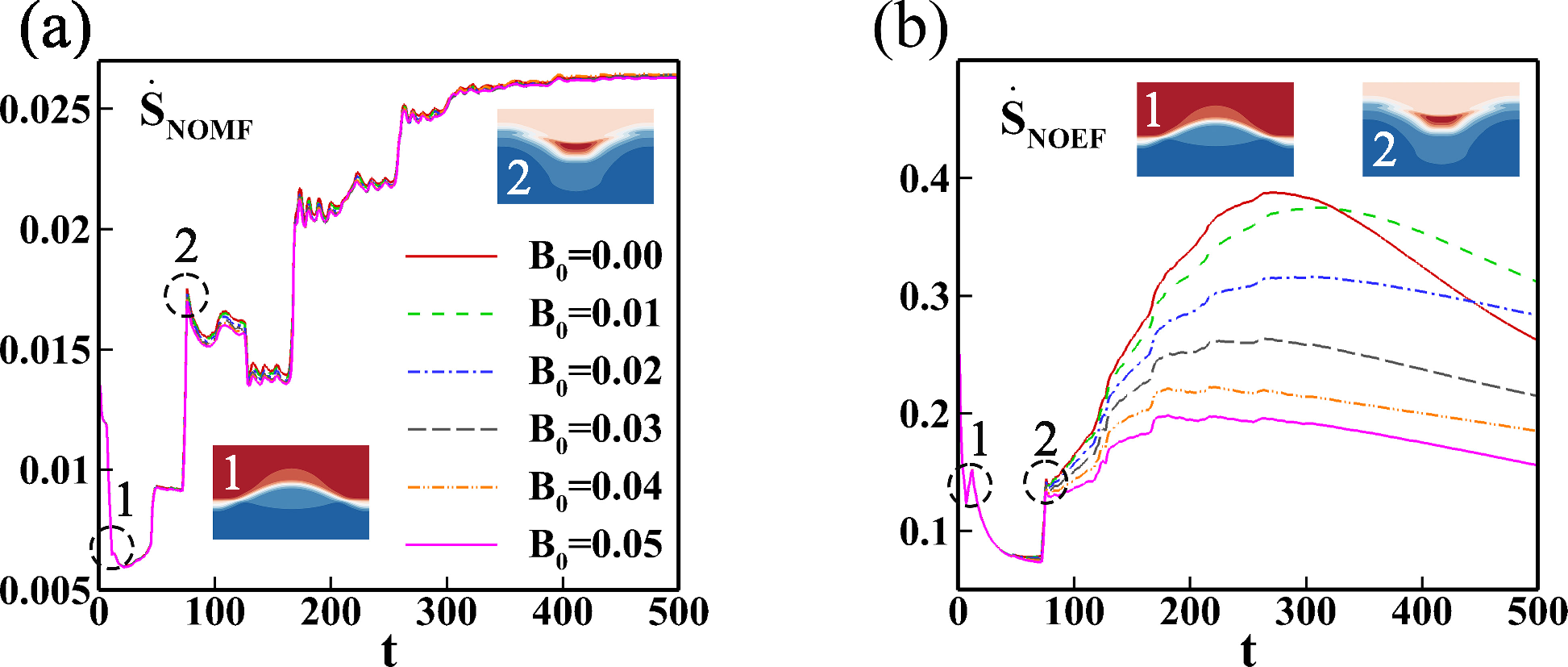}
\caption{  (From Ref. ~\cite{Song2023POF} Fig. 14, with permission.)
Evolution of entropy production rates from $t = 0$ to 500 with magnetic fields ranging from 0.01 to 0.05.
} \label{fig031}
\end{figure*}

\begin{figure*}[htbp]
\center\includegraphics*
[width=0.8\textwidth]{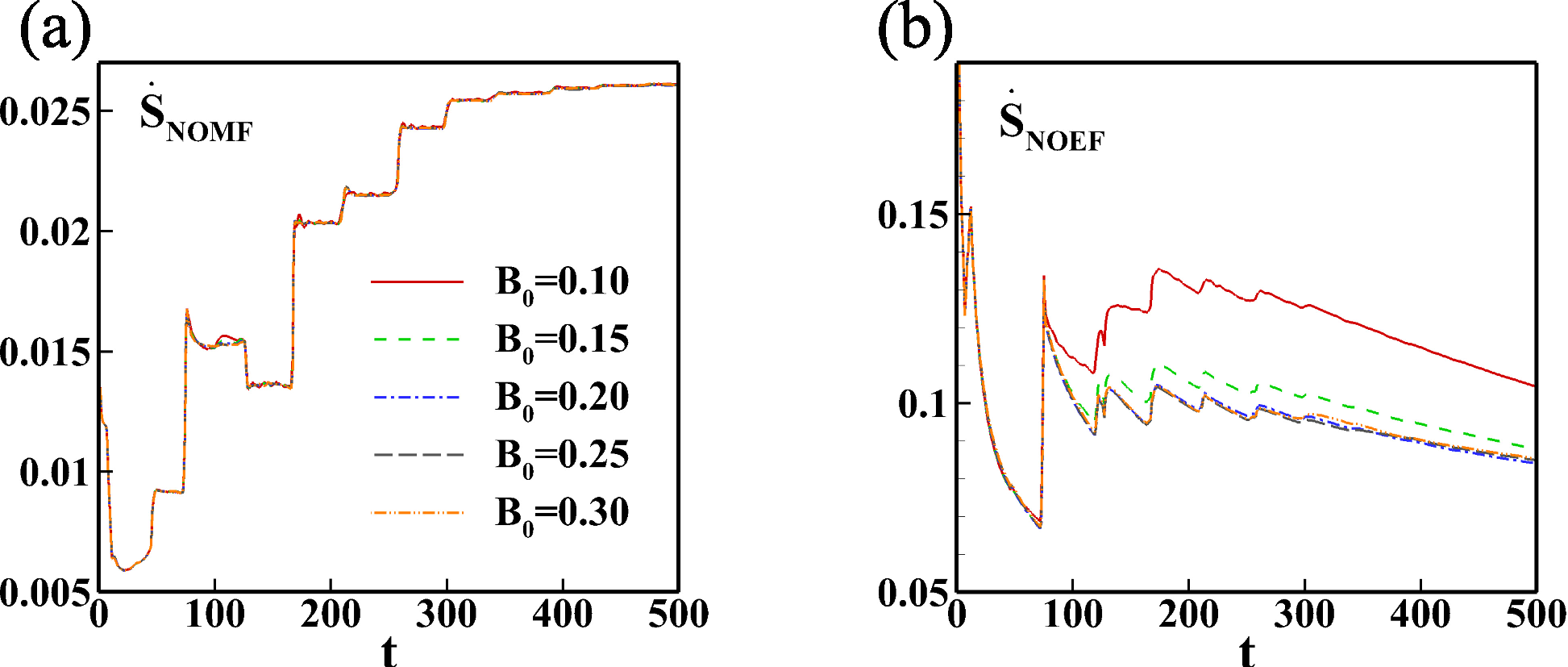}
\caption{  (From Ref. ~\cite{Song2023POF} Fig. 15, with permission.)
Evolution of entropy production rates from $t = 0$ to 500 with magnetic fields ranging from 0.10 to 0.30.
} \label{fig0046}
\end{figure*}

The influence of magnetic field on KHI has been extensively studied since the 1960s, where transport characteristics such as viscosity and heat conduction are the focus of attention. 
The physical models used in previous studies are mainly macroscopic magnetohydrodynamic (MHD) models based on the continuum assumption. However, the MHD is valid only when the time- and length-scales of particle collisions are sufficiently small compared to those of the system behaviors under consideration.
Rinderknecht \emph{et al.} \cite{Rinderknecht2018PPCF} pointed out that the kinetic behaviors occurring on the time- and length-scales of particle collisions may cause the hydrodynamic results to deviate from the experimental results.
Song \emph{et al.} \cite{Song2023POF} studied RMI kinetics using DBM modeling and analysis method. 
As a preliminary study, only the case of first order discrete/TNE effect is considered in this work. A few typical results are as follows: (i) As shown in  Fig. \ref{fig030}, before the RMI interface inversion, the magnetic field indirectly enhances the TNE strength $\overline{D}_{T}$ by inhibiting the interface development. 
After the interface inversion, the TNE strength $\overline{D}_{T}$ is significantly reduced. 
The minimum value of the global average TNE strength can be used as a physical criterion to determine the critical magnetic field strength under which the RMI interface amplitude could be reduced to 0. 
(ii) As shown in Figs. \ref{fig031} and \ref{fig0046}, the magnetic field has little effect on the entropy production rate contributed by the NOMF $\dot{S}_{NOMF}$, but has a strong inhibition effect on the entropy production rate contributed by the NOEF $\dot{S}_{NOEF}$. 
The minimum value of $\dot{S}_{NOEF}$ can be used as a physical criterion to determine the critical magnetic field strength under which the RMI interface amplitude could be reduced to 0.

\subsection{  Plasma KHI }
Song \emph{et al.} \cite{Song2024-in-preparation} are introducing tracer particles into DBM to study the influence of magnetic field on the transport and mixing characteristics of plasma KHI, as well as the influence of magnetic field on various kinetic behaviors. Preliminary results are as follows:
(i) For the cases of weak magnetic fields, the magnetic field could inhibit the development of KHI at an early stage, but will lead to the formation of small-scale vortex structure and cause the KHI to enter the mixing stage earlier. 
The degree of mixing $\bar \chi _p $ decreases in the early stage but increases in the later stage. 
(ii) For the cases of strong magnetic fields, the development of KHI is significantly inhibited, and the mixing degree is greatly reduced. 
(iii) With the increase of Knudsen number, the TNE strength, entropy production rate and mixing degree all decrease.

The following two subsections continue to introduce the application of TNE effects in probing complex systems. 
Just as a fingerprint is a person's identity. TNE feature is the identity feature of a system or behavior.
The techniques described below can be used to study various complex flows, not just the types listed above.

\subsection{ Recovery of actual distribution function }

\begin{figure*}[htbp]
\center\includegraphics*
[width=0.7\textwidth]{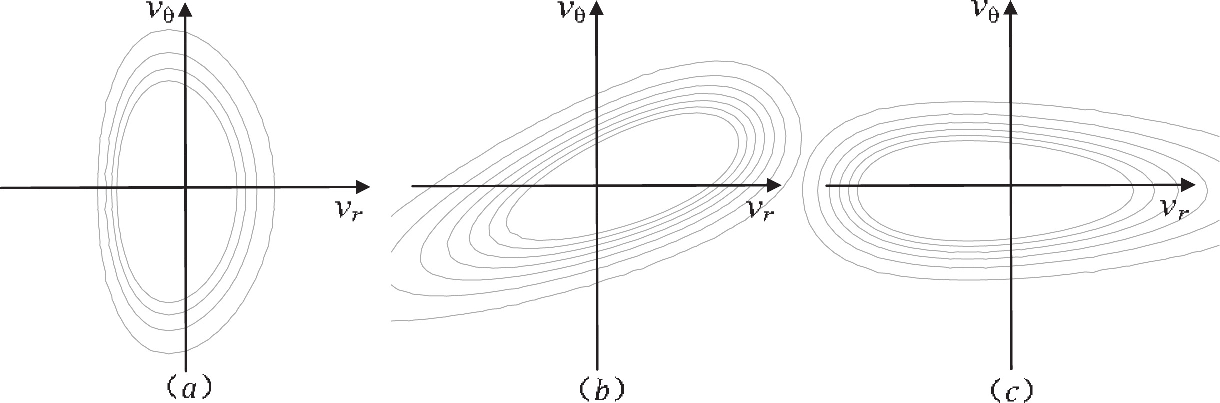}
\caption{   (From Ref. ~\cite{Lin2014PRE} Fig. 26, with permission.) The sketch of the two-dimensional contours of actual distribution function in velocity space ($v_{r}$,$v_{\theta}$) at rarefaction front, material interface, and shock front, respectively.
} \label{fig0030}
\end{figure*}

\begin{figure*}[htbp]
\center\includegraphics*
[width=0.7\textwidth]{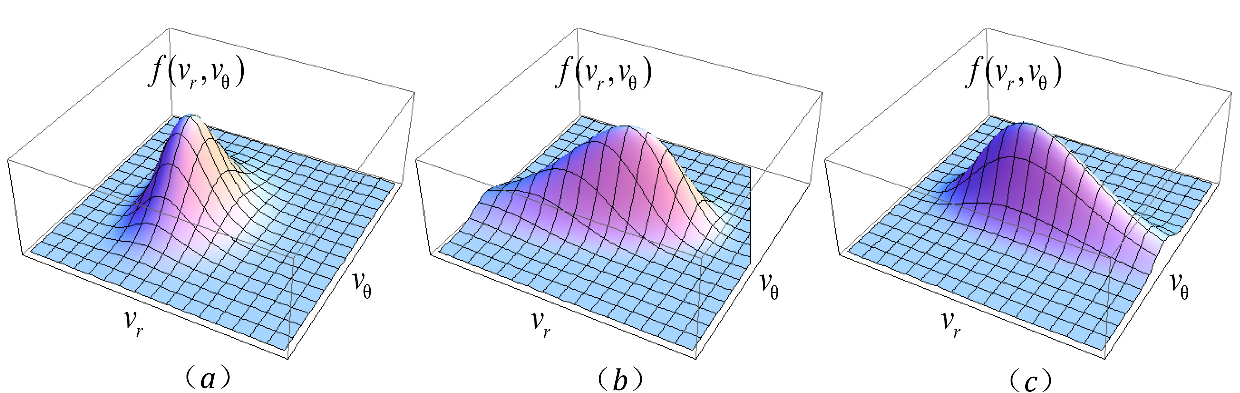}
\caption{   (From Ref. ~\cite{Lin2014PRE} Fig. 27, with permission.)
The three-dimensional contour graph of the actual distribution function in velocity space ($v_{r}$,$v_{\theta}$) at rarefaction front, material interfaces, and shock fronts, respectively.
} \label{fig0031}
\end{figure*}

The TNE feature of a flow state is its identity.
The kinetic theory uses the distribution function $f$ to describe the system state, while the discrete distribution function $f_i$ itself does not have a clear physical meaning in LBM and unsteady DBM. Specifically, $f_i$ is by no means the actual distribution function of velocity ${\mathbf{v}_i}$. What physically meaningful are the kinetic moments, such as ${\mathbf{M}}_n^* \left( {f_i } \right)$ and 
$ {\bm{\Delta }}_n^*  = {\mathbf{M}}_n^* \left( {f_i - f_i^{eq} } \right)$.
The equilibrium distribution function $f^{eq}$ can be obtained from the density $\rho$, velocity $\mathbf{v}$ and temperature $T$. Therefore,
DBM can use $f^{eq}$ and $ {\bm{\Delta }}_n^* $ to recover the main features of the actual distribution function $f$ of the target region~\cite{Lin2014PRE,Zhang2018FOP}.
Figure \ref{fig0030} is the two-dimensional contours of the actual distribution function in velocity space ($v_{r}$,$v_{\theta}$) at the rarefaction wave, the material interface and the shock front, respectively, and Fig. \ref{fig0031} displays  the three-dimensional contours of actual distribution function at the corresponding positions.
It can be seen that the distribution functions of different interfaces show different morphological appearances.
DBM simulations and theoretical derivations allow for the representation of the morphological features of the real distribution functions.
In literatures \cite{Zhang2018FOP,Su2022CTP,Su2023CTP}, an alternative method 
for recovering the main characteristics of the velocity distribution quantitatively is presented and developed through the macroscopic quantities and their spatial and temporal derivatives.

\subsection{ Identification of various interfaces }

\begin{figure*}[htbp]
\center\includegraphics*
[width=0.7\textwidth]{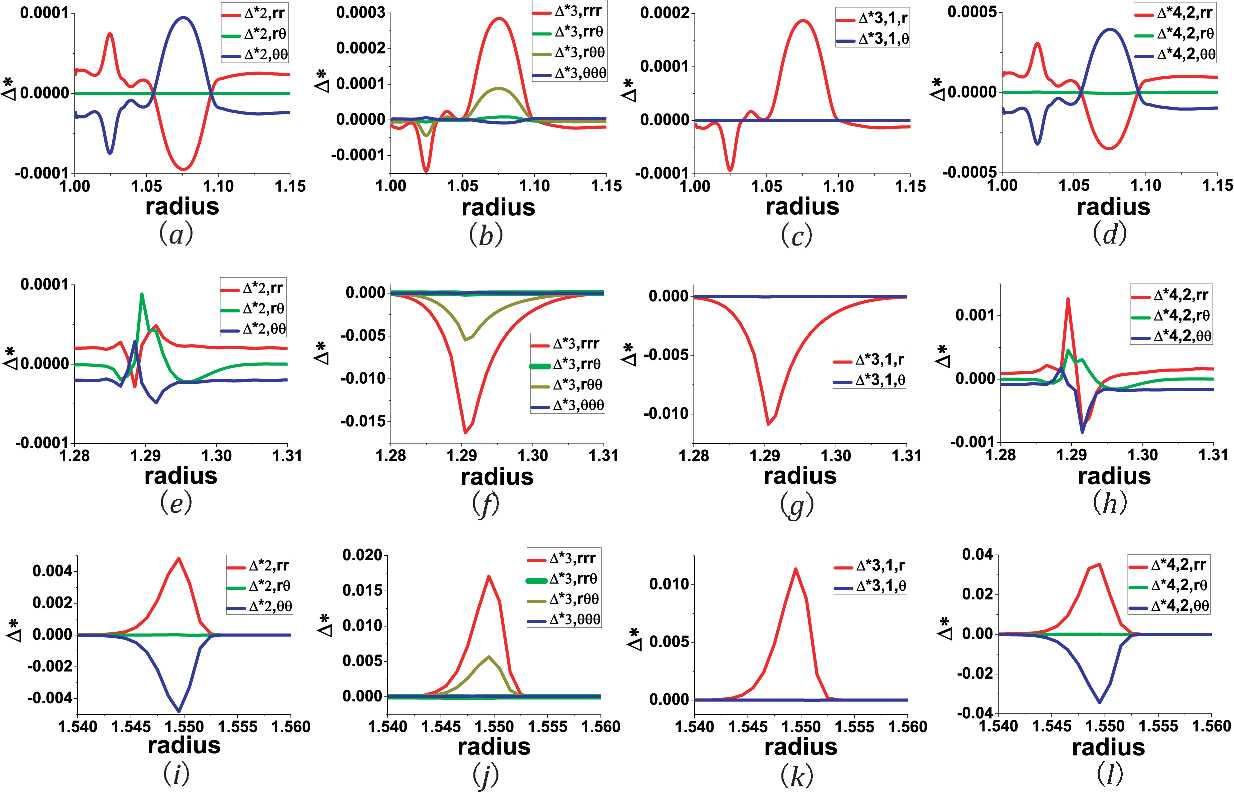}
\caption{   (From Ref. ~\cite{Lin2014PRE} Fig. 23, with permission.)
Profiles of four types of TNE quantities.
Each column of the figure represent different TNE quantities, and each row indicate the case with various positions.
} \label{fig0035}
\end{figure*}

\begin{figure*}[htbp]
\center\includegraphics*
[width=0.5\textwidth]{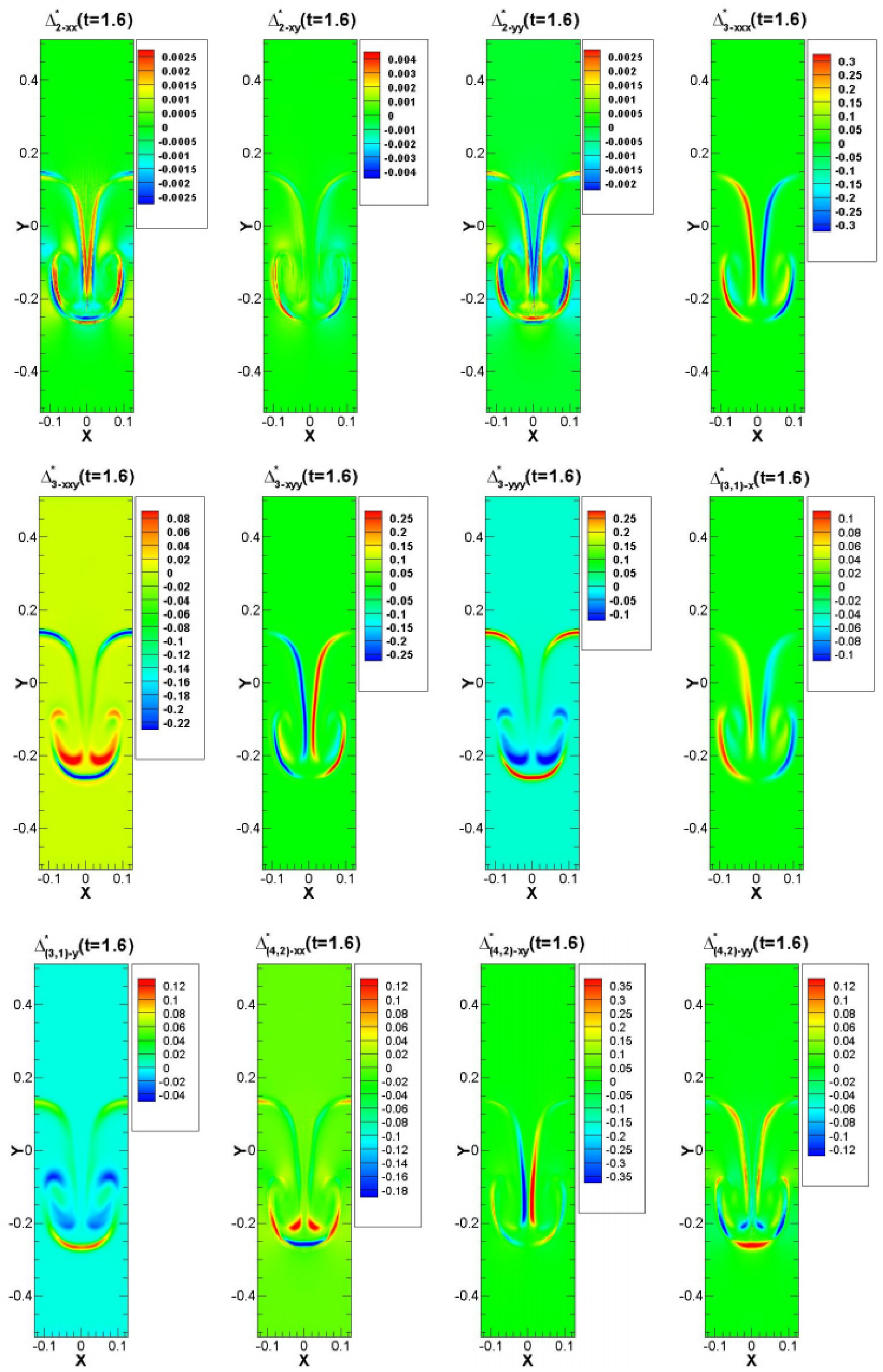}
\caption{  (From Ref. ~\cite{Lai2015} Fig. 4.3, with permission.) Contours of various TNE quantities at a typical moment during the RTI evolution.
} \label{fig0036}
\end{figure*}

\begin{figure*}[htbp]
\center\includegraphics*
[width=0.5\textwidth]{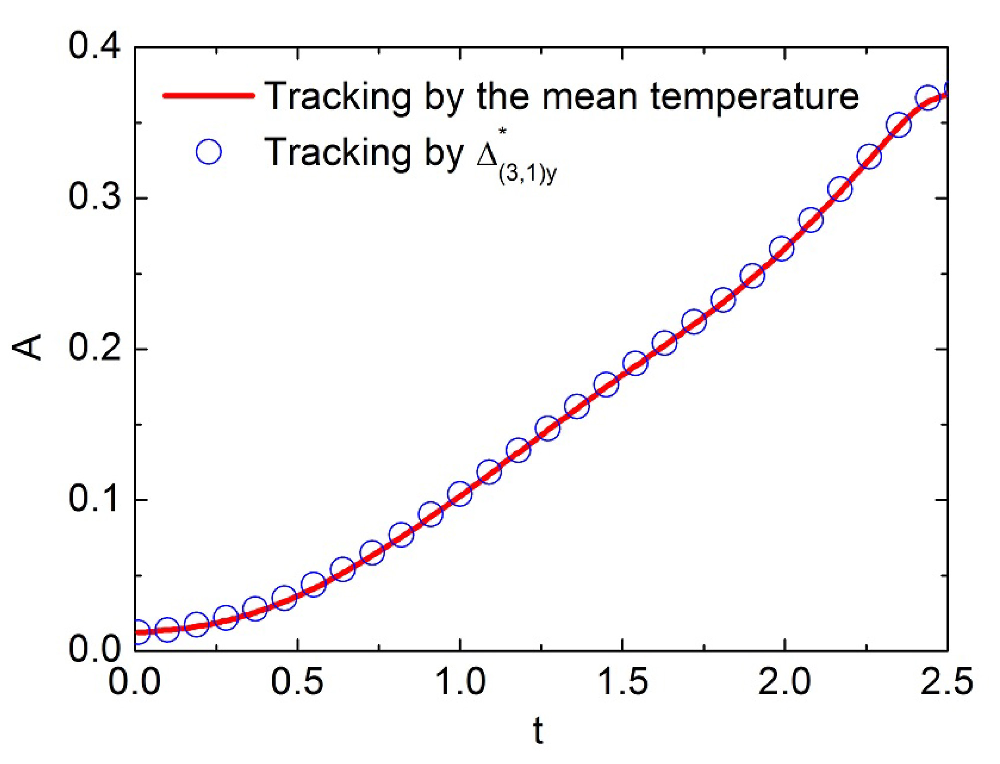}
\caption{ (From Ref. ~\cite{Lai2016PRE} Fig. 5, with permission.)
Evolution curve of interface amplitude obtained from two kinds of tracking techniques.
} \label{fig0037}
\end{figure*}

There generally more than one kinds of interfaces in complex flows. 
The TNE feature of an interface is its identity.
By analyzing the TNE effect, the finer physical structure of the interface can be obtained, which can be used for the design of physical identification and tracking technology of different interfaces in the system.
Figure \ref{fig0035} shows the characteristics of non-equilibrium quantities at the rarefaction wave, the material interface, and the shock front when the shock wave propagates from the heavy fluid to the light fluid.
These distinct characteristics of these quantities can be used to discriminate different types of interfaces~\cite{Lin2014PRE}.
Figure \ref{fig0036} shows the contour of TNE quantity at the interface of the RTI system~\cite{Lai2015}.
It is evident that TNE quantity is highly correlated with the macroscopic quantity gradient at the interface.
Therefore, the distribution of TNE can well represent the characteristics of the interface and then can be used to identify the interface.
As Fig. \ref{fig0037} shows, the evolution of interface amplitude captured with the $\Delta_{(3,1)y}^{*}$ matches the interface captured with the average temperature~\cite{Lai2016PRE}.

\section{Summary and prospect}\label{Summary and prospect}

The research content presented in this article is primarily motivated by two sets of scientific problems.
These two sets of scientific problems can also be said to be  ``two dark clouds'' faced in this cross-over field at that time. Each dark cloud is composed of a set of scientific problems.
The first set of scientific questions are related to the LBM method: 
what exactly can LBM do as a system of method other than restore known governing equations (or as a revolutionary discrete format for known governing equations)? Can it be really beyond traditional fluid theory? If so, where and how is it beyond? If still not, in which way can it be hopeful to be truly beyond?
The second set of scientific problems, independent of LBM and originated from the study of complex medium kinetics, correspond to the set mentioned in section \ref{sec:Problems-challenges}.
Thinking about the first set of scientific problems becomes one of the methodological backgrounds and basis for the authors to seek solutions to the second set of scientific problems.

Compared to macroscopic (large-scale and slow-changing) behavior, the description of mesoscale behavior requires more physical quantities.
The typical characteristics of mesoscale behavior are  significant discrete effects and significant thermodynamic non-equilibrium effects.
Therefore, before simulation, the ability to describe these discrete effects and non-equilibrium effects is the basic requirement for selecting/developing physical models.
After simulation, it is the basic requirement of the analysis method to construct appropriate physical quantities to detect and identify these discrete effects and non-equilibrium effects, and then to give an intuitive geometric image correspondence.
The discrete Boltzmann method is a physical model construction method and complex physical field analysis method developed based on the discrete Boltzmann equation for a series of fundamental scientific problems involved in the study of mesoscale behavior. It is the specific application and further development of coarse-grained description method, non-equilibrium behavior description method, and phase space description method  in statistical physics under the framework of discrete Boltzmann equation.
From a historical perspective, DBM was developed from the physical modeling branch of LBM, undergoing a process of taking, abandoning and adding.
It aims at the two major problems of how to model and how to analyze in numerical experimental research.
It no longer relies on the continuity assumption and near equilibrium approximation of traditional fluid modeling, and it abandons the ``lattice gas'' images of standard LBM.
It adds detection, presentation, description, and analysis solutions based on phase space for discrete/TNE states and effects.
Over time and according to the research needs, more information extraction techniques and complex physical field analysis schemes will be introduced.

Although the flow behavior targeted is complex, the DBM approach is inherently simple. As far as the modeling method is concerned, it only needs to check the kinetic moments of $f^{(0)}$ involved in the more accurate constitutive relationships according to the orders of tensors, and determine them as the kinetic moments that needs to preserve value, instead of deriving extremely complex extended hydrodynamic equations like the KMM approach. As far as the analysis methods are concerned, they are simply a concrete application and development of the basic principles of statistical physics.  They have clear physical images, are easy to understand and convenient to use. The model equations of DBM include the system behavior evolution equation(s) and discrete velocity constraint equations, where the evolution equations, in addition to discrete Boltzmann equation, depending on the specific situation, may also include phase field evolution equation, chemical reaction evolution equation, electromagnetic field evolution equation, etc. With the model equations, if further with the concrete discrete scheme, numerical simulation can be carried out. With the model equations, if without specific discrete format, the Chapman-Enskog multi-scale analysis and subsequent analytical studies can be performed.

The ``mesoscopic" characteristics of DBM  can be summarized as follows: (i) In terms of scale, it connects micro and macro scales. (ii) In terms of functionality, it is beyond the NS but is weaker than molecular dynamics in describing the discrete behavior and non-equilibrium behavior.
It should be pointed out that, compared with traditional fluid modeling, what the kinetic modeling brings are the improvement of understanding level, the broadening of application range, and the supplement of technology, but not a replacement.
For the cases where the conventional fluid modeling is convenient, effective, and adequate, it remains the first choice for most practical applications.
Because the associated theory, method, software and so on are more mature.
The user's own convenience is also an important factor in decision-making process.

The mesoscale behaviors promote the emergence of new technologies.
The studies on mesoscale behavior have made gratifying progress, but there is still a long way to go.
Fortunately, experimental studies on non-equilibrium flow are attracting increasing attention~\cite{Chen2022R, Bao2023R, Zhu2023R, Ai2023R, Chen2021PoG, Chen-Hu-Wang2023R, Chen-Translation,Danehy2013}.
Due to various limitations of measurements for non-equilibrium effects, progress in experimental study has been relatively slower.
This is precisely why numerical simulation studies have surged ahead of experiments, and it is also why numerical simulation should be accorded priority  in many studies of non-equilibrium flow.
Some practical applications of discrete/non-equilibrium effects have first been realized in numerical experimental studies. There will be more of these discoveries and applications over time. Numerical research provides cognitive basis for experimental research and engineering applications, which is the value of numerical simulation.

In addition to the relative discreteness caused by rarefaction or small structure and the fast flow in the case of no reaction or slow reaction, both the real gas effect at high temperature and fast chemical reaction can also cause the thermodynamic non-equilibrium flow.  In addition to the further study of discrete and non-equilibrium effects, DBM modeling in the latter two cases and related non-equilibrium regulation techniques are meaningful research directions that need to be explored in the near future. 

Astrophysics is one of the important application fields of fluid mechanics. Since many of the extreme conditions, behaviors, or physical images of high energy density physics techniques like ICF can be found in astrophysics, in addition to pure astrophysicists, scientists in fields related to high energy density physics techniques (such as ICF-related scientists) are often interested in astrophysics related research. Using fluid mechanics to model astrophysical processes has many specific applications, such as the evolution of stars, planets, accretion disks, galaxies, etc. Star behaviors are divided into atmospheric and internal convection regions, and accretion disk behaviors are divided into protostellar disks, black hole disks, binary disks, and so on. At present, the fluid mechanics simulations in astrophysics are mainly the NS simulation, Smoothed Particle Hydrodynamics (SPH) simulation and Particle In Cell (PIC) simulation. SPH and PIC simulations have a certain degree of kinetics simulation characteristics, but mainly focus on physics images of NS. The NS continuous description corresponds to the fact that we are far enough away from that celestial structure so that the field of view is large enough, and consequently the distance between the stars is negligible relative to the scale of our field of view. As our spacecraft's observing instruments get closer and the scale of the field of view decreases, we will gradually observe larger and larger gaps and more inhomogeneity between the stars including different sizes of the stars, different translational and rotational speeds, etc. The existence of gravitation, etc., makes these asymmetries often imply non-equilibrium. Discreteness and non-equilibrium begin to be observed, which is the benefit of improved observation accuracy. The gradual approach of our spacecraft's observing instruments corresponds to a gradual increase in the Knudsen number, because the scale corresponding to our observing field of view is gradually decreasing. DBM simulations can show changes in the view as our spacecraft's observing instruments get closer. Of course, it is presented in the form of statistical features to the extent that DBM theory is valid.

Dust gas kinetics and granular flow kinetics have important engineering application background and prospect. “Particles” in dust gases and granular flows are “macroscopic” particles made up of a large number of molecules. Collisions between such particles are not elastic, but dissipative. Therefore, dust gas kinetics and granular flow kinetics are more complex than the usual complex fluids. The study of this kind of complex fluid system is also divided into macroscopic, mesoscopic and microscopic levels, facing the meso-scale dilemma of“insufficient physical function of traditional fluid modeling and impossibility of molecular dynamics simulation in terms of scale”. As long as the generalized BGK-like Boltzmann equation including particle collision dissipation effect can be obtained, the DBM modeling and analysis method introduced in this review can be generalized and applied. The theory of dust kinetics and granular flow kinetics, although more complex and beyond the discussion scope of the current review, has a history of nearly 30 years. Interested readers are referred to more specialized literature. Everything flows. As long as traditional fluid mechanics has applications, there will be kinetic behaviors that traditional fluid mechanics fails to well describe,and DBM can find a reasonable application. Although the former can also be studied using the long-established NS simulation, DBM can achieve a certain degree ofcross-scale simulation under the same framework,thereby avoiding the problems that arise when different scale methods are interfaced.

The previous sections of this article do not cover the discussion of specific discrete format, because that is another field. 
Finally, just present a brief comment from the perspective of application. Any form of discretization of an originally continuous space will inevitably bring about symmetry breaking. In numerical simulation, 
symmetry breakings from different dimensions are coupled together to make effects.
In the study of specific discrete format, not only the numerical accuracy but also the numerical stability should be guaranteed. In many practical applications, numerical stability is more important than numerical accuracy. 
The numerical stability depends not only on the discrete scheme, but also on the specific flow behavior.
Because the traditional fluid modeling works in time-position space, the specific discrete format of time-position space has been deeply and systematically studied in the traditional computational fluid dynamics. With the emergence of kinetic methods using discrete particle velocities such as lattice gas method, LBM and DBM, the study of stable discrete format in time-position-velocity space has become an urgent research topic in computational mathematics. For more details on discrete format study, please refer to the more specialized literature.

\acknowledgements{
The authors thank Academician Shigang Chen for his academic guidance and training for so many years, thank Academician Jinghai Li and Professor Limin Wang, et al. for their enlightening comments, discussions and suggestions. Dr. Jiahui Song organized some materials for the current review article.
This work was supported by the National Natural Science Foundation of China  (Grant Nos. 12172061 and 11875001),  the opening project of State Key Laboratory of Explosion Science and Technology (Beijing Institute of Technology) (Grant No. KFJJ23-02M), Foundation of National Key Laboratory of Shock Wave and Detonation Physics (Grant No. JCKYS2023212003), Foundation of National Key Laboratory of Computational Physics, 
Hebei Outstanding Youth Science Foundation (Grant No. A2023409003), Hebei Natural Science Foundation (Grant No. A2021409001), Central Guidance on Local Science and Technology Development Fund of Hebei Province (Grant No. 226Z7601G), and “Three, Three and Three Talent Project” of Hebei Province (Grant No. A202105005).}

\section*{Appendix A: Rationality, accuracy, and efficiency of DBM results} \label{sec:AppendixesA}

The accuracy of the simulation results depends on both the accuracy of the physical model and the accuracy of the numerical algorithm. Problems in physical model cannot be solved by improving the accuracy of numerical algorithm. 
Given the authors' background in physics, the accuracy discussed herein is predominantly focuses on the accuracy of the physical model. 
The comparison of results presented in this paper primarily evaluates the reasonableness of the physical model, rather than the accuracy of the numerical algorithm. Physically real results are stable and invariable within the range of numerical error for any reasonable numerical algorithm.

The comparison of computation cost needs to be based on the premise that both models possess identical physical function. This paper focuses on the \emph{mesoscale} dilemmas in which the macroscopic continuous model falters, but the molecular dynamics method is unable to simulate because of the scale problem.

How to test the rationality of such models?

First of all, just as the theory of special relativity, which mainly addresses macroscopic high-speed situations, needs to be able to smoothly revert to Newtonian mechanics theory in macroscopic low-speed situations, the DBM, which mainly focuses on the aforementioned mesoscale dilemma, needs to smoothly return to traditional fluid Navier-Stokes theory in near continuous and near equilibrium situations. This is the first test for DBM.

As the degree of discreteness/non-equilibrium increases, the theoretical basis of Navier-Stokes begins to be challenged. How can the rationality of DBM results be tested or guaranteed? Given that engineering applications generally use macroscopic continuous modeling, yet encounter some dissatisfaction, our first consideration is: how can mesoscale modeling be ``seamless" with macroscopic continuous modeling? Therefore, the starting point for DBM is the side of the mesoscale range near the macro. Due to the stage of development, DBM primarily considers cases where Chapman-Enskog (CE) multi-scale analysis theory remains valid. So, CE multi-scale analysis theory is the effective mathematical guarantee for this set of ideas and methods.

In the absence of experimental or simulation result comparisons, the rationality test largely relies on theoretical analysis. 
In practical scenarios, a common testing method involves identifying the physical mechanism behind characteristic behaviors and examining whether these behaviors exhibit self-consistency from various perspectives. The theoretical basis is that any fundamental flaw will inevitably lead to inconsistencies in some way.

Just as in situations where Newtonian mechanics is sufficient, special relativity theory can be used, but is generally no longer the first choice because it is more complex, DBM can also be used in situations where traditional fluid Navier-Stokes theory is convenient and sufficient, but is no longer necessary. As for the final choice, it depends on the user's own convenience.

The computation cost depends on the chosen discrete format and algorithm. Different formats and algorithms can significantly alter the actual computational workload. In general, however, DBM simulations might demand more computationally resources in situations where traditional fluid Navier-Stokes theory suffices due to its expanded physical capability. Its physical capability is in between Navier-Stokes and molecular dynamics. In the cases where all three of them are physically reasonable and available, its computational cost is also in between them two. In short, the function is ``mesoscopic" and the cost is also ``mesoscopic".  That is fair and reasonable.

DBM considers Navier-Stokes as a special case in its quasi-continuous near equilibrium situation. However, its primary target is not to replace Navier-Stokes, but to reveal the kinetic behavior characteristics missed by Navier-Stokes, provide more methods for analyzing complex physical fields, and explore the discrete, non-equilibrium complex flows that Navier-Stokes is no longer accurate and/or effective.

Even when traditional Navier-Stokes theory suffices, DBM's complex physical field analysis method can still be utilized to analyze simulation data.
For example, one can open phase space based on different behavior characteristics to provide an intuitive geometric image corresponding to the set of behavior characteristics, and can also construct non-equilibrium strength vector, each of whose component representing a non-equilibrium strength from a perspective, to multi-channel cross-locate the non-equilibrium strength of the complex flow system.

\section*{Appendix B: Nonlinear constitutive relations} \label{sec:AppendixesB}

For clarity, the first two order constitutive relations (i.e., the expressions of $\bm{\Delta}_{2}^{*}$ and $\bm{\Delta}_{3,1}^{*}$) based on the BGK collision model and derived from CE multi-scale analysis are given below.
The first order expressions of $\bm{\Delta}_{2}^{*(1)}$ and $\bm{\Delta}_{3,1}^{*(1)}$ are
\begin{equation}
{\bm{\Delta }}_2^{*(1)}  =  - \mu \left[ {\bm{\nabla} {\mathbf{u}} + (\bm{\nabla} {\mathbf{u}})^{\text{T}}  - \frac{2}
{{l + 2}}{\mathbf{I}}\bm{\nabla}  \cdot {\mathbf{u}}} \right]
,
\end{equation}
\begin{equation}
{\bm{\Delta }}_{3,1}^{*(1)}  =  - \kappa \bm{\nabla} T
,
\end{equation}
where $\mu=\tau p$ represents the dynamic viscosity coefficient, $\kappa$ denotes the heat conduction coefficient, and the scalar quantity $l$ indicates number of additional degree of freedom.
For convenience, we refer $\bm{\Delta}_{2}^{*(j)}$ to the $j$-th order term of $\bm{\Delta}_{2}^{*}$.
When a DBM considers up to the $j$-th order non-equilibrium effect [i.e., calculating $\bm{\Delta}_{2}^{*}$ up to $\bm{\Delta}_{2}^{*(j)}$ and $\bm{\Delta}_{3,1}^{*}$ up to $\bm{\Delta}_{3,1}^{*(j)}$], the retained DBM is termed the $j$-th order DBM.
The following part gives the expressions of $\bm{\Delta}_{2}^{*(2)}$ and $\bm{\Delta}_{3,1}^{*(2)}$, with detailed derivation available in Ref. ~\cite{Gan2018PRE}.
The expressions are:
\begin{equation}
\begin{gathered}
  \Delta _{2xx}^{ * (2)}  = {\text{ }}2l_2^{ - 2} \tau ^2 \{ \rho RT[l_{ - 2} l_1 \left( {\partial _x u_x } \right)^2  + l_1 l_2 \left( {\partial _y u_x } \right)^2  \\
   - 4I\,\partial _x u_x \partial _y u_y  - l_2 \left( {\partial _x u_y } \right)^2  - l_{ - 2} \,\left( {\partial _y u_y } \right)^2 ] \\
   + \rho R^2 [l_1 l_2 \left( {\partial _x T} \right)^2  - l_2 \left( {\partial _y T} \right)^2 ] - R^2 T^2 [l_1 l_2 \frac{{\partial ^2 }}
{{\partial x^2 }}\rho  - l_2 \frac{{\partial ^2 }}
{{\partial y^2 }}\rho ] \\
   + \frac{{R^2 T^2 }}
{\rho }[l_1 l_2 (\partial _x \rho )^2  - l_2 (\partial _y \rho )^2 ]\},  \\
\end{gathered}
\end{equation}
\begin{equation}
\begin{gathered}
  \Delta _{2xy}^{ * (2)}  = 2\tau ^2 [\,l_2^{ - 1} \rho T(l\partial _x u_x \partial _x u_y  + l\partial _y u_x \partial _y u_y  - 2\partial _x u_y \partial _y u_y  \\
  - 2\partial _x u_x \partial _y u_x ) + \rho R^2 \partial _x T\partial _y T \\
  - R^2 T^2 \frac{{\partial ^2 }}
{{\partial x\partial y}}\rho  + \frac{{R^2 T^2 }}
{\rho }\partial _x \rho \partial _y \rho ], \\
\end{gathered}
\end{equation}
\begin{equation}
\begin{gathered}
  \Delta _{2yy}^{ * (2)}  =  - 2l_2^{ - 2} \tau ^2 \{ \rho RT[l_{ - 2} \left( {\partial _x u_x } \right)^2  + l_2 \left( {\partial _y u_x } \right)^2  \\
  + 4I\,\partial _x u_x \partial _y u_y  \\
   - l_1 l_2 \left( {\partial _x u_y } \right)^2  - l_{ - 2} l_1 \,\left( {\partial _y u_y } \right)^2 ] \\
   + \rho R^2 [l_2 \left( {\partial _x T} \right)^2  - l_1 l_2 \left( {\partial _y T} \right)^2 ] - R^2 T^2 [l_2 \frac{{\partial ^2 }}
{{\partial x^2 }}\rho  - l_1 l_2 \frac{{\partial ^2 }}
{{\partial y^2 }}\rho ] \\
   + \frac{{R^2 T^2 }}
{\rho }[l_2 (\partial _x \rho )^2  - l_1 l_2 (\partial _y \rho )^2 ]\},  \\
\end{gathered}
\end{equation}
\begin{equation}
\begin{gathered}
  \Delta _{3,1x}^{ * (2)}  = l_2^{ - 1} \tau ^2 \{ \rho R^2 T^2 [l_{ - 2} \frac{{\partial ^2 }}
{{\partial x^2 }}u_x  + l_2 \frac{{\partial ^2 }}
{{\partial y^2 }}u_x  - 4\,\frac{{\partial ^2 }}
{{\partial x\partial y}}u_y ] \\
   + \rho R^2 T[(l_2^2  + 4l)\partial _x u_x \partial _x T + l_2 l_6 \partial _y u_x \partial _y T \\
- 2l_6 \partial _y u_y \partial _x T + 2l_2 \partial _x u_y \partial _y T\,]\},  \\
\end{gathered}
\end{equation}
\begin{equation}
\begin{gathered}
  \Delta _{3,1y}^{ * (2)}  = l_2^{ - 1} \tau ^2 \{ \rho R^2 T^2 [l_2 \frac{{\partial ^2 }}
{{\partial x^2 }}u_y  + l_{ - 2} \frac{{\partial ^2 }}
{{\partial y^2 }}u_y  - 4\,\frac{{\partial ^2 }}
{{\partial x\partial y}}u_x ] \\
   + \rho R^2 T[(l_2^2  + 4l)\partial _y u_y \partial _y T + l_2 l_6 \partial _x u_y \partial _x T - 2l_6 \partial _x u_x \partial _y T\, \\
   + 2l_2 \partial _y u_x \partial _x T]\},  \\
\end{gathered}
\end{equation}
where $l_a = l +a$.
The formulations of $\bm{\Delta}_{2}^{*(j)}$ and $\bm{\Delta}_{3,1}^{*(j)}$ with $j > 2$ are so complex, presenting significant complexity in their derivation.
This indicates that the KMM modeling and simulation approach rapidly becomes infeasible as the degree of discreteness/non-equilibrium increases.



\begin{small}

\end{small}

\end{document}